\documentclass[12pt,reqno]{amsart}
 
\usepackage{amssymb,amsxtra}
\usepackage{amsfonts}
\usepackage{amsmath} 
\usepackage{graphicx} 
\usepackage{layout}

\newcommand{\ba}{\begin{eqnarray}} 
\newcommand{\ea}{\end{eqnarray}}
\newcommand{\be}{\begin{equation}}
 \newcommand{\ee}{\end{equation}}
\newcommand{\bdm}{\begin{displaymath}}
\newcommand{\edm}{\end{displaymath}} 
\newcommand{\brr}{\begin{array}}
\newcommand{\err}{\end{array}}

\newcommand{\bml}{\begin{gather}} \newcommand{\eml}{\end{gather}}

\newcommand{\spaz}{\vspace{.5cm} \noindent}

\renewcommand{\(}{\left(}          \renewcommand{\)}{\right)}
\renewcommand{\[}{\left[}          \renewcommand{\]}{\right]}
\newcommand{\R}{{\mathbb R}}         
\newcommand{\var}{\varepsilon}
\newcommand{\ud}{\mathrm{d}}
\newcommand{\pa}{\partial}

\begin{document}
\begin{center}
{\Large\textbf{Mean-field limit and Semiclassical Expansion of a Quantum Particle System}}
\end{center}
\vspace{0.5cm}
\begin{center}
Federica Pezzotti\footnote{Dipartimento di Matematica Pura ed Applicata, Universit$\grave{\text{a}}$ dell'Aquila, Italy - 

\emph{e-mail:} federica.pezzotti@univaq.it},
Mario Pulvirenti\footnote{Dipartimento di Matematica ''G. Castelnuovo'', Universit$\grave{\text{a}}$ di Roma ''La Sapienza'', Italy - 

\emph{e-mail:} pulvirenti@mat.uniroma1.it }
\end{center}
%\begin{center}
% {\footnotesize Dipartimento di Matematica Pura
%ed Applicata, Universit\`{a} degli Studi de l'Aquila, Italy, federica.pezzotti@univaq.it; Dipartimento di Matematica ''Guido Castelnuovo'', Universita' degli Studi di Roma ''La Sapienza'', pulvirenti@mat.uniroma1.it}
%\end{center}
\vspace{0.5cm}
\begin{center}
\textbf{Abstract}
\end{center}

{\footnotesize We consider a quantum system constituted by $N$ identical particles interacting by means of a mean-field Hamiltonian. It is well known that, in the limit $N\to\infty$, the one-particle state obeys to the Hartree equation. Moreover, propagation of chaos holds. In this paper, we take care of the $\hbar$ dependence by considering the semiclassical expansion of the $N$-particle system. We prove that each term of the expansion agrees, in the limit $N\to\infty$, with the corresponding one associated with the Hartree equation. We work in the classical phase space by using the Wigner formalism, which seems to be the most appropriate for the present problem.}
\tableofcontents
\section{Introduction}
\setcounter{equation}{0}    
\def\theequation{1.\arabic{equation}}
%{\bf Remark:}\\
%statistics (vedi Spohn) + law of large numbers \\\\
The Hartree equation is the following nonlinear one-particle Schr$\ddot{\text{o}}$dinger equation:
\begin{equation}\label{eq: hartree}
i\hbar\pa_t\psi=-\frac{\hbar^{2}}{2}\Delta\psi + \left(\phi \ast \vert\psi\vert^{2}\right)\psi,
\end{equation}
where the mass of the particle is chosen equal to one, $\phi:\R^3\rightarrow \R$ is the two-body potential and
\begin{equation}\label{eq: hartreeINT}
\left(\phi \ast \vert\psi\vert^{2}\right)(x)=\int_{\R^3}\ud y \phi(x-y)\vert \psi(y)\vert^2
\end{equation}
is the effective self-consistent interaction.
\\
Equation (\ref{eq: hartree}) arises as the reduced description of a system of $N$ identical particles  interacting by means of the mean-field potential:
\begin{equation}\label{eq: meanINT}
U_N\left(X_N\right)=\frac{1}{N}\sum_{\substack{1\leq l<j\leq N}}^{N}\phi(x_l-x_j),
\end{equation}
(where $X_N=\{x_1,\dots,x_N\},\ x_j\in\R^3,\ j=1,\dots, N $) in the limit $N\to\infty$.\\ In fact, consider the $N$-particle wave function $\Psi_N=\Psi_N\left(X_N;t\right)$ solution of the Schr$\ddot{\text{o}}$dinger equation:
\begin{equation}\label{eq: mean}
i\hbar\pa_t\Psi_N=-\frac{\hbar^{2}}{2}\sum_{i=1}^{N}\Delta_{x_i}\Psi_N + U_N\Psi_N,
\end{equation}
with a completely factorized initial state given by:
\begin{equation}\label{eq: meanINDAT}
\Psi_N(X_N;0)=\Psi_{N,0}(X_N)=\prod_{j=1}^{N}\psi_0(x_j).
\end{equation}
Then, it is well known that
%, under suitable assumptions on the interaction potential $\phi$, 
the $j$-particle reduced density matrices, defined as:
\begin{equation}\label{eq: redDENSmatrN}
\rho_j^N\left(X_j,Y_j;t\right)=\int_{\R^{3(N-j)}} \ud X_{N-j} \Psi_N\left(X_j, X_{N-j};t\right)\overline{\Psi}_N\left(Y_j, X_{N-j};t\right),
\end{equation}
converges, in the limit $N\to\infty$ and for any fixed $j=1,\dots,N$, to the factorized state:
\begin{equation}\label{eq: redDENSmatrJlim}
\rho_j\left(X_j,Y_j;t\right)=\prod_{k=1}^{j}\psi\left(x_k;t\right)\overline{\psi}\left(y_k;t\right),
\end{equation}
where $\psi(x;t)$ solves the one-particle 
%(but non linear)
Hartree equation (\ref{eq: hartree}) with initial datum $\psi_0$. This feature is usually called ''propagation of chaos''.
 
The previous result was originally obtained for sufficiently smooth potentials (see \cite{SPOHN}, \cite{HEPP}, \cite{GV}); 
%\cite{GV1}
 then it has been %successively 
generalized to include Coulomb interactions (see \cite{BGM}, \cite{EY}, \cite{BEGMY}). Furthermore, some results concerning the speed of convergence of the mean-field evolution to the Hartree dynamics (for all fixed times), have been proven more recently (see \cite{BENJIROD}, \cite{ErdBenji}). 

The limit $N\to\infty$ for a classical system interacting by means of the same mean-field interaction (\ref{eq: meanINT}), can be considered as well (see \cite{McK}, \cite{BH}, \cite{Neun}, \cite{Sp},\cite{DOB} for the case of smooth potential, and \cite{JABIN} for more singular interaction).
%, \cite{Sp}, \cite{DOB}). 
In fact, considering as initial state of the system a probability measure $F_{N,0}=F_{N,0}\left(X_N,V_N\right)$ in the $N$-particle phase space which is completely factorized, namely:
\begin{equation}\label{eq: classicINDAtN}
F_{N,0}\left(X_N,V_N\right)=
%F_{N,0}\left.\left(X_N,V_N;t\right)\right\vert_{t=0}=
\prod_{j=1}^{N}f_0\left(x_j, v_j\right),
\end{equation}
where $f_0$ is a given one-particle probability density, it is known that its evolution at time $t>0$, denoted by $F_{N}\left(X_N,V_N;t\right)$, is obtained by solving the Liouville equation:
\begin{equation}\label{eq: LiouvilleEQ}
\left(\pa_t +V_N\cdot \nabla_{X_N}\right)F_{N}=\nabla_{X_N}U_N\cdot \nabla_{V_N}F_{N}.
\end{equation}
Then, the $j$-particle distribution at time $t>0$, defined as 
\begin{equation}\label{eq: redCLASS}
f_j^N\left(X_j,V_j;t\right)=\int_{\R^{3(N-j)}\times \R^{3(N-j)}} \ud X_{N-j}\ud V_{N-j} F_N\left(X_j, V_j, X_{N-j}, V_{N-j};t\right),
\end{equation}
%$F_N\left(\cdot\ ;t\right)$ being the time evolved $N$-particle probability measure (solution of the Liouville equation with initial datum $F_{N,0}$), 
converges, as $N\to\infty$, to the product state:
\begin{equation}\label{eq: prodCLASS}
f_j\left(X_j,V_j;t\right)=\prod_{k=1}^{j}f\left(x_k, v_k;t\right),
\end{equation}
where $f(x,v ;t)$ is the solution of the Vlasov equation:
\begin{equation}\label{eq: Vlasov}
\left(\pa_t+v\cdot \nabla_x\right)f=\left(\nabla\phi\ast f \right)\cdot\nabla_v f,
\end{equation}
(the convolution above is with respect to both the variables $x$ and $v$) with initial datum $f_0$.\\ Equation (\ref{eq: Vlasov}) is the classical analogous of the Hartree equation (\ref{eq: hartree}).
%(see \cite{DOB} CONTROLLA SE MANCA UNA REF DI Dob).\\

Although the mean-field limit $N\to\infty$ is well understood for both classical and quantum systems, 
%(at least for smooth potentials)
there is a question which seems to be still open, namely, does that limit hold for quantum systems uniformly in $\hbar$, at least for systems having a reasonable classical analogue? 
%Except for the one presented in \cite{FGS},\cite{FPK} CONTROLLA for a restricted class of two-body interactions, 
\\The proofs which are available up to now exhibit an error vanishing when $N\to\infty$ but diverging as $\hbar\to 0$, although in \cite{NS}, \cite{Graffi}, \cite{FGS}, \cite{FPK} some efforts in the direction of a better control of the error term have been done. 
%On the other hand, the same limit holds for $\hbar=0$, then it is natural to look for a better error bound as regard to the dependence on $\hbar$.\\

If one wants to deal with the classical and quantum case simultaneously, it is natural to work in the classical phase space by using the Wigner formalism. 

The one-particle Wigner function associated with the wave function $\psi(x;t)$ is given by:
\begin{equation}\label{eq: WignerUNA}
f^{\hbar}\left(x, v;t\right)=(2\pi)^{-3}\int_{\R^{3}} \ud y\ e^{iy\cdot v}\overline{\psi}\left(x+\frac{\hbar y}{2};t\right)\psi\left(x-\frac{\hbar y}{2};t\right),
\end{equation}
and, similarly, the $N$-particle Wigner function associated with the wave function $\Psi_N(X_N;t)$ is defined as:
\begin{equation}\label{eq: WignerN}
W_N^{\hbar}\left(X_N, V_N;t\right)=(2\pi)^{-3N}\int_{\R^{3N}} \ud Y_N \ e^{iY_N\cdot V_N}\overline{\Psi}_N\left(X_N+\frac{\hbar Y_N}{2};t\right)\Psi_N\left(X_N-\frac{\hbar Y_N}{2};t\right).
\end{equation}
Then, by using that $\psi(x;t)$ and $\Psi_N(X_N;t)$ solve equations (\ref{eq: hartree}) and (\ref{eq: mean}) respectively, we find the equations:
\begin{equation}\label{eq: WigLIOUhartree1}
\left(\pa_t+v\cdot \nabla_x\right)f^{\hbar}=T^{\hbar}f^{\hbar}
\end{equation}
and
\begin{equation}\label{eq: WigLIOUhartreeN}
\left(\pa_t+V_N\cdot \nabla_{X_N}\right)W_N^{\hbar}=T_N^{\hbar}W_N^{\hbar},
\end{equation}
where $T^{\hbar}$ and $T_N^{\hbar}$ are suitable pseudodifferential operators. 

The initial data for equations (\ref{eq: WigLIOUhartree1}) and (\ref{eq: WigLIOUhartreeN}) are
\begin{equation}\label{eq: WignerUNAindat}
%f^{\hbar}\left.\left(x, v;t\right)\right\vert_{t=0}=
f_0^{\hbar}\left(x, v\right)=(2\pi)^{-3}\int_{\R^{3}} \ud y\ e^{iy\cdot v}\overline{\psi}_0\left(x+\frac{\hbar y}{2}\right)\psi_0\left(x-\frac{\hbar y}{2}\right),
\end{equation}
and
\begin{eqnarray}\label{eq: WignerNindat}
%W_N^{\hbar}\left.\left(X_N, V_N;t\right)\right\vert_{t=0}=
W_{N,0}^{\hbar}\left(X_N, V_N\right)&&=(2\pi)^{-3N}\int_{\R^{3N}} \ud Y_N \ e^{iY_N\cdot V_N}\overline{\Psi}_{N,0}\left(X_N+\frac{\hbar Y_N}{2}\right)\Psi_{N,0}\left(X_N-\frac{\hbar Y_N}{2}\right)\nonumber\\
&& = \prod_{j=1}^{N}f_0^{\hbar}\left(x_j, v_j\right),
\end{eqnarray}
respectively.

One can easily rephrase the result of \cite{SPOHN} by showing that
%, provided that the two-body interaction $\phi$ is sufficiently smooth, 
the $j$-particle Wigner function
\begin{equation}\label{eq: WignerJN}
W_{N,j}^{\hbar}\left(X_j, V_j;t\right)=\int_{\R^{3(N-j)}\times\R^{3(N-j)}} \ud X_{N-j} \ud V_{N-j}W_N^{\hbar}\left(X_j, X_{N-j},V_j,V_{N-j};t\right)
\end{equation}
converges, in a suitable sense, to
\begin{equation}\label{eq: WignerJlim}
f_j^{\hbar}\left(X_j, V_j;t\right)=\prod_{k=1}^{j}f^{\hbar}\left(x_k, v_k;t\right)\ \ \text{for any}\ t>0.
\end{equation}
%(see (\cite{FGS}, \cite{FPK}, \cite{NS} CONTROLLa GLI ultimi DUE...)). 
However, the convergence error is diverging when $\hbar\to 0$ (for example, for sufficiently small times $t<t_0$ it is of the form $\frac{C_j}{N}e^{\frac{c}{\hbar}}$). The reason is that the operators $T^{\hbar}$ and $T_N^{\hbar}$ appearing in (\ref{eq: WigLIOUhartree1}) and (\ref{eq: WigLIOUhartreeN}) are bounded as operators acting on the space in which we can prove the convergence of (\ref{eq: WignerJN}), but their norm diverge as $\frac{c}{\hbar}$ when $\hbar\to 0$. On the other hand, the classical counterpart of this problem has been solved,
so that it seems natural to look for an asymptotic expansion for the $j$-particle distributions $W_{N,j}^{\hbar}$, namely:
\begin{equation}\label{eq: WignerJNexp}
W_{N,j}^{\hbar}\left(t\right)=W_{N,j}^{(0)}\left(t\right)+\hbar W_{N,j}^{(1)}\left(t\right)+\hbar^2 W_{N,j}^{(2)}\left(t\right)+\dots,
\end{equation}
and for an analogous expansion for the $j$-fold product of solutions of the equation (\ref{eq: WigLIOUhartree1}), namely:
\begin{equation}\label{eq: WignerJlimexp}
f_j^{\hbar}\left(t\right)=\left(f^\hbar\right)^{\otimes j}(t)=f_j^{(0)}\left(t\right)+\hbar f_j^{(1)}\left(t\right)+\hbar^2 f_j^{(2)}\left(t\right)+\dots
\end{equation}
The zero order term in (\ref{eq: WignerJNexp}) corresponds properly to what we previously denoted by $f_j^N$, while the function $f_j^{(0)}$ appearing in (\ref{eq: WignerJlimexp}) is exactly the $j$-fold product of what we called $f$ (see (\ref{eq: redCLASS}) and (\ref{eq: Vlasov})). Therefore, at zero order in $\hbar$ we obtained the classical quantities, as expected, and we know that the convergence of $W_{N,j}^{(0)}$ to $f_j^{(0)}$ is well established. Then, it looks natural to try to show the convergence
\begin{equation}\label{eq: convergence}
W_{N,j}^{(k)}\left(t\right)\to f_j^{(k)}\left(t\right),\ \ \text{as $N\to\infty$, for any}\ k>0.
\end{equation}
This is the goal of the present paper. 

A complete proof of the uniformity in $\hbar$ of the limit $N\to\infty$ would require a control of the remainder of the expansion (\ref{eq: WignerJNexp}), but we are not able to do this. However the error term of order $k$ in the expansion (\ref{eq: WignerJlimexp}) can be proven to be $O(\hbar^{k+1})$ by adapting the proof in \cite{PULVIRENTI} for the linear case to the present context, under suitable smoothness assumptions.
%Nonetheless, the term by term convergence of the expansion is interesting by itself because it implies that:
%, as a consequence of that, we are able to affirm that:
%\begin{equation}\label{eq: convergenceNUOVA1}
%\text{lim}_{N\to\infty}\text{lim}_{\hbar\to 0}W_{N,j}(t)= f_j^{(0)}\left(t\right),\ \ \text{as $N\to\infty$},
%\end{equation}
%and
%\begin{equation}\label{eq: convergenceNUOVA}
%\text{lim}_{N\to\infty}\text{lim}_{\hbar\to 0}\frac{W_{N,j}(t)-\sum_{n=0}^{k-1}\hbar^n W_{N,j}^{(n)}\left(t\right)}{\hbar^{k}}= f_j^{(k)}\left(t\right),\ \ \text{as $N\to\infty$, for}\ k\geq 1.
%\end{equation}
%\\

The plan of the paper is the following. In the next two sections we present a semiclassical expansion of the Hartree equation and of the $N$-particle system. After a brief discussion of the hierarchical structures and their inadequacy as a technical tool for the present problem, we introduce the classical mean-field limit which is the basis of our analysis. After that, we explain the strategy of the proof of the convergence (\ref{eq: convergence}). The last two sections are devoted to the precise statement of our result, its proof and supplementary comments. Three Appendices contain technicalities.

\section{The Hartree dynamics }
\setcounter{equation}{0}    
\def\theequation{2.\arabic{equation}}
The Wigner-Liouville equation associated with the Hartree equation (\ref{eq: hartree}) reads as:
\begin{equation}\label{eq: WigLIOUhartree}
\left\{
\begin{aligned}
&\left(\pa_t+v\cdot \nabla_x\right)f^{\var}=T_{f^{\var}}^{\var}f^{\var},\\
&\left.f^{\var}(x,v;t)\right\vert_{t=0}=f_0^{\var}(x,v),
\end{aligned}
\right.
\end{equation}
where, from now on, we set $\var=\hbar$. Furthermore $(x,v)\in \R^3\times\R^3$ and $f^{\var}(x, v;t)$ is defined as in (\ref{eq: WignerUNA}) and it is a real (but non necessarily positive) function on the classical phase space.
We remark that in the previous section we used the notation $T^{\var}f^{\var}$ for the right hand side of (\ref{eq: WigLIOUhartree}) 
%in order to underline the $\hbar$ dependence of the operator $T$; 
but now we use $T_{f^{\var}}^{\var}f^{\var}$ to stress the nonlinearity of equation (\ref{eq: WigLIOUhartree}).
%that the dependence of $T$ on $\hbar$ is due both to its structure than to its nonlinearity, namely, to its dependence on the function $f^{\hbar}$ 
%(we remind that the Hartree equation (\ref{eq: hartree}) is nonlinear, then, the corresponding Wigner-Liouville equation is nonlinear as well). 
%CONTROLLO DEF DI FOURIER TRANSFORM MA NON CI DOVREBBERO ESSERE $\pi$ DI TROPPO!! FORSE DEF SENZA NIENTE MA CONTROLLO! SICURAMENTE NELL'ESPRESSIONE DEI $T^{(n)}$ con le derivate non ci vanno $\pi$ (WIGNER usa def di Fourier transform senza nulla)Ma in quella che segue ci va!\\

Here $T^\var$ acts as follows: 
%CONTROLLO $\phi$!!! SEMBRA CHE CI VADA MA NON SONO SICURA PERCHE' A LIVELLO DI OPERATORE SI', SICURAMENTE!! MA SE LO PENSO APPLICATO A $f^{\hbar}$ CHE HA GIA' UN $(2\pi)^{-3}$....E IL $(2\pi)^{-3}$ che e' nella $g$???Credo non conti...
\begin{equation}\label{eq: operatorT}
T_g^{\var} f^{\var}(x, v)=(2\pi)^{-3}i\int_{-1/2}^{1/2}\ud \lambda\int_{\R^3}\ud k \hat{\phi}(k)\hat{\rho}_g(k) e^{i \ k\cdot x}(k\cdot \nabla_v)f^{\var}(x, v+\var\lambda k),
\end{equation}
where, as usual, $\phi=\phi(x)$ denotes the two-body interaction potential and
\begin{equation}\label{eq: rho}
\rho_g(x)=\int_{\R^3}\ud v \ g(x, v).
\end{equation}
%is the spatial density associated with the distribution $g$. 
Finally, we denoted by $\hat{\rho}_g$ and $\hat{\phi}$ the Fourier transforms of $\rho_g$ and $\phi$ respectively, namely:
\begin{equation}\label{eq: fourier}
\hat{\rho}_g(k)=\int_{\R^3}\ud x\ e^{-i\ k\cdot x} \rho_g(x) \ \ \text{and}\ \ \hat{\phi}(k)=\int_{\R^3}\ud x\ e^{-i\ k\cdot x} \phi(x).
\end{equation}

Following \cite{PULVIRENTI}, for a fixed $g$, $T_g^{\var}$ can be expanded as
\begin{equation}\label{eq: operatorTexp}
T_g^{\var}=T^{(0)}_g+\var T^{(1)}_g+\var^2 T^{(2)}_g+\dots
\end{equation}
where
\begin{equation}\label{eq: operatorT^nEVEN}
T^{(n)}_g =c_n(2\pi)^{-3}i\int_{\R^3}\ud k \hat{\phi}(k)\hat{\rho}_g(k) e^{i \ k\cdot x}(k\cdot \nabla_v)^{n+1},
\end{equation}
\begin{equation}\label{eq: operatorT^nCOSTANTE}
c_n=\frac{1}{2^n (n+1)!},
\end{equation}
for $n$ even and
\begin{equation}\label{eq: operatorT^nODD}
T^{(n)}_g =0,
\end{equation}
for $n$ odd. The operator $T^{(n)}_g$, for $n$ even, can be also written as
\begin{equation}\label{eq: operatorT^nEVEN1}
T^{(n)}_g =(-1)^{n/2}c_n\left(D_x^{n+1}\phi\ast g\right)\cdot D_v^{n+1},
\end{equation}
where, as in (\ref{eq: Vlasov}), $\ast$ denotes the convolution with respect to both $x$ and $v$ and we used the notation:
\begin{eqnarray}\label{eq: derivate PRODOTTO}
&&D_x^{n}\nu\cdot D_v^{n}\zeta=\sum_{\substack{n_1,n_2,n_3\in \mathbb{N}:\\\sum_j n_j=n}}\frac{\pa^n \nu}{\pa^{n_1}{x^1}\pa^{n_2}{x^2}\pa^{n_3}{x^3}}\frac{\pa^n \zeta}{\pa^{n_1}{v^1}\pa^{n_2}{v^2}\pa^{n_3}{v^3}},\\
&&\text{with}\ \ x=(x^1,x^2,x^3)\in\R^3\ \text{and}\ v=(v^1,v^2,v^3)\in\R^3\nonumber
\end{eqnarray}
for the one-particle functions $\nu$ and $\zeta$.

We want to determine a semiclassical expansion of $f^{\var}(x,v;t)$, solution of equation (\ref{eq: WigLIOUhartree}), namely:
\begin{equation}\label{eq: fEXP}
f^{\var}(t)=f^{(0)}(t)+\var f^{(1)}(t)+\var^2 f^{(2)}(t)+\dots
\end{equation}
First of all, we observe that interesting physical states, such as coherent states (see e.g. the forthcoming Section 6), have a similar expansion even at time $t=0$, namely:
\begin{equation}\label{eq: fEXPtimeZERO}
f_0^{\var}=f_0^{(0)}+\var f_0^{(1)}+\var^2 f_0^{(2)}+\dots
\end{equation}
Inserting (\ref{eq: fEXP}) in (\ref{eq: operatorTexp}) and setting:
\begin{equation}\label{eq: opTnot}
T_k^{(n)}=T_{f^{(k)}}^{(n)},
\end{equation}
we readily arrive to the following sequence of problems for the coefficients $ f^{(k)}(t)$ of the expansion (\ref{eq: fEXP}):
\begin{equation}\label{eq: WigLIOUseqk=0}
\left\{
\begin{aligned}
&\left(\pa_t+v\cdot \nabla_x\right)f^{(0)}=T_{0}^{(0)}f^{(0)},\\
&\left.f^{(0)}(x,v;t)\right\vert_{t=0}=f_0^{(0)}(x,v),
\end{aligned}
\right.
\end{equation}
and
\begin{equation}\label{eq: WigLIOUseq}
\left\{
\begin{aligned}
&\left(\pa_t+v\cdot \nabla_x\right)f^{(k)}=L(f^{(0)})f^{(k)}+\Theta^{(k)},\\
&\left.f^{(k)}(x,v;t)\right\vert_{t=0}=f_0^{(k)}(x,v),
\end{aligned}
\right.
\end{equation}
for $k\geq1$, where
\begin{equation}\label{eq: opL}
L(h)f=T_{h}^{(0)}f+T_{f}^{(0)}h=\left(\nabla_x\phi\ast h\right)\cdot \nabla_v f+\left(\nabla_x\phi\ast f\right)\cdot \nabla_v h,
\end{equation}
and
\begin{equation}\label{eq: Teta}
\Theta^{(k)}=\sum_{\substack{r,s,l:\\s+r+l=k\\l<k,r<k}}T_r^{(s)}f^{(l)}.
\end{equation}
Note that equation (\ref{eq: WigLIOUseqk=0}) is nothing else than the classical Vlasov equation (see (\ref{eq: Vlasov})) associated with the interaction $\phi$ (which will assume to be smooth). It is well known that the Vlasov equation can be solved by means of characteristics and fixed point. Moreover, the problems (\ref{eq: WigLIOUseq}) are linear and can be solved by a recursive argument (see Section 6 below). Indeed, the source terms $\Theta^{(k)}$ involve only those coefficients $f^{(n)}$ with $n<k$, so that they are known by the previous steps. We shall give a sense to the solutions $f^{(k)}(t)$, for $k\geq 1$, once we will have established precise assumptions on $\phi$ and $f_0^{(k)}$.
\section{The $N$-particle dynamics }
\setcounter{equation}{0}    
\def\theequation{3.\arabic{equation}}
Consider a quantum system constituted by $N$ identical particles. Its time evolution in the classical phase space is given by the Wigner-Liouville equation:
\begin{equation}\label{eq: WigLIOUhartreeNsec3}
\left(\pa_t+V_N\cdot \nabla_{X_N}\right)W_N^{\var}=T_N^{\var}W_N^{\var},
\end{equation}
where $W_N^{\var}(X_N,V_N;t)$ is the Wigner function describing the state of the system, 
$$
X_N=(x_1,\dots,x_N)\in \R^{3N}, \ \ V_N=(v_1,\dots,v_N)\in \R^{3N},
$$
and the pair $Z_N:=(X_N, V_N)$ denotes the generic point in the classical $N$-particle phase space. Moreover,
\begin{eqnarray}\label{eq: opT_N}
&&\(T_N^{\var}W_N^{\var}\)(Z_N)=\frac{i}{(2\pi)^{3N}}\int_{-1/2}^{1/2}\ud \lambda\int \ud K_N \hat{U}_N(K_N) e^{iK_N\cdot V_N}\left(K_N\cdot \nabla_{V_N}\right)W_N^{\var}(X_N,V_N+\lambda \var K_N),\nonumber\\
&&
%\begin{equation}\label{eq: opT_N}
%\begin{aligned}
%&\(T_N^{\var}W_N^{\var}\)(X_N,V_N)=\frac{i}{(2\pi)^{3N}}\int_{-1/2}^{1/2}\ud \lambda\int_{\R^{3N}}\ud K_N \hat{U}_N(K_N) e^{iK_N\cdot V_N}\\
%&\qquad\qquad\qquad\qquad\qquad\qquad\quad \left(K_N\cdot \nabla_{V_N}\right)W_N^{\var}(X_N,V_N+\lambda \var K_N),
%\end{aligned}
\end{eqnarray}
where $K_N=(k_1,\dots,k_N)\in \R^{3N}$ and $U_N$ is the (mean-field) interaction potential given by:
\begin{equation}\label{eq: meanFIELDpot}
U_N\left(X_N\right)=\frac{1}{N}\sum_{\substack{1\leq l<j\leq N}}^{N}\phi(x_l-x_j).
\end{equation}

We choose, as initial datum, the factorized state:
\begin{equation}\label{eq: meanFIELDinCOND}
W_{N,0}^{\var}(X_N,V_N)=\prod_{j=1}^{N}f_0^{\var}(x_j,v_j),
\end{equation}
where $f_0^{\var}$ is the initial datum of the equation (\ref{eq: WigLIOUhartree}). Following \cite{PULVIRENTI}, we expand
\begin{equation}\label{eq: expT_N}
T_N^{\var}=T_N^{(0)}+\var T_N^{(1)}+\var^2 T_N^{(2)}+\dots
\end{equation}
where, for $n$ even we have
\begin{equation}\label{eq: T_NcoeffEVEN}
T_N^{(n)}=i(2\pi)^{-3N}C_n\int_{\R^{3N}}\ud K_N \hat{U}(K_N)e^{iK_N\cdot X_N}\left(K_N \cdot \nabla_{V_N}\right)^{n+1},
\end{equation}
$C_n$ being constants depending on $n$, and, for $n$ odd, we find
\begin{equation}\label{eq: T_NcoeffODD}
T_N^{(n)}=0.
\end{equation}
%By the notation $\left(K_N\bullet \nabla_{V_N}\right)^{n+1}$ we used in (\ref{eq: T_NcoeffEVEN}), we denoted the following ''weighted''scalar product:
%\begin{eqnarray}\label{eq: T_NcoeffEVENscalarPROD}
%&&\left(K_N\bullet \nabla_{V_N}\right)^{n+1}:=\sum_{\substack{ \alpha_{j,i}=1,2,3\\i=1,2,3\\j=1,\dots,N}}\sum_{\substack{s_{j,i}:\\  0\leq s_{j,i}\leq n+1\\i=1,2,3\\j=1,\dots,N\\\sum_{j,i}s_{j,i}=n+1}}C(s_{1,1}\dots, s_{N,3})
%%%%%%\sum_{\substack{j_1,\dots, j_N\in\mathbb{N}:\\\sum_l  j_l=n+1}}\sum_{\substack{j_{l,i}
%\\ \dots\\j_{N,1},j_{N,2},j_{N,3}
%%%%%%}}C_{j_{1,1},\dots, j_{N,1}}
%\(k_1^{\alpha_{1,1}}\)^{s_{1,1}}\dots \(k_N^{ \alpha_{N,3}}\)^{s_{N,3}}
%D^{ s_{1,1}}_{{v_1}^{\alpha_{1,1}}}\dots D^{ s_{N,3}}_{{v_N}^{ \alpha_{N,3}}},\nonumber\\
%&&
%\end{eqnarray} 
%where $k_j=\(k_j^1,k_j^2,k_j^3\)$, $v_j=\(v_j^1,v_j^2,v_j^3\)$ and $C(s_{1,1}\dots, s_{N,3})$ is a constant depending on $s_{1,1}\dots,s_{N,3}$.\\
%where the second sum is over $l=1,\dots, N$ and over all integers $j_{l,i}$, $i=1,2,3$, such that, for each $l$, $j_{l,1}+j_{l,2}+j_{l,3}=j_l$. Furthermore, $k_l^{ j_l}=\(k_{l}^1\)^{j_{l,1}}\(k_{l}^2\)^{j_{l,2}}\(k_{1}^3\)^{j_{l,3}}$, and
%$$k^{ j}D_{v}^{ j}$ means that the $ j  $-th power is ''distributed'' over the three components of $k$ in the same way in which the derivative of order $j$ is ''distributed'' over the three components of $v$. \\

Looking for a semiclassical expansion
\begin{equation}\label{eq: expW_N}
W_N^{\var}(t)=W_N^{(0)}(t)+\var W_N^{(1)}(t)+\var^2 W_N^{(2)}(t)+\dots,
\end{equation}
we first expand the initial datum
\begin{equation}\label{eq: expW_Nindat}
W_{N,0}^{\var}=W_{N,0}^{(0)}+\var W_{N,0}^{(1)}+\var^2 W_{N,0}^{(2)}+\dots.
\end{equation}
The coefficients $W_{N,0}^{(k)}$ are determined by (\ref{eq: meanFIELDinCOND}) and (\ref{eq: fEXPtimeZERO}) as
\begin{eqnarray}
&&W_{N,0}^{(0)}=\prod_{j=1}^{N}f_0^{(0)}(x_j, v_j),\label{eq: W_Nindatcoeff0}\\
&&\dots\nonumber\\
&& W_{N,0}^{(k)}=\sum_{\substack{s_1\dots s_N\\0\leq s_j\leq k\\ \sum_j s_j=k}}\prod_{j=1}^{N}f_0^{(s_j)}(x_j, v_j)\label{eq: W_Nindatcoeffk}.
\end{eqnarray}
Note that $W_{N,0}^{(k)}$ is factorized only for $k=0$. 

By (\ref{eq: expW_N}) and (\ref{eq: expT_N}), we arrive to the sequence of problems:
\begin{equation}\label{eq: WigLiouNk=0}
\left\{
\begin{aligned}
&\left(\pa_t+V_N\cdot \nabla_{X_N}\right)W_N^{(0)}=T_{N}^{(0)}W_N^{(0)},\\
&\left.W_N^{(0)}(X_N,V_N;t)\right\vert_{t=0}=W_{N,0}^{(0)}(X_N,V_N),
\end{aligned}
\right.
\end{equation}
and
\begin{equation}\label{eq: WigLiouN}
\left\{
\begin{aligned}
&\left(\pa_t+V_N\cdot \nabla_{X_N}\right)W_N^{(k)}=T_N^{(0)}W_N^{(k)}+\Theta_N^{(k)},\\
&\left.W_N^{(k)}(X_N,V_N;t)\right\vert_{t=0}=W_{N,0}^{(k)}(X_N,V_N),
\end{aligned}
\right.
\end{equation}
for $k\geq1$, where
\begin{equation}\label{eq: Teta_N}
\Theta_N^{(k)}=\sum_{\substack{0\leq l<k}}T_N^{(k-l)}W_N^{(l)}.
\end{equation}
Note that $T_N^{(0)}=\nabla_{X_N}U_N\cdot \nabla_{V_N}=\frac{1}{N}\sum_{1\leq i< j\leq N}\nabla_x\phi(x_i - x_j)\cdot\nabla_{v_i}$ is the classical Liouville operator, while the source terms $\Theta_N^{(k)}$, at each order $k$, are known by the previous steps. Note also that, under reasonable assumptions on the interaction potential $\phi$, equation (\ref{eq: WigLiouNk=0}) can be solved by considering the Hamiltonian flow $\Phi^t(X_N, V_N)$, solution of the problem
\begin{equation}\label{eq: Newton}
\left\{
   \begin{aligned}
           &\dot{x}_i=v_i\\
           &\dot{v}_i=-\frac{1}{N}\sum_{j\neq i}^N \nabla_{x} \phi(x_i -x_j).
   \end{aligned}
\right.
\end{equation}
Indeed
\begin{equation}\label{eq: Liouville}
W_N^{(0)}(X_N,V_N;t)=S_N(t)W_{N,0}^{(0)}(X_N,V_N)=W_{N,0}^{(0)}\left(\Phi^{-t}\left(X_N,V_N\right)\right),
\end{equation}
where, from now on, we denote by $S_N$ the flow generated by the Liouville operator $T_N^{(0)}$,
while equations (\ref{eq: WigLiouN}) can be solved by recurrence thanks to the Duhamel formula:
\begin{equation}\label{eq: DuhamelN}
W_N^{(k)}(t)=S_N(t)W_{N,0}^{(k)}+\int_{0}^{t}\ud t_1\ S_N(t-t_1)\Theta_N^{(k)}(t_1).
\end{equation}

We conclude this section by expressing the operators $T_N^{(n)}$ ($n$ even) in terms of the variables $X_N, V_N$. From (\ref{eq: T_NcoeffEVEN}), we find that:
\begin{equation}\label{eq: restoOPt}
T_N^{(n)}=\hat{T}_N^{(n)}+R_N^{(n)},
\end{equation}
where
\begin{equation}\label{eq: diagOPt}
\hat{T}_N^{(n)}=c_n\frac{(-1)^{n/2}}{N}\sum_{1\leq l<j\leq N}D_x^{n+1}\phi(x_l - x_j)\cdot D_{v_l}^{n+1},
\end{equation}
where $c_n$ is the same of (\ref{eq: operatorT^nCOSTANTE}), and
\begin{eqnarray}\label{eq: OFFdiagOPt}
%%%&&D_G^{k}=\sum_{\substack{\alpha_1\dots\alpha_k:\\ \alpha_j=1,\dots,6}}\ \sum_{\substack{s_1\dots s_k:\\ 0\leq s_j\leq k\\s_1+\dots +s_k=k}}C_G(s_1\dots s_k)\frac{\pa^k}{\pa^{s_1} z^{\alpha_1}\dots \pa^{s_k}z^{\alpha_k}},\nonumber\\
%%%%&&\\
&&R_N^{(n)}=\frac{1}{N}\sum_{1\leq l<j\leq N}
%%%%%\sum_{\substack{\alpha_{l,i},\alpha_{j,i}:\\ \alpha_{l,i},\alpha_{j,i}=1,2,3\\i=1,2,3}} \sum_{\substack{s_{j,i},s_{l,i}:\\ 0\leq s_{l,i},s_{j,i}\leq n+1\\i=1,2,3\\ \sum_{i=1,2}s_{l,i}+\sum_{i=1,2}s_{j,i}=n+1}}C(s_{l,1},s_{l,2},s_{j,1},s_{j,2})
\sum_{\substack{k_1,k_2\in \mathbb{N}^{3}\\ \vert k_1\vert+\vert k_2\vert=n+1}}C_{k_1, k_2}\frac{\pa^{n+1}}{\pa_{x_l}^{\vert k_1\vert}\pa_{x_j}^{\vert k_2\vert}}\phi(x_l - x_j)\cdot\frac{\pa^{n+1}}{\pa_{v_l}^{\vert k_1\vert}\pa_{v_j}^{\vert k_2\vert}},
\end{eqnarray}
where,\ for $i=1,2$, $k_i=\(k_{i,1},k_{i,2},k_{i,3}\)$, $\vert k_i\vert=k_{i,1}+k_{i,2}+k_{i,3}$ , and
\begin{eqnarray}\label{eq: DICOder}
&&\frac{\pa^{n+1}}{\pa_{x_l}^{\vert k_1\vert}\pa_{x_j}^{\vert k_2\vert}}=\frac{\pa^{\vert k_1\vert}}{\pa^{ k_{1,1}}x_l^1\pa^{ k_{1,2}}x_l^2\pa^{k_{1,3}} x_l^3}\ \frac{\pa^{\vert k_2\vert }}{\pa^{k_{2,1}}x_j^1\pa^{ k_{2,2}}x_j^2\pa^{ k_{2,3}}x_j^3}, \\
&&\text{with}\ x_j=\(x_{j}^1,x_{j}^2,x_{j}^3\), \ x_l=\(x_{l}^1,x_{l}^2,x_{l}^3\),\nonumber
\end{eqnarray}
while $C_{k_1,k_2}$ are suitable coefficients. The same holds for the derivatives with respect to the velocities.

We observe that, by the expression (\ref{eq: OFFdiagOPt}), we mean that the derivative of order $\vert k_1\vert$ is distributed over the three components of $x_l$ in the same way in which it is distributed over the three components of $v_l$, and the same holds for the derivative of order $\vert k_2\vert$.
\section{Hierarchies}
\setcounter{equation}{0}    
\def\theequation{4.\arabic{equation}}
One way to investigate the behavior of the $N$-particle system in the limit $N\to \infty$, is to consider the hierarchy associated with equation (\ref{eq: WigLIOUhartreeNsec3}). More precisely, introducing the $j$-particle functions:
\begin{equation}\label{eq: marginaliJ}
W_{N,j}^{\var}(X_j, V_j;t)=\int_{\R^{3(N-j)}\times\R^{3(N-j)}} \ud X_{N-j}\ud V_{N-j}W_{N}^{\var}(X_j, X_{N-j}, V_j, V_{N-j};t),
\end{equation}
a straightforward computation yields the following sequence of equations:
\begin{eqnarray}\label{eq: hierarchyN}
&&\left(\pa_t +V_j\cdot \nabla_{X_j}\right)W_{N,j}^{\var}=T_j^{\var}W_{N,j}^{\var}+\left(\frac{N-j}{N}\right)C_{j+1}^{\var}W_{N,j+1}^{\var},\ \ j=1,2,\dots, N,\nonumber\\
&&\\
&&\text{with}\ \ \ W_{N,N}^{\var}=W_N^{\var}\ \ \text{and}\ \ C_{N+1}^{\var}\equiv 0,\nonumber
\end{eqnarray}
which, as usual, is called ''hierarchy'' because each equation is linked to the subsequent one.\\ 
%(by virtue of the presence of the $j+1$-th marginal in the equation for the $j$-th one).\\
The operator $T_j^{\var}$ (for a fixed $j$) describes the interaction of the first $j$ particles 
(we recall that we are dealing with identical particles, thus, in order to refer to any group of $j$ particles we can say ''the first $j$ particles'' because we can rearrange them as we want) among themselves, while the operator $C_{j+1}^{\var}$ describes the interaction of the first $j$ particles with the remaining $N-j$. The explicit form of such operators is:
\begin{eqnarray}\label{eq: opT_j}
&&\left(T_j^{\var}W_{N,j}^{\var}\right)(X_j, V_j)=\nonumber\\
&&=\frac{i(2\pi)^{-3N}}{N}\sum_{1\leq l<r\leq j}\int_{-1/2}^{1/2}\ud \lambda \int_{\R^3}\ud k\  \hat{\phi}(k)e^{ik\cdot (x_l-x_r)}(k\cdot \nabla_{v_l})W_{N,j}^{\var}(X_j,V_{l-1},v_l+\lambda \var k,V_{j-l}),\nonumber\\
&&
\end{eqnarray}
and
\begin{eqnarray}\label{eq: opC_j+1}
&&\left(C_{j+1}^{\var}W_{N,j+1}^{\var}\right)(X_j, V_j)=\nonumber\\
&&=i(2\pi)^{-3N}\sum_{l=1}^{j}\int_{-1/2}^{1/2}\ud \lambda \int_{\R^3}\ud k \ \hat{\phi}(k)\int_{\R^3\times\R^3}\ud x_{j+1}\ud v_{j+1}\ e^{ik\cdot( x_l-x_{j+1})}\nonumber\\
&&\qquad\qquad\qquad\qquad(k\cdot \nabla_{v_l})W_{N,j+1}^{\var}(X_j,x_{j+1},V_{l-1},v_l+\lambda \var k,V_{j-l},v_{j+1}).\nonumber\\
&&
\end{eqnarray}
Note now that $T_j^{\var}=O\left(\frac{j^2}{N}\right)$, thus we expect its action to be negligible in the limit. On the other hand, if we consider the sequence $\{f_j^{\var}(t)\}_{j}$, where $f_j^{\var}(t)=f_j^{\var}(X_j, V_j;t)$ is given by:
 \begin{eqnarray}\label{eq: factJ}
&&f_j^{\var}(X_j, V_j;t)=\prod_{k=1}^{j}f^{\var}(x_k, v_k;t)
\end{eqnarray}
and $f^{\var}(t)$ is the solution of the Wigner-Liouville equation associated with the Hartree dynamics (see (\ref{eq: WigLIOUhartree}) ), we easily deduce the following hierarchy:
\begin{eqnarray}\label{eq: hierarchyINF}
&&\left(\pa_t +V_j\cdot \nabla_{X_j}\right)f_{j}^{\var}=C_{j+1}^{\var}f_{j+1}^{\var}.
\end{eqnarray}
Therefore, we expect that
\begin{eqnarray}\label{eq: conv}
&&W_{N,j}^{\var}(t)\rightarrow f_{j}^{\var}(t),\ \text{as} \ N\to\infty,
\end{eqnarray}
for any $t>0$, provided that the same convergence holds at time $t=0$. As we have already recalled in Section 1, such a result can indeed be proven under various assumptions on the interaction potential $\phi$, both in the reduced density matrix formalism and in the Wigner one.
%(see Fr, PULVI BRESSA 2005 e poi chi altro lo fa solo limite cinetico senza semiclassico?NO ALLA LUCE DI QUELLO CHE DICO DOPO QUA CI VANNO: spohn, Erdosc/Yau...ecc...GINIBRE/VELO e HEPP?) 
This constitutes a validation of the Hartree equation in the mean-field limit. 
%Actually, the mentioned results are obtained in the reduced density matrix formalism, but they can be easily rephrased in classical phase space. 
Nevertheless, as we have already remarked, a common features of these results is that the limit is singularly behaving when $\var\to 0$. In fact, the operators involved in the above hierarchies are bounded in the norm appropriate to study the convergence, both in the reduced density matrix formalism and in the Wigner one, but their norm is diverging when $\var$ goes to zero. This suggests to consider the semiclassical equations described in Sections 1 and 2. In this way, considering equations at each order in $\var$ and analyzing the hierarchies associated with each of those equation, we have to deal with operators which are clearly independent of $\var$ ($T_N^{(n)}$, for the $N$-particle case, and $T_k^{(n)}$ for the expansion associated with the Hartree dynamics), and we have to investigate only the limit $N\to\infty$ without any dependence on $\var$. The price we have to pay is that now those operators are unbounded, as it comes out for the classical mean-field limit we are going to discuss in the next section. Therefore, if we want to prove that the coefficient of order $\var^k$ of the expansion of $W_{N,j}^{\var}$, namely:
\begin{eqnarray}\label{eq: coeffN}
&&W_{N,j}^{(k)}(X_j, V_j;t)=\int_{\R^{3(N-j)}\times \R^{3(N-j)}}\ud X_{N-j}\ud V_{N-j}W_{N}^{(k)}(X_j,X_{N-j}, V_j,V_{N-j};t),\nonumber\\
&&
\end{eqnarray}
converges to the corresponding object for the Hartree flow, which is:
\begin{eqnarray}\label{eq: coeffHAR}
&&f_{j}^{(k)}(X_j, V_j;t)=\sum_{\substack{s_1\dots s_j:\\  0\leq s_r\leq k\\ \sum_r s_r=k}}\prod_{r=1}^{j}f^{(s_r)}(x_r, v_r;t),
\end{eqnarray}
(where the one-particle functions $f^{(s_r)}(x_r, v_r;t)$ solve equation (\ref{eq: WigLIOUseqk=0}), if $s_r=0$, and (\ref{eq: WigLIOUseq}), for $s_r>0$), the use of the hierarchy solved by $W_{N,j}^{(k)}(t)$ does not seem a good idea. In fact, even at level zero, when we have to deal with the classical mean-field limit, the hierarchy is very difficult to handle with. Such a hierarchy is given by:
\begin{eqnarray}\label{eq: LiouvHIER}
&&\left(\pa_t+V_j\cdot \nabla_{X_j}\right)W_{N,j}^{(0)}=T_j^{(0)}W_{N,j}^{(0)}+\frac{N-j}{N}C_{j+1}^{(0)}W_{N,j+1}^{(0)},
\end{eqnarray}
with
\begin{eqnarray}\label{eq: opT_jCLASS}
&&\left(T_j^{(0)}W_{N,j}^{(0)}\right)(X_j, V_j)=\frac{1}{N}\sum_{1\leq l<r\leq j}\nabla_{x_l}\phi(x_l-x_r)\cdot \nabla_{v_l}W_{N,j}^{(0)}(X_j,V_{j}),
\end{eqnarray}
and
\begin{eqnarray}\label{eq: opC_j+1CLASS}
&&\left(C_{j+1}^{(0)}W_{N,j+1}^{(0)}\right)(X_j, V_j)=\nonumber\\
&&=\sum_{l=1}^{j}\int_{\R^3\times\R^3}\ud x_{j+1}\ud v_{j+1}\nabla_{x_l}\phi(x_l-x_{j+1})\cdot  \nabla_{v_l}W_{N,j+1}^{(0)}(X_j,x_{j+1},V_j,v_{j+1}).
\end{eqnarray}
Clearly these operators are unbounded (unless to make them act on analytical functions) because they involve derivatives with respect to the velocity variables. For this reason, to deal with the hierarchy is quite difficult and the obstacle which occurs in facing the higher order terms is precisely the same. However, in the classical case we can treat the convergence in a more natural way, avoiding to use the hierarchy. The idea is to control the $j$-particle marginals $W_{N,j}^{(0)}$ in terms of the expectation of the $j$-fold product of empirical measures with respect to the initial $N$-particle probability distribution (see Section 5 below). In the present paper we follow a similar strategy in dealing with the convergence of the higher order terms of the expansion. More precisely, we will express $W_{N,j}^{(k)}$ in terms of the expectation, with respect to the initial $N$-particle zero order coefficient (which is known to be a probability distribution), of suitable operators acting on empirical measures. The control of these objects will be obtained thanks to some estimates of the derivatives of the classical flow with respect to the initial data (see Proposition 5.1).
\section{The classical mean-field limit}
\setcounter{equation}{0}    
\def\theequation{5.\arabic{equation}}
The semiclassical expansion of the $N$-particle system leads us to consider the sequence of problems (\ref{eq: WigLiouNk=0})-(\ref{eq: WigLiouN}). The zero order equation (\ref{eq: WigLiouNk=0}) is purely classical and well understood.
%, at least in case of smooth pair interaction $\phi$. 
At this regard, let us remind some basic facts concerning the case of smooth potentials, which will be crucial in what follows. 

The Vlasov equation (\ref{eq: Vlasov}) makes sense even for a generic probability measure $\nu$ because $\nabla\phi \ \ast\  \nu\in C^{\infty}(\R^3)$ (thanks to the smoothness of $\phi$) and, by using the characteristic flow associated with the equation and a fixed point argument, it is possible to prove the existence and uniqueness of the solution. Furthermore, introducing the Wasserstein distance $\mathcal{W}$ (e.g. \cite{DOB}) based on a bounded metric in $\R^6$ to avoid unnecessary boundedness assumptions on the moments of the measures we deal with, it is possible to prove the following continuity property:
%not only that given a probability measure $\nu_0$, the Vlasov equation has a unique weak solution $\nu_t$ with $\nu_t\vert_{t=0}=\nu_0$ (it can be proven also by using the characteristic flow associated with the equation and a fixed point argument (see DOB)), but also that the following continuity property holds:
\begin{eqnarray}\label{eq: DOBineq}
&&\mathcal{W}(\nu^1_t,\nu^2_t)\leq e^{Ct}\mathcal{W}(\nu^1_0,\nu^2_0)
\end{eqnarray}
where $\nu^1_0$ and $\nu^2_0$ are two probability measures and $\nu^1_t$ and $\nu^2_t$ are the weak solutions of the Vlasov equation with initial data given by $\nu^1_0$ and $\nu^2_0$ respectively (see again \cite{DOB}). Moreover, for a configuration $Z_N=\{z_1,\dots,z_N\}$, where $z_j=(x_j, v_j)\in \R^6$, consider the Hamiltonian flow
\begin{eqnarray}\label{eq: hamFLOW}
\Phi^t(X_N,V_N)=Z_N(t)=Z_N(t;Z_N),
\end{eqnarray}
with initial datum $Z_N$ (see (\ref{eq: Newton})), and construct the empirical measure $\mu_N(t)$ as follows:
\begin{eqnarray}\label{eq: emp}
&&\mu_N(t):=\mu_N(z\vert Z_N(t))=\frac{1}{N}\sum_{j=1}^{N}\delta\left(z-z_j(t)\right).
\end{eqnarray}

The basic remark is that $\mu_N(t)$ is a weak solution of the Vlasov equation, so that, by (\ref{eq: DOBineq}), we have:
\begin{eqnarray}\label{eq: empCONV}
&&\mu_N(t)\rightarrow f^{(0)}(t),\ \text{as}\ N\to\infty,
\end{eqnarray}
provided that
\begin{eqnarray}\label{eq: empCONVt=0}
&&\mu_N \rightarrow f^{(0)}(0)=f_0^{(0)},\ \text{as}\ N\to\infty,
\end{eqnarray}
where 
\begin{eqnarray}\label{eq: empTEMPO0}
&&\mu_N:=\mu_N(z\vert Z_N)=\frac{1}{N}\sum_{j=1}^{N}\delta\left(z-z_j\right)
\end{eqnarray}
is the empirical distribution at time $t=0$. Moreover, $f_0^{(0)}$ is a (possibly smooth) probability distribution and $f^{(0)}(t)$ is the solution of the Vlasov equation with initial datum $f_0^{(0)}$.
Clearly, the convergences (\ref{eq: empCONV}) and (\ref{eq: empCONVt=0}) hold with respect to the metric induced by $\mathcal{W}$ on the space of probability measures on $\R^6$ and this is equivalent to the weak topology of probability measures.
%% (see \cite{DOB} e/o altri??). \\

Next, 
%we consider a system constituted by $N$ (classical) particles interacting by a mean-field potential $U_N$ as seen previously. L
%%let $Z_N$ be the initial configuration in the $N$-particle phase space, 
let us consider the (factorized) $N$-particle probability distribution $W_{N,0}^{(0)}(Z_N)=\prod_{k=1}^{N}f_0^{(0)}(x_k, v_k)$ and let $W_{N}^{(0)}(t)$ be its time evolution according to the Liouville equation (\ref{eq: Liouville}). We want to investigate the behavior of the $j$-particle marginals $W_{N,j}^{(0)}(t)$. Denoting by $\mathbb{E}_N$ the expectation with respect to $W_{N,0}^{(0)}(Z_N)$, after straightforward computations, we obtain:
\begin{eqnarray}\label{eq: EprodEMP}
\mathbb{E}_N\left[\mu_N\(z'_1\vert Z_N(t)\)\dots\mu_N\(z'_j\vert Z_N(t)\)\right]&&=\frac{N(N-1)\dots(N-j+1)}{N^j}\ W_{N,j}^{(0)}(Z'_j;t)+\nonumber\\
&&+\ O\left(\frac{1}{N}\right),\nonumber\\
&&
\end{eqnarray}
where $Z'_j=(z'_1\dots z'_j)$ and $W_{N,j}^{(0)}(Z'_j;t)=W_{N,j}^{(0)}(Z'_j(t))$ (see (\ref{eq: hamFLOW})).

Consider now a typical sequence $Z_N$ with respect to $f_0^{(0)}$, namely such that (\ref{eq: empCONVt=0}) holds. By the strong law of large numbers this happens a.e. with respect to $\(f_0^{(0)}\)^{\otimes \infty}$ and by (\ref{eq: empCONV}) and (\ref{eq: EprodEMP}) we have:
\begin{eqnarray}\label{eq: prodEMP}
&&\lim_{N\to\infty}\mathbb{E}_N\left[\mu_N\(z'_1\vert Z_N(t)\)\dots \mu_N\(z'_j\vert Z_N(t)\)\right]=\lim_{N\to\infty}W_{N,j}^{(0)}(Z'_j;t)=\left(f^{(0)}\right)^{\otimes j}(Z'_j;t),\nonumber\\
&&
%\ \text{in the weak topology of probability measures.}
\end{eqnarray}
in the weak topology of probability measures. Thus propagation of chaos is proven, and, this is the remarkable fact, it has been done without using the hierarchy. 

For fixed $\var>0$, the quantum hierarchy is, in a certain sense, easier. In fact, in that situation the operators involved are bounded (as operators acting on the spaces appropriate for that context)
%, namely, the space of the trace class operators) 
and it is possible to realize the limit by using the hierarchy. On the contrary, in the quantum context we cannot use any characteristic flow
%, because the concept of ''trajectories'' has no sense, 
and there is not any object analogous to the empirical measure.
% (because of the uncertainty principle!). \\
Nevertheless, if we consider the semiclassical expansion of the time evolved Wigner function, the higher order terms can be viewed as quantum corrections to the classical dynamics. 
%%Therefore, looking at them in this perspective, what we have to do, to understand their behavior in the limit $N\to\infty$, is precisely to catch on how the classical dynamics is modified at each order in $\var$. At that point, we will have a characterization of these new dynamics by means of completely classical objects and we will show how to control them by using what we recalled previously about the classical system.\\

We now explain heuristically our approach fully exploited in Section 7.

The first correction to the Vlasov equation in the Hartree dynamics satisfies (see (\ref{eq: fEXP}) and (\ref{eq: fEXPtimeZERO})):
\begin{equation}\label{eq: WigLIOUseqONE}
\left\{
\begin{aligned}
&\left(\pa_t+v\cdot \nabla_x\right)f^{(1)}=L(f^{(0)})f^{(1)},\\
&\left.f^{(1)}(x,v;t)\right\vert_{t=0}=f_0^{(1)}(x,v),
\end{aligned}
\right.
\end{equation}
(looking at the expression (\ref{eq: Teta}) for the source terms $\Theta^{(k)}$, we verify that $\Theta^{(1)}\equiv 0$). As we shall see in detail in the following section, our choice for the initial one-particle datum is a mixture of coherent states and each coefficient of the expansion it is given by suitable derivatives of the zero order distribution. In particular, the explicit form for $f_0^{(1)}$ is:
\begin{equation}\label{eq: PRIMOcoef}
f_0^{(1)}(x, v)=D^2_G f_0^{(0)}(x, v),
\end{equation}
where $D_G^2$ is a suitable second order derivation operator (see formula (\ref{eq: operatorD_G}) below in the case $k=2$) involving derivatives with respect to the initial variables $z_1,\dots, z_N$.

As regard to the $N$-particle dynamics, looking at (\ref{eq: W_Nindatcoeffk}) in the case $k=1$, we know that the initial datum for the coefficient of order one in $\var$
%, namely $W_{N,0}^{(1)}$, 
is:
\begin{eqnarray}\label{eq: PRIMOcoefN}
W_{N,0}^{(1)}(Z_N)&&=\sum_{j=1}^{N}f_0^{(1)}(z_j)\prod_{l\neq j}^{N}f_0^{(0)}(z_l)=\mathcal{D}^2W_{N,0}^{(0)}(Z_N),
\end{eqnarray}
where
\begin{eqnarray}\label{eq: opDstorto}
\mathcal{D}^2=\sum_{j=1}^{N}D_{G,j}^2,
\end{eqnarray}
and $D_{G,j}^2$ is the operator $D_G^2$ relative to the variable $z_j\in \R^6$ . Let us now define $\mathcal{D}^2\mu_N(t)$ as the distribution acting on a test function $u$
%$\in C_b^{\infty}(\R^6)$ 
in the following way:
\begin{eqnarray}\label{eq: opDstortoMu}
\(u,\mathcal{D}^2\mu_N(t)\)&&=\mathcal{D}^2\(\frac{1}{N}\sum_{l=1}^{N}u(z_l(t))\)=\frac{1}{N}\sum_{l,j=1}^{N}D_{G,j}^2 u(z_l(t)).
\end{eqnarray}
We remind that the operator $D_{G,j}^2$ involves derivatives with respect to the initial variables $z_1,\dots,z_N$, thus, if at time $t=0$ we have $\mu_N\to f_0^{(0)}$ when $N\to\infty$ (in the weak sense of probability measures), it follows that:
\begin{eqnarray}\label{eq: opDstortoMut=0}
\(u,\mathcal{D}^2\mu_N\)&&=\mathcal{D}^2\frac{1}{N}\sum_{l=1}^{N}u(z_l)= \frac{1}{N}\sum_{l,j=1}^{N}D_{G,j}^2 u(z_l)=\frac{1}{N}\sum_{j=1}^{N}D_{G,j}^2 u(z_j)=\nonumber\\
&& =\(D_{G}^2 u,\mu_N \)\ \rightarrow\ \(D_{G}^2 u,f_0^{(0)}\)= \(u,D_{G}^2f_0^{(0)}\)=\(u,f_0^{(1)}\)\nonumber\\
&&
\end{eqnarray}
as $N\to\infty$.
%The previous computation shows that $\mathcal{D}^2\mu_N$ converges to $f_0^{(1)}$, as distributions acting on a suitable class of functions. We do not want to discuss about that point now and we will be more precise in the following sections, but, from (\ref{eq: opDstortoMut=0}), we can argue that we will need to do derivatives of these test functions, and then, they will be not only bounded and continous (as it is in the case of the weak convergence of probability measures). Furthermore, we remark that
%%, as it is clarified by (\ref{eq: opDstortoMut=0}), 
%the convergence of $\mathcal{D}^2\mu_N$ to $f_0^{(1)}$ follows straightforward from that of $\mu_N$ to $f_0^{(0)}$ in the weak sense of probability measures. Therefore, if the last one is obtained for any configuration $Z_N$ which is typical with respect to the measure $W_{N,0}^{(0)}$ (as it is established by the strong law of large numbers), it follows that the convergence of $\mathcal{D}^2\mu_N$ to $f_0^{(1)}$, as suitable distributions, holds for any configuration $Z_N$ verifying the same property. Thanks to the previous considerations
Moreover, by (\ref{eq: PRIMOcoefN}) and (\ref{eq: opDstortoMut=0}), we can conclude that:
\begin{eqnarray}\label{eq: marg1ord1t=0}
\(u,W_{N,1}^{(1)}(t)\vert_{t=0}\)=\(u,\mathbb{E}_N\left[\mathcal{D}^2\mu_N\right]\) \ \rightarrow\  \(u,f_0^{(1)}\)\ \text{as}\ N\to\infty.\nonumber\\
&&
\end{eqnarray}
%Now, we look at the one-particle marginal $W_{N,1}^{(1)}(t)$, with $t>0$,
%%, for $t\geq 0$, 
%as a distribution acting on a suitable function $\varphi$. We find that:
%\begin{eqnarray}\label{eq: W_N,1sufunztest}
%<W_{N,1}^{(1)}(t),\varphi>&&=\int_{\R^6}\ud z_1 W_{N,1}^{(1)}(z_1;t)\varphi(z_1)=\nonumber\\
%&&=\int_{\R^6}\ud z_1\int_{\R^{3(N-1)}\times \R^{3(N-1)}}\ud Z_{N-1}W_{N}^{(1)}(Z_N;t)\varphi(z_1)=\nonumber\\
%&&=\int_{\R^{3N}\times \R^{3N}}\ud Z_{N}W_{N}^{(1)}(Z_N;t)\varphi(z_1)=\nonumber\\
%&&=\int_{\R^{3N}\times \R^{3N}}\ud Z_{N}W_{N}^{(1)}(Z_N;t)\frac{1}{N}\sum_{l=1}^{N}\varphi(z_l)=\nonumber\\
%&&=\int_{\R^{3N}\times \R^{3N}}\ud Z_{N}W_{N}^{(1)}(Z_N;t)<\mu_N,\varphi>,\nonumber\\
%&&
%\end{eqnarray}
%where we made use of the symmetry of the coefficient $W_{N}^{(1)}(Z_N;t)$ with respect to any permutation of the variables. Moreover, writing the equation 

By equation (\ref{eq: WigLiouN}) for $k=1$, we have:
\begin{eqnarray}\label{eq: W_N^1eq}
&&\left(\pa_t +V_N\cdot \nabla_{X_N}\right)W_{N}^{(1)}=\nabla_{X_N}U_N\cdot \nabla_{V_N}W_{N}^{(1)},\nonumber\\
&&W_{N}^{(1)}(Z_N;t)\vert_{t=0}=W_{N,0}^{(1)}(Z_N),
\end{eqnarray}
namely, the classical Liouville equation. Therefore:
\begin{eqnarray}\label{eq: W_N^1espress}
&&W_{N}^{(1)}(Z_N;t)=S_N(t)W_{N,0}^{(1)}(Z_N).
\end{eqnarray}
Finally, 
%putting (\ref{eq: W_N^1espress}) in (\ref{eq: W_N,1sufunztest}) and 
by virtue of (\ref{eq: W_N^1espress}) and (\ref{eq: PRIMOcoefN}), we obtain
\begin{eqnarray}\label{eq: W_N,1sufunztestA}
\(u,W_{N,1}^{(1)}(t)\)&&=\int_{\R^{3N}\times \R^{3N}}\ud Z_{N}S_N(t)W_{N,0}^{(1)}(Z_N)\(u,\mu_N\)=\nonumber\\
&&=\int_{\R^{3N}\times \R^{3N}}\ud Z_{N}W_{N,0}^{(1)}(Z_N)\(u,\mu_N(t)\)=\nonumber\\
&&=\int_{\R^{3N}\times \R^{3N}}\ud Z_{N}\mathcal{D}^2W_{N,0}^{(0)}(Z_N)\(u,\mu_N(t)\)=\nonumber\\
&&=\int_{\R^{3N}\times \R^{3N}}\ud Z_{N}W_{N,0}^{(0)}(Z_N)\(u,\mathcal{D}^2\mu_N(t)\)=\nonumber\\
&&=\(u,\mathbb{E}_N\left[\mathcal{D}^2\mu_N(t)\right]\).
%\ \ \text{for any}\ t\geq 0.\nonumber\\
%&&
\end{eqnarray}
Therefore, the behavior of $W_{N,1}^{(1)}(t)$ is determined by that of $\mathcal{D}^2\mu_N(t)$ for any initial configuration $Z_N$ which is typical with respect to $f_0^{(0)}$.
%the $N$-particle probability measure $W_{N,0}^{(0)}$. 
Finally, since $\mu_N(t)$ solves:
\begin{equation}\label{eq: VlasovWEAK}
\left\{
\begin{aligned}
&\left(\pa_t+v\cdot \nabla_x\right)\mu_N(t)=\left(\nabla \phi \ast \mu_N(t)\right)\cdot \nabla_v \mu_N(t)\\
&\left. \mu_N(t)\right\vert_{t=0}=\mu_N,
\end{aligned}
\right.
\end{equation}
applying $\mathcal{D}^2$, we get:
\begin{equation}\label{eq: VlasovWEAKconD^2}
\left\{
\begin{aligned}
&\left(\pa_t + v\cdot \nabla_x \right)\mathcal{D}^2\mu_N(t)=L\left(\mu_N(t)\right) \mathcal{D}^2\mu_N(t)+R_N,\\
&\left. \mathcal{D}^2\mu_N(t)\right\vert_{t=0}=\mathcal{D}^2\mu_N,
\end{aligned}
\right.
\end{equation}
where $R_N$ is a term involving objects of the form $\sum_j \(D_{G,j}\mu_N(t)\)\( D_{G,j}\mu_N(t)\)$ which, as we shall see later, are of order $1/N$ when tested versus smooth functions.
The equation (\ref{eq: VlasovWEAKconD^2}) is similar to (\ref{eq: WigLIOUseqONE}), except for the presence of the term $R_N$ and for the fact that we have $L\left(\mu_N(t)\right) $ instead of $L\left(f^{(0)}\right)$. Therefore, the proof of the convergence of $W_{N,1}^{(1)}(t)$ to $f^{(1)}(t)$ reduces to that of a stability property for the solution of (\ref{eq: WigLIOUseqONE}) with respect to suitable weak topologies. Propositions $6.1$ and $6.2$ below will provide us such property.
%Indeed, thanks to the considerations done previously and to the approximation Proposition 6.2 below, we will show that $\mathcal{D}^2\mu_N(t)\to f^{(1)}(t)$ in a suitable weak sense, for any initial configuration $Z_N$ which is typical with respect to $W_{N,0}^{(0)}$. Clearly, by virtue of (\ref{eq: W_N,1sufunztestA}), this fact is equivalent to the convergence of the one-particle coefficient $W_{N,1}^{(1)}(t)$ to the corresponding one arising from the Hartree expansion, namely $ f^{(1)}(t)$. \\
%%From the convergence of $\mathcal{D}^2\mu_N(t)\to f^{(1)}(t)$ for any initial configuration $Z_N$ which is typical with respect to $W_{N,0}^{(0)}$, it is easy to prove also the convergence $\mathcal{D}^2\(\mu_N(t)\)^{\otimes j}\to f_j^{(1)}(t)$ for any $j$ (essentially by virtue of the Proposition 5.1 below). Finally, thanks to the convergence of $\mathcal{D}^2\(\mu_N(t)\)^{\otimes j}$ for any $j$ and to that of $W_{N,1}^{(1)}(t)$, it is possible to prove the convergence of the $j$-particle coefficient $W_{N,j}^{(1)}(t)$, for any $j$, to $f_j^{(1)}(t)$ (see Section 7). \\
%As we shall see in Section 7, 

The general case $k>1$ is only technically more complicated because of the presence of source terms, but the main ideas are those presented here. 

We conclude by establishing a Proposition controlling the size of the derivatives of the Hamilton flow (\ref{eq: Newton}) with respect to the initial data. 

From now on we shall denote by $C$ a positive constant, independent of $N$, possibly changing from line to line.
\\\\
{\bf Proposition 5.1}\\\\
\emph{Let $z_i(t)=\left(x_i(t),v_i(t)\right),\ i=1,\dots,N$ be the solution of equations (\ref{eq: Newton}) with initial datum $z_i=\left(x_i,v_i\right),\ i=1,\dots,N$. %given at a certain time $t_0<t$. 
Let $z_{i}^{\beta}\ \forall\ \beta=1,\dots,6$ be the $\beta$-th component of $z_i\in \R^6$. If the pair interaction potential $\phi$ is $ \mathcal{C}^{\infty}(\R^3)$ and the derivatives of any order of $\phi$ are uniformly bounded, then, for each $k\in \mathbb{N}$: }
\begin{equation}\label{eq: derivateK}
%\max_{i,j_1,\dots,j_k}
%\frac{1}{N}\sum_{i=1}^{N}
\left\vert\frac{\pa^k z_i^{\beta}(t)}{\pa z_{j_1}^{\alpha_{1}}\dots \pa z_{j_k}^{\alpha_{k}}} \right\vert\leq
\frac{C}{N^{d_k^{(i)}}},
%C_k\(\frac{1}{N^s}+\prod_{m=1}^{s}\delta_{i j_m}\)
%\pa^r_{z_i^{\alpha}}
\end{equation}
\emph{where $I:=(j_1,\dots, j_k)$ is any sequence of possibly repeated indices and $d_k^{(i)}$ is the number of different indices in $I$ which are also different from \ $i$}.\\
%which are different from $i$ (counted without their multiplicity)}. \\\\
%%%%%%$r+k_1 +\dots +k_s=k$, $j_m\neq i\ \forall\ m=1,\dots, s$ and $j_m\neq j_n$ if $m\neq n$}.\\\\
%$j_m\neq i\ \forall\ m=1,\dots, s$, $j_m\neq j_n$ if $m\neq n$, 
%and $\alpha,\alpha_{1},\dots,\alpha_{s}=1,\dots, 6$}.\\\\

The physical significance of (\ref{eq: derivateK}) is obvious. In the mean-field context, the quantity $z_i(t)$ depends weakly on $z_j$ if $j\neq i$ for each $t>0$. Actually $\frac{\pa z_i^\beta(t)}{\pa z_j^\alpha}=O\(\frac{1}{N}\)$ while $\frac{\pa z_i^\beta(t)}{\pa z_i^\alpha}=O\(1\)$ and these two estimates give rise to (\ref{eq: derivateK}) in the case $k=1$. Estimate (\ref{eq: derivateK}) says that for each derivative of any order with respect to some $z_j$ of 
%the function $1/N\sum_{i=1}^{N}
$z_i(t)$ , we gain a factor $1/N$.
We have also the following corollary whose straightforward proof will be omitted.\\\\
{\bf Proposition 5.2}\\\\
\emph{Let $U=U(Z_N(t))$ be a function of the time evolved configuration $Z_N(t)$ of the form:
$$
U(Z_N(t))=\frac{1}{N}\sum_{i=1}^{N}u(z_i(t)),
$$
where $u\in C_b^{\infty}(\R^3\times\R^3)$. Then, if the pair interaction potential $\phi$ satisfies the assumptions of Proposition 5.1, the following estimate holds:
\begin{equation}\label{eq: corollario}
%\max_{i,j_1,\dots,j_k}
\left\vert\frac{\pa^k U(Z_N(t))}{\pa z_{j_1}^{\alpha_{1}}\dots \pa z_{j_k}^{\alpha_{k}}} \right\vert\leq
\frac{C}{N^{d_k}},
%C_k\(\frac{1}{N^s}+\prod_{m=1}^{s}\delta_{i j_m}\)
%\pa^r_{z_i^{\alpha}}
\end{equation}
where $d_k$ is the number of different indices in the sequence $I=(j_1,\dots, j_k)$}.\\\\
%, with $j\neq i$, we gain $1/N$.\\\\
The proof of Proposition 5.1 will be given in Appendix A.
%\newline \spaz \hspace{1cm} \hfill $\square$ \newline
%\\
\section{Results and technical preliminaries}
\setcounter{equation}{0}    
\def\theequation{6.\arabic{equation}}
We choose, as initial condition for the one-particle Wigner function, a mixture of coherent states. The Wigner function associated with a pure coherent state centered at the point $(x_0, v_0)$ is given by:
\begin{equation}\label{eq: coeherent}
w(x,v\vert x_0,v_0)=\frac{1}{(\pi\var)^{3}}e^{-\frac{(x-x_0)^{2}}{\var}}e^{-\frac{(v-v_0)^{2}}{\var}}.
\end{equation}

Let now $g=g(x,v)$ be a smooth probability density on the one-particle phase space independent of $\var$ 
%we will specify in the sequel which are the regularity properties we assume on $g$, 
(see Hypotheses H below) . Then we define:
\begin{equation} \label{eq: mix}
f_{0}(x, v)=\int_{\R^3\times\R^3}\ud x_0 \ud v_0 \ w(x, v\vert x_0,v_0)g(x_0, v_0).
\end{equation}
Using the standard notation $z=(x,v)$ and $z_0=(x_0,v_0)$, (\ref{eq: mix}) is equivalent to:
\begin{eqnarray} \label{eq: mix2}
f_{0}(z)&&=\frac{1}{(\pi\var)^{3}}\int_{\R^6}\ud z_0\  e^{-\frac{(z-z_0)^{2}}{\var}} g(z_0)=\nonumber\\
&&=\frac{1}{(\pi)^{3}}\int_{\R^6}\ud \zeta \ e^{- \zeta^{2}} g(z-\sqrt{\var}\zeta).
\end{eqnarray}
Expanding
\begin{eqnarray} \label{eq: expG}
 g(z-\sqrt{\var}\zeta)&&=g(z)-\left(\zeta \cdot \nabla_z \right)g(z)\sqrt{\var}+\left(\zeta \cdot \nabla_z \right)^2g(z)\frac{(\sqrt{\var})^2}{2}+\dots\nonumber\\
&& \dots  -\left(\zeta \cdot \nabla_z \right)^{2n-1}g(z)\frac{(\sqrt{\var})^{2n-1}}{(2n-1)!}+\left(\zeta \cdot \nabla_z \right)^{2n}g(z)\frac{(\sqrt{\var})^{2n}}{(2n)!}+\dots,\nonumber\\
&&
\end{eqnarray}
and performing the gaussian integrations (which cancels the terms with the odd powers of $\sqrt{\var}$), we readily arrive to the following expansion for the Wigner function $f_0$:
\begin{eqnarray} \label{eq: expf_0}
&& f_0=f_0^{(0)}+\var f_0^{(1)}+\dots +\var^n f_0^{(n)}+\dots,
\end{eqnarray}
where 
\begin{eqnarray} 
&& f_0^{(0)}=g,\label{eq: expf_0COEFF}\\
&& f_0^{(n)}=D_G^{2n}f_0^{(0)}\ \ \text{for}\ n\geq 1,\label{eq: expf_0COEFFk}
\end{eqnarray}
and $D_G^{k}$ ($G$ stands for ''Gaussian''), for each $k>0$, is the following derivation operator with respect to the variable $z=(x,v)$:
\begin{eqnarray} \label{eq: operatorD_G}
&&D_G^{k}=\sum_{\substack{\alpha_1\dots\alpha_k:\\ \alpha_j=1,\dots,6}}\ 
%\sum_{\substack{s_1\dots s_k:\\ 0\leq s_j\leq k\\s_1+\dots +s_k=k}}
C_G(\alpha_1\dots \alpha_k)\frac{\pa^k}{\pa z^{\alpha_1}\dots \pa z^{\alpha_k}},
\end{eqnarray}
where
\begin{eqnarray} \label{eq: operatorD_Gcost}
&&C_G(\alpha_1\dots \alpha_k)=\frac{1}{k!}\int_{\R^6}\ud \zeta\ e^{-\zeta^2} \prod_{j=1}^{k} \zeta^{\alpha_j}.
\end{eqnarray}
Therefore, $C_G(\alpha_1\dots \alpha_k)$ is equal to zero for each sequence $\alpha_1\dots \alpha_k$ in which at least one index appears an odd number of times.\\
\\
{\bf Hypotheses H}: 

In the present paper we assume that the probability density $g=f_0^{(0)}\in \mathcal{S}(\R^3\times\R^3)$, thus 
%(\ref{eq: expf_0}) and 
(\ref{eq: expf_0COEFFk}) make sense for any $n\geq 1$. As regard to the pair interaction potential $\phi$, we assume that $\phi\in \mathcal{C}^{\infty}(\R^3)$, that any derivative of $\phi$ is uniformly bounded (in order to be able to apply Proposition 5.1) and that $\phi$ is spherically symmetric.
\\\\\\\\
{\bf Remark 6.1}:

In this paper we consider a completely factorized $N$-particle initial state. Furthermore the one-particle state is a mixture and this automatically excludes the Bose statistics.\\\\
{\bf Remark 6.2}:

We made the choice to expand fully the initial state $f_0$ according to equation (\ref{eq: expf_0}). Another possibility is to assume the ($\var$ dependent) state $f_0$ (which is a probability measure in the present case) as initial condition for the Vlasov problem and, consequently, $f_0^{(k)}=0$ for the problems (\ref{eq: WigLIOUseq}). Now the coefficients $f^{(k)}(t)$ are $\var$ dependent but this does not change deeply our analysis because $f_0$ is smooth, uniformly in $\var$.\\
% namely, $\phi(x)=\phi(\vert x \vert)$ $\forall\ x\in \R^3$.\\

Under hypotheses H, we can give a sense to the linear problem (\ref{eq: WigLIOUseq}) for any $k\geq 1$, by virtue of the following proposition, whose (straightforward) proof will be given in Appendix B.\\\\
{\bf Proposition 6.1}\\
\emph{
%Let $T>0$ be fixed. 
Consider the following initial value problem:}
\begin{equation}\label{eq: Prop6.1}
\left\{
\begin{aligned}
&\left(\pa_t+v\cdot \nabla_x\right)\gamma=L(h)\gamma+\Theta,\\
%\ \ \text{for}\ t\in [0,T]\\
&\left. \gamma(x,v;t)\right\vert_{t=0}=\gamma_{0}(x,v),
\end{aligned}
\right.
\end{equation}
\emph{with $\gamma_{0}\in L^1(\R^3\times\R^3)$, $h=h(x,v;t)$ is such that $\left \vert \nabla_v h\right\vert \in \mathcal{C}^0\(L^1(\R^3\times\R^3), \R^{+} \)$, $\Theta=\Theta(x,v;t)$ is such that $\Theta\in \mathcal{C}^0\(L^1(\R^3\times\R^3),\R^{+}\)$.\\
% and also $\nabla \gamma_{0}\in \mathcal{C}\(\R^{+}, L^1(\R^3\times\R^3)\)$. 
Then, there exists a unique solution $\gamma=\gamma(x,v;t)$ of (\ref{eq: Prop6.1}), such that $\gamma\in \mathcal{C}^0\(L^1(\R^3\times\R^3), \R^{+} \)$, given by an explicit series expansion.}\\
\emph{Furthermore},\ \emph{denoting by $\Sigma_h$ the flow generated by $L(h)$, we have that $\Sigma_h(t,0)\gamma_0\in \mathcal{C}^{k}\(\R^3\times \R^3\)$ provided that $\nabla_v h\in \mathcal{C}^{k}\(\R^3\times \R^3\)$ and $\gamma_0 \in \mathcal{C}^{k}\(\R^3\times \R^3\)$}.
%the flow associated with the operator $L(h)$, which we denote by $\Sigma_h$, preserves the regularity properties which are common to $\nabla_v h(t)$ (for any $t$) and to the initial datum. Namely, computing $u(t):=\Sigma_h(t,0)u(x,v)$, for some function $u$ which is in a certain space $X$, if $\nabla_v h(t)$ (for any $t$) is in $X$, $u(t)$ is also in $X$ for any $t$.}
\\\\

The main goal of the present paper is to compare the $j$-particle semiclassical expansion associated with the $N$-particle flow, namely $W_{N,j}^{(k)}(t)$, $k=0,1,2,\dots$, with the corresponding coefficients $f_j^{(k)}(t)$ of the expansion:
\begin{equation}\label{eq: expHartreeJ}
f_j^{\var}(t)=f_j^{(0)}(t)+\var f_j^{(1)}(t)+\dots +\var^k f_j^{(k)}(t)+\dots,
\end{equation}
where $f_j^{(k)}(t)$ is given by (\ref{eq: coeffHAR}). The main result is the following.\\\\
{\bf Theorem 6.1}\\
\emph{Under the Hypotheses H, for all $t> 0$, for any integers $k$ and $j$, the following limit holds in $\mathcal{S}'(\R^{3j}\times\R^{3j})$:}
\begin{equation}\label{eq: Th6.1}
W_{N,j}^{(k)}(t)\rightarrow f_j^{(k)}(t).
\end{equation}
as $N\to \infty$.
\\\\
{\bf Remark 6.3:}

As we shall see in the sequel, the convergence (\ref{eq: Th6.1}) is slightly stronger than the convergence in $\mathcal{S}'(\R^{3j}\times\R^{3j})$. Indeed, the sequence $W_{N,j}^{(k)}(t)$ converges also when it is tested on functions in $\mathcal{C}_b^{\infty}(\R^{3j}\times\R^{3j})$, namely, the space of functions which are uniformly bounded and infinitely differentiable. Such kind of convergence, which is natural in the present context, will be called $\mathcal{C}_b^{\infty}$-weak convergence.\\
\\
{\bf Proposition 6.2}\\
\emph{Let $\gamma_N(x,v;t)$ be a sequence in $\mathcal{S}'(\R^{3}\times\R^{3})$ (for each $t$) satisfying:}
\begin{equation}\label{eq: Prop6.2}
\left\{
\begin{aligned}
&\left(\pa_t+v\cdot \nabla_x\right)\gamma_N=L(h_N)\gamma_N+\Theta_N,\\
%\ \ \text{for}\ t\in [0,T]\\
&\left. \gamma_N(x,v;t)\right\vert_{t=0}=\gamma_{N,0}(x,v),
\end{aligned}
\right.
\end{equation}
where $\gamma_{N,0}$, $\Theta_N$ are sequences in $\mathcal{S}'(\R^{3}\times\R^{3})$. We assume that:\\
{\bf i)}\ \emph{$h_N(x,v;t)$ is a sequence of probability measures converging, as $N\to\infty$, to a measure $h(t)\ud x\ud v$ with a density $h(t)\in \mathcal{C}^{\infty}_b\(\R^3\times\R^3\)$ and such that
%is a distribution in $\mathcal{S}'(\R^{3}\times\R^{3})$ such that $h_N(t)\stackrel{N\to\infty}{\rightarrow}h(t)$ in the weak sense of probability measures, and $h(t)=h(x,v;t)$ is a function such that $h(t)\in\mathcal{C}_b^{\infty}(\R^{3}\times\R^{3})$ for any $t$, and 
$\left\vert \nabla_v h\right\vert \in \mathcal{C}^0\( L^1(\R^{3}\times\R^{3}),\R^+\)$.}\\
{\bf ii)}\ \emph{for all $u_1,u_2$ in $\mathcal{C}_b^{\infty}(\R^{3}\times\R^{3})$ , there exists a constant $C=C(u_1, u_2)>0$, not depending on $N$, such that}:
\begin{equation}\label{eq: Prop6.2hyp}
\left\|  u_1\ast \(u_2\gamma_N\)\right\|_{L^{\infty}(\R^3\times\R^3)}< C<+\infty\ \ \ for \ any \  t.
\end{equation}
{\bf iii)}\ \emph{$\gamma_{N,0}\rightarrow \gamma_0\text{,\ as}\ \ N\to\infty$,\ \  $\mathcal{C}_b^{\infty}$-weakly , $\gamma_0=\gamma_0(x,v)$ is a function in $L^1(\R^{3}\times\R^{3})$.}
\\
{\bf iv)}\ \emph{$\Theta_N \rightarrow \Theta\text{,\ as}\ \ N\to\infty$,\  \ $\mathcal{C}_b^{\infty}$-weakly , $\Theta=\Theta(x,v;t)$ is a function in \ $\mathcal{C}^0\(L^1(\R^{3}\times\R^{3}), \R^+\)$.}
\\
\emph{Then:}
\begin{eqnarray}\label{eq: Prop6.2thesis}
&&\text{\emph{$\gamma_N \rightarrow\gamma\text{,\ as}\ \ N\to\infty$ \ \ $\mathcal{C}_b^{\infty}$-weakly}},\nonumber\\
&&
\end{eqnarray}
\emph{where $\gamma$ is the unique solution of the problem (\ref{eq: Prop6.1}) in $\mathcal{C}^0\(L^1\(\R^{3}\times\R^{3}\),\R^+\)$}.
%\text{\emph{ $\gamma=\gamma(x,v;t)$ is in $\mathcal{C}\([0,T],L^1(\R^{3}\times\R^{3})\)$}},
\\\\
For the proof, see Appendix B.
\section{Convergence}
\setcounter{equation}{0}    
\def\theequation{7.\arabic{equation}}
This section is devoted to the proof of Theorem 6.1. \\\\
By (\ref{eq: DuhamelN}) and (\ref{eq: Teta_N}), for $k\geq 0$ we have:
\begin{eqnarray}\label{eq: expW_N^k}
W_N^{(k)}(Z_N;t)&&=\sum_{n\geq 0}\sum_{r=0}^{k}\sum_{\substack{r_1\dots r_n:\\r_j>0\\\sum r_j=k-r}}\int_{0}^{t}\ud t_1\int_{0}^{t_1}\ud t_2\dots \int_{0}^{t_{n-1}}\ud t_n \nonumber\\
&&\quad \quad \quad S_N(t-t_1)T_N^{(r_1)}S_N(t_1-t_2)\dots T_N^{(r_n)}S_N(t_n)W_{N,0}^{(r)}(Z_N).
\end{eqnarray}
%with the convention $t_0=t$.

It is useful to remind that the only non-vanishing terms in (\ref{eq: expW_N^k}) are those for which all $r_1,\dots,r_n$ are even.

According to (\ref{eq: W_Nindatcoeffk}) and (\ref{eq: expf_0COEFFk}),
\begin{eqnarray}\label{eq: W_NindatcoeffkSEC7}
&&W_{N,0}^{(r)}(Z_N)=\sum_{\substack{s_1\dots s_N\\0\leq s_j\leq r\\ \sum_j s_j=r}}\prod_{j=1}^{N}\(D_{G,j}^{2 s_j}f_0^{(0)}(z_j)\),
\end{eqnarray}
where $D_{G,j}^{k}$ is defined in (\ref{eq: operatorD_G}) and the extra symbol $j$ means that this operator acts on the variable $z_j$.
%The sum over $r_1,\dots,r_n$ appearing in (\ref{eq: expW_N^k}) is indeed a sum over all $even$ integers with the properties specified above, because there are no odd terms in the expansion of the operator $T_N^\var$. We have not made this feature explicit in the above expression but it is useful to remind it for what follows. \\\\Moreover, $W_{N,0}^{(r)}$ is given by (\ref{eq: W_Nindatcoeffk}). In (\ref{eq: opDstorto}) we defined the operator $\mathcal{D}^{2}$ such that $W_{N,0}^{(1)}=\mathcal{D}^{2}W_{N,0}^{(0)}$; now, in order to characterize the $r$-th order coefficient $W_{N,0}^{(r)}$ (for any $r$), we define 
Defining the operator $\mathcal{D}^{2r}$ as:
% in the case $r=1$ (see (\ref{eq: opDstorto})); for any $r\geq 0$, we have:
\begin{eqnarray}\label{eq: opDstortoGEN}
&&\mathcal{D}^{0}=\texttt{1},\nonumber\\
&&\mathcal{D}^{2r}=\sum_{\substack{s_1\dots s_N:\\0\leq s_j\leq r\\\sum_j s_j=r}}\prod_{j=1}^{N}D_{G,j}^{2s_j},\ \ \ r\geq 1,
\end{eqnarray}
%where the simbol $D_{G,j}^{2s_j}$ denotes, as specified in Section 5, the action of the operator $D_G^{2 s_j}$ with respect to the variable $z_j$ (at time $t=0$) and $D_G^{2 s_j}$ acts as it is described in (\ref{eq: operatorD_G}) and (\ref{eq: operatorD_Gcost}). Then, from (\ref{eq: W_Nindatcoeffk}) and (\ref{eq: opDstortoGEN}), we find that:
we have:
\begin{eqnarray}\label{eq: W_N,0^r}
W_{N,0}^{(r)}(Z_N)=\mathcal{D}^{2r}W_{N,0}^{(0)}(Z_N)\ \ \forall\ r\geq 0.
\end{eqnarray}

In order to investigate the behavior of the $j$-particle functions $W_{N,j}^{(k)}(Z_j;t)$ when $N\to\infty$, we consider the following object, for a given configuration $Z'_j=(z'_1\dots z'_j)$:
%%%can be conveniently expressed in terms of the empirical distribution $\mu_N$. Indeed, we find the following equality between
% $j$-particle 
%%%distributions in $\mathcal{S}'\(\R^{3j}\times\R^{3j}\)$: 
%ALL'INIZIO SONO IN Schwartz???????
%%%%%TOGLI LE FORMULE CON SCRITTE IN MEZZO ALLA FINE!!!!!!!!!!!!!!!!!!!
%%%%%FORSE CAMBIA NOTAZIONE INdat??? CI METTO SUBSCRIPT $in$.....\\
\begin{eqnarray}\label{eq: W_N,j^k}
&&\omega_{N,j}^{(k)}(Z'_j;t)=\int_{\R^{6N}}\ud Z_N\ W_{N}^{(k)}(Z_N;t)\mu_N(z'_1\vert Z_N)\dots \mu_N(z'_j\vert Z_N).
%+E_{N}^{1},\nonumber\\
\end{eqnarray}
In the end of the section, we will show that (\ref{eq: W_N,j^k}) is asymptotically equivalent to $W_{N,j}^{(k)}(Z'_j;t)$.

%%and we will characterize its limit. Thus, we will show that, 
%$W_{N,j}^{(k)}(Z_j;t)$ and (\ref{eq: W_N,j^k}) are asymptotically equivalent when $N\to\infty$, indeed, 
%%to conclude the proof of Theorem 6.1, it is enough to know the behavior of (\ref{eq: W_N,j^k}) for any $j$.\\
%=1,2,\dots,N$. \\
%%where, from now on, $E_N^p,\ p=1,2,\dots$ denote errors vanishing as $N\to\infty$. Estimates of these errors will be proven in Appendix C. 
%%%%%%%%%%%According to the decomposition
%%%%%%%%%%%\begin{eqnarray}\label{eq: opT_Ndec}
%%%%%%%%%%%&&T_N^{(r)}=\hat{T}_N^{(r)}+R_N^{(r)},
%%%%%%%%%%%\end{eqnarray}
%%%%%%%%%%%(see (\ref{eq: restoOPt})), 
From (\ref{eq: expW_N^k}), (\ref{eq: W_N,0^r}) and (\ref{eq: W_N,j^k}), it follows that:
\begin{eqnarray}\label{eq: expW_N^kBIS}
&&\omega_{N,j}^{(k)}(Z'_j;t)=\sum_{n\geq 0}\sum_{r=0}^{k}\sum_{\substack{r_1\dots r_n:\\r_j>0\\\sum r_j=k-r}}\int_{0}^{t}\ud t_1\int_{0}^{t_1}\ud t_2\dots \int_{0}^{t_{n-1}}\ud t_n \int_{\R^{6N}}\ud Z_N \mu_{N,j}(Z'_j\vert Z_N)\nonumber\\
&&\qquad \quad  
%\mu_N(z_1\vert Z_N)\dots \mu_N(z_j\vert Z_N)
S_N(t-t_1)T_N^{(r_1)}S_N(t_1-t_2)\dots T_N^{(r_n)}S_N(t_n)\mathcal{D}^{2r}W_{N,0}^{(0)}(Z_N),
%&&\qquad \quad  S_N(t-t_1)\hat{T}_N^{(r_1)}S_N(t_1-t_2)\dots \hat{T}_N^{(r_n)}S_N(t_n)\mathcal{D}^{2r}W_{N,0}^{(0)}(Z_N)+\nonumber\\
\end{eqnarray}
where
\begin{eqnarray}\label{eq: muperJ}
&&\mu_{N,j}(Z'_j\vert Z_N)=\mu_{N}(z'_1\vert Z_N)\dots \mu_{N}(z'_j\vert Z_N).
\end{eqnarray}
%%Moreover, $E_{N}^{1}$ is the error term arising from approximation of $W_{N,j}^{(k)}$ through the empirical measure $\mu_N$ and $E_{N}^{2}$ is the error term arising from the decomposition of the operators $T_N^{(r_j)}$. More precisely, $E_N^2$ arises from the terms, in the ''streak'' of operators $T_N^{(r_j)}$, in which some operator $R_N^{(r_j)}$ is involved. \\
%(in Appendix C we will see that their contribution is vanishing in the limit). \\
Integrating by parts, 
%in (\ref{eq: expW_N^kBIS}), 
reminding that each $r_j$ is even and that each $T_N^{(r_j)}$ involves derivatives of order $r_j +1$, 
%(namely, odd), 
we have:
\begin{eqnarray}\label{eq: expW_N^kTRIS}
\omega_{N,j}^{(k)}(Z'_j;t)&&=\sum_{n\geq 0}(-1)^n\sum_{r=0}^{k}\ \sum_{\substack{ {\bf \underline{r}}_n:\ r_j>0\\\vert {\bf \underline{r}}_n\vert =k-r}}\ \int_{ord}^{t}\ud  {\bf \underline{t}}_n\nonumber\\
&&\qquad \mathbb{E}_N\left[\mathcal{D}^{2r}T_N^{(r_n)}(t_n)T_N^{(r_{n-1})}(t_{n-1})\dots T_N^{(r_{1})}(t_{1}) \mu_{N,j}(Z'_j\vert Z_N(t))\right],
%+\nonumber\\&&+E_{N}^{1}+E_{N}^{2},
\end{eqnarray}
where ${\bf \underline{r}}_n$ is the sequence of positive integers $r_1,\dots ,r_n$, $\vert {\bf \underline{r}}_n\vert=\sum_{j=1}^{n}r_j$ and $Z_N(t)$ is the Hamiltonian flow defined in (\ref{eq: hamFLOW}). Moreover ${\bf \underline{t}}_n=t_1\dots t_n$ and $\int_{ord}^{t}\ud  {\bf \underline{t}}_n$ denotes the integral over he simplex $0<t_n<t_{n-1}<\dots <t_1<t$. Finally, $\mathbb{E}_N$ stands for the expectation with respect to the $N$-particle density $W_{N,0}^{(0)}$ and
\begin{eqnarray}\label{eq: opTtempo}
T_N^{(r)}(t)=S_N(-t)T_N^{(r)}S_N(t).
\end{eqnarray}
%Finally
%\begin{eqnarray}\label{eq: muperJ}
%&&\mu_{N,j}(Z_j\vert Z_N(t))=\mu_{N}(z_1\vert Z_N(t))\dots \mu_{N}(z_j\vert Z_N(t)).
%\end{eqnarray}
Therefore, the objects we have to investigate in the limit $N\to\infty$ are:
\begin{eqnarray}\label{eq: senza MEDIA}
\nu_j^{(k)}(Z'_j;t)&&=\sum_{n\geq 0}(-1)^n\sum_{r=0}^{k}\ \sum_{\substack{{\bf \underline{r}}_n:\ r_j>0\\\vert {\bf \underline{r}}_n\vert =k-r}}\ \int_{ord}^{t}\ud  {\bf \underline{t}}_n\eta_j(Z'_j;t,r,{\bf \underline{r}}_n,{\bf \underline{t}}_n,Z_N),\nonumber\\
&&
\end{eqnarray}
(for any configuration $Z_N$, typical with respect to $f_0^{(0)}$),
%$W_{N,0}^{(0)}$)
where $\eta_j$ is given by:
\begin{eqnarray}\label{eq: eta_j}
&&\eta_j(Z'_j;t,r,{\bf \underline{r}}_n,{\bf \underline{t}}_n,Z_N)=\mathcal{D}^{2r}T_N^{(r_n)}(t_n)T_N^{(r_{n-1})}(t_{n-1})\dots T_N^{(r_{1})}(t_{1}) \mu_{N,j}(Z'_j\vert Z_N(t)).\nonumber\\
&&
\end{eqnarray}

Note that:
\begin{eqnarray}\label{eq: senza MEDIA k=0}
\nu_j^{(0)}(Z'_j;t)=
%\eta_j(Z_j;t,0,{\bf \underline{r}}_0,{\bf \underline{t}}_0,Z_N)=
\mu_{N,j}\(Z'_j\vert Z_N(t)\).
%\ \ j=1,2,\dots
\nonumber\\
&&
\end{eqnarray}

We start by analyzing the behavior of $\nu_{j}^{(k)}$ in the cases $j=1,2$, thus we are lead to consider:
% $\nu_1^{(k)}(z_1;t)$ and $\nu_2^{(k)}(z_1, z_2;t)$ and the corresponding quantities
\begin{eqnarray}
&&\eta_1(z'_1;t,r,{\bf \underline{r}}_n,{\bf \underline{t}}_n,Z_N)=\mathcal{D}^{2r} T_N^{(r_n)}(t_n) T_N^{(r_{n-1})}(t_{n-1})\dots T_N^{(r_{1})}(t_{1}) \mu_{N}(z'_1\vert Z_N(t)),\label{eq: aUnCorpo}\nonumber\\
&&
\end{eqnarray}
and
\begin{eqnarray}
&&\eta_2(z'_1,z'_2;t,r,{\bf \underline{r}}_n,{\bf \underline{t}}_n,Z_N)=\mathcal{D}^{2r}T_N^{(r_n)}(t_n)T_N^{(r_{n-1})}(t_{n-1})\dots T_N^{(r_{1})}(t_{1}) \mu_{N,2}\(Z'_2\vert Z_N(t)\).\nonumber\\
&&\label{eq: aDueCorpi}
\end{eqnarray}
%where we used the notation $\mu_t(Z_j\vert Z_N):=\mu_N(Z_j\vert Z_N(t))$ for $j=1,2$. 

It is useful to stress that the operators $T_N^{(r_j)}(t_j)$  ($j=1,\dots,n$) and $\mathcal{D}^{2r}$ act as suitable distributional derivatives with respect to the variables $Z_N$. To evaluate $\eta_1$, let us first analyze the action of $T_N^{(r)}(\tau)$. By (\ref{eq: opTtempo}) and (\ref{eq: restoOPt}), for any function $G=G(Z_N)$, we have:
\begin{eqnarray}\label{eq: opTtemposuG}
&&\left(T_N^{(r)}\(\tau\)G\right)\(Z_N\)=
%S_N\(-\tau\)T_N^{(r)}S_N\(\tau\)G\(Z_N\)=\nonumber\\
S_N\(-\tau\)\(\hat{T}_N^{(r)}
%\left(S_N\(-\tau\)G\right)\(Z_N\(\tau\)\)
+R_N^{(r)}\)\(S_N\(\tau\)G\)\(Z_N\)=\nonumber\\
&&=(-1)^{r/2}\frac{c_r}{N}\sum_{j,l}S_N\(-\tau\)D_x^{r+1}\phi(x_j-x_l)\cdot D_{v_j}^{r+1}\(S_N\(\tau\)G\)\(Z_N\)+\nonumber\\
&&+\frac{1}{N}\sum_{l,j=1}^{N}\sum_{\substack{k_1,k_2\in \mathbb{N}^{3}\\ \vert k_1\vert+\vert k_2\vert=r+1}}C_{k_1, k_2}S_N\(-\tau\)\frac{\pa^{r+1}}{\pa_{x_l}^{\vert k_1\vert}\pa_{x_j}^{\vert k_2\vert}}\phi(x_l - x_j)\cdot\frac{\pa^{r+1}}{\pa_{v_l}^{\vert k_1\vert}\pa_{v_j}^{\vert k_2\vert}}\(S_N\(\tau\)G\)\(Z_N\).\nonumber\\
&&
\end{eqnarray}
Note that the derivatives involved here are done with respect to the variables at time $t=0$.

Denoting by $D_{z_j}^{r}$ any derivative of order $r$ with respect to a variable $z_j$ at time $t=0$, we observe that:
\begin{eqnarray}\label{eq: derTEMPsec7}
&& S_N(-t)D_{z_j}^{r}G(Z_N)=\(D_{z_j}^{r}G\)(Z_N(t))=D_{z_j}^{r}(t)\(S_N(-t)G\)(Z_N),
\end{eqnarray}
where, by $D_{z_j}^{r}(t)$, we denote the same derivative of order $r$ with respect to the variable $z_j(t)$. Then, by (\ref{eq: derTEMPsec7}) and (\ref{eq: opTtemposuG}):
\begin{eqnarray}\label{eq: opTtemposuGbis}
&&\left(T_N^{(r)}\(\tau\)G\right)\(Z_N\)=S_N\(-\tau\)\(\hat{T}_N^{(r)}
%\left(S_N\(-\tau\)G\right)\(Z_N\(\tau\)\)
+R_N^{(r)}\)S_N\(\tau\)G\(Z_N\)=\nonumber\\
&&=(-1)^{r/2}\frac{c_r}{N}\sum_{j,l}\(D_x^{r+1}\phi\)(x_j(\tau)-x_l(\tau))\cdot D_{v_j}^{r+1}(\tau)G\(Z_N\)+\nonumber\\
&&+\frac{1}{N}\sum_{l,j=1}^{N}\sum_{\substack{k_1,k_2\in \mathbb{N}^{3}\\ \vert k_1\vert+\vert k_2\vert=r+1}}C_{k_1, k_2}\(\frac{\pa^{r+1}}{\pa_{x_l}^{\vert k_1\vert}\pa_{x_j}^{\vert k_2\vert}}\phi\)(x_l(\tau) - x_j(\tau))\cdot\frac{\pa^{r+1}}{\pa_{v_l}^{\vert k_1\vert}\pa_{v_j}^{\vert k_2\vert}}(\tau)G\(Z_N\).\nonumber\\
&&
\end{eqnarray}
%&&\qquad=(-1)^{r/2}\frac{c_r}{N}\sum_{j,l}D_x^{r+1}\phi(x_j(\tau)-x_l(\tau))\cdot D_{v_j}^{r+1}\left(G\(Z_N\)\right).
Therefore, in computing the action of $T_N^{(r)}(\tau)$, we have to consider derivatives with respect to the variables at time $\tau$. 
%In particular, in evaluating 
%$$
%D_{v_j}^{r+1}\left(S_N(-\tau)G\right)(Z_N(-\tau)),
%$$
As a consequence, we have to deal with a complicated function of the configuration $Z_N$ which, however, we do not need to make explicit, as we shall see in a moment.

On the basis of the previous considerations,
we compute the time derivative of $\eta_1$ by applying the operators $\mathcal{D}^{2r}T_N^{(r_n)}(t_n)T_N^{(r_{n-1})}(t_{n-1})\dots T_N^{(r_{1})}(t_{1}) $ to the Vlasov equation:
\begin{eqnarray}\label{eq: VlasovWEAKcap7}
&&\left(\pa_t+v'_1\cdot \nabla_{x'_1}\right)\mu_N(t)=\(\nabla_{x'_1}\phi\ast\mu_N(t)\)\cdot\nabla_{v'_1}\mu_N(t).
\end{eqnarray}
In doing this we have to compute 
\begin{eqnarray}\label{eq: VlasovWEAKcap7BIS}
&&\mathcal{D}^{2r}T_N^{(r_{n})}(t_{n})T_N^{(r_{n-1})}(t_{n-1})  \dots T_N^{(r_{1})}(t_{1}) \mu_N(z'_1\vert Z_N(t))\mu_N(z'_2\vert Z_N(t)).
\end{eqnarray}
Now we select the contribution in which each $T_N^{(r_{\ell})}(t_{\ell})$ and $\mathcal{D}^{2r}$ apply either on $ \mu_N(z'_1\vert Z_N(t))$ or to $\mu_N(z'_2\vert Z_N(t))$. The other contribution involves terms in which are present products of derivatives with respect to the same variable. By Proposition 5.1 and Proposition 5.2 we expect those terms to be negligible (in the $\mathcal{C}^{\infty}_{b}$-weak sense) in the limit $N\to\infty$.
Therefore we obtain the following equation:
\begin{eqnarray}\label{eq: eqPEReta}
&&\left(\pa_t+v'_1\cdot \nabla_{x'_1}\right)\eta_1(z'_1,t,r,{\bf \underline{r}}_n,{\bf \underline{t}}_n,Z_N)=L(\mu_N(t))\eta_1(z'_1,t,r,{\bf \underline{r}}_n,{\bf \underline{t}}_n,Z_N)+\nonumber\\
&&\quad +\sum_{0\leq \ell\leq r}\ \sum_{0\leq m\leq n}\sum_{\substack{I\subset I_n:\\ \vert I\vert=m,\\0<\vert{\bf  \underline{r}}_I\vert+\ell<k}}\(\nabla_{x'_1}\phi\ast\eta_1(\cdot,t,\ell,{\bf \underline{r}}_I,{\bf \underline{t}}_I,Z_N)\)\cdot\nabla_{v'_1}\eta_1(z'_1,t,r-\ell,{\bf \underline{r}}_{I_n\setminus I},{\bf \underline{t}}_{I_n\setminus I},Z_N)+\nonumber\\
&&\quad+E_N^1,
\end{eqnarray}
where $E_N^1$ is an error term which will be proven to be negligible in the limit $N\to\infty$ in Appendix C.
%arising from the action of $\mathcal{D}^{2r}\hat{T}_N^{(r_n)}(t_n)\hat{T}_N^{(r_{n-1})}(t_{n-1})\dots\hat{T}_N^{(r_{1})}(t_{1})$ on the the right hand side of (\ref{eq: VlasovWEAKcap7}) (which ''works'' as a tensor product of two empirical measures), and in Appendix C we will prove that it is vanishing in the limit by using essentially properties (\ref{eq: opTtemposuG1G2}) and (\ref{eq: opDstorto su G_1G_2}). 
In (\ref{eq: eqPEReta}) we used the notations:
\begin{eqnarray}\label{eq: NOTAZIONI}
&& I_n=\{1,2,\dots,n\},\ I\ \text{is any subset of}\ I_n,\  {\bf \underline{r}}_I=\{r_j\}_{j\in I}, \  {\bf \underline{t}}_I=\{t_j\}_{j\in I}.\nonumber\\
&&
\end{eqnarray}

Next, we compute the time derivative of $\nu_1^{(k)}(t)$. We have:
\begin{eqnarray}\label{eq: eqPERnu}
&&\left(\pa_t+v'_1\cdot \nabla_{x'_1}\right)\nu_1^{(k)}=\nonumber\\
&&=\sum_{n\geq 0}(-1)^n\sum_{r=0}^{k}\sum_{\substack{\vert {\bf \underline{r}}_n\vert:\\r_j>0\\\vert {\bf \underline{r}}_n\vert =k-r}}\left.\int_{0}^t \ud t_2\int_0^{t_2}\ud t_3\dots \int_0^{t_{n-1}}\ud t_n \eta_1\(z'_1;t,r,{\bf \underline{r}}_n,{\bf \underline{t}}_n,Z_N\)\right \vert_{t_1=t}+\nonumber\\
&&+\sum_{n\geq 0}(-1)^n\sum_{r=0}^{k}\sum_{\substack{\vert {\bf \underline{r}}_n\vert:\\r_j>0\\ \vert {\bf \underline{r}}_n\vert =k-r}}\int_{ord}^{t}\ud  {\bf \underline{t}}_n \left(\pa_t+v'_1\cdot \nabla_{x'_1}\right)\eta_1(z'_1;t,r,{\bf \underline{r}}_n,{\bf \underline{t}}_n,Z_N).
\end{eqnarray}
%where $E_N^2$ is obtained by $E_N^1$.\\
%, multiplying by $(-1)^n$ and summing over $n$, $r$ and ${\bf \underline{r}}_n$.\\ 
In evaluating the first term on the right hand side of (\ref{eq: eqPERnu}), we are lead to consider $\eta_1$ evaluated in $t=t_1$. Thus, according to the expression of $\eta_1$ (see (\ref{eq: aUnCorpo})), we have to deal with:
\begin{eqnarray}\label{eq: sorgPERnu}
T_N^{(r_1)}(t)\mu_N(z'_1\vert Z_N(t))=S_N(-t)T_N^{(r_1)}\mu_N(z'_1\vert Z_N).
\end{eqnarray}
%since the term appearing inside the integral at the first line of (\ref{eq: eqPERnu}) is evaluated for $t_1=t$. 
Therefore:
%&&+\sum_{j=1}^{N}\sum_{\substack{k_1,k_2\in \mathbb{N}^{3}\\ \vert k_1\vert+\vert k_2\vert=r+1}}C_{k_1, k_2}D_{x_j}^{\vert k_2\vert}\(\phi\ast D_{x}^{\vert k_1\vert}\mu_N(t)\)(x_j(t))\cdot\frac{\pa^{r+1}}{\pa_{v_l}^{\vert k_1\vert}\pa_{v_j}^{\vert k_2\vert}}\mu_N(z\vert Z_N(t))\nonumber\\
\begin{eqnarray}\label{eq: sorgPERnuCONTO}
&&T_N^{(r_1)}(t)\mu_N(z'_1\vert Z_N(t))
%(-1)^{r_1/2}\frac{c_{r_1}}{N}\sum_{j,l}\(D_x^{r_1 +1}\phi\)(x_j(t)-x_l(t))\cdot D_{v_j}^{r_1 +1}(t)\mu_N(z\vert Z_N(t))+\nonumber\\
%&&+\frac{1}{N}\sum_{j,l}\sum_{\substack{k_1,k_2\in \mathbb{N}^{3}\\ \vert k_1\vert+\vert k_2\vert=r_1 +1}}C_{k_1, k_2}\(\frac{\pa^{r_1 +1}}{\pa_{x_l}^{\vert k_1\vert}\pa_{x_j}^{\vert k_2\vert}}\phi\)(x_l(t) - x_j(t))\cdot\frac{\pa^{r_1 +1}}{\pa_{v_l}^{\vert k_1\vert}\pa_{v_j}^{\vert k_2\vert}}(t)\mu_N(z\vert Z_N(t))+\nonumber\\
%&& = (-1)^{r_1/2}c_{r_1}\sum_{j}\left(D_x^{r_1 +1}\phi\ast\mu_N(t)\right)(x_j(t))\cdot D_{v_j}^{r_1 +1}(t)\mu_N(z\vert Z_N(t))+\nonumber\\
%&&+\sum_{j}\sum_{\substack{k_1,k_2\in \mathbb{N}^{3}\\ \vert k_1\vert+\vert k_2\vert=r+1}}C_{k_1, k_2}D_{x_j}^{\vert k_2\vert}\left(D_x^{\vert k_1\vert}\phi\ast\mu_N(t)\right)(x_j(t))\cdot\frac{\pa^{r_1 +1}}{\pa_{v_l}^{\vert k_1\vert}\pa_{v_j}^{\vert k_2\vert}}(t)\mu_N(z\vert Z_N(t))=\nonumber\\
%&&=(-1)^{r_1/2}\frac{c_{r_1}}{N}\sum_{j}\left(D_x^{r_1 +1}\phi\ast\mu_N(t)\right)(x_j(t))\cdot D_{v_j}^{r_1 +1}(t)\delta(z-z_j(t))=\nonumber\\
= (-1)^{r_1/2}c_{r_1}\left(D_{x'_1}^{r_1 +1}\phi\ast\mu_N(t)\right)(x'_1)\cdot D_{v'_1}^{r_1 +1}\mu_N(z'_1\vert Z_N(t))=\nonumber\\
&& = (-1)^{r_1/2}c_{r_1}\int
%_{\R^3\times\R^3}
\ud x'_2\ \ud v'_2\ D_{x'_1}^{r_1 +1}\phi(x'_1-x'_2)\cdot D_{v'_1}^{r_1 +1}\mu_N(x'_1,v'_1\vert Z_N(t))\mu_N(x'_2,v'_2\vert Z_N(t)),\nonumber\\
&&
\end{eqnarray}
where the term involving off-diagonal derivatives, namely $R_N^{(r_1)}$ (see (\ref{eq: OFFdiagOPt})),
%with respect to $v_j$ and $v_l$ 
disappears because both the derivatives and the empirical distribution are evaluated at time $t$.
%Then, thanks to (\ref{eq: sorgPERnuCONTO}) and to (\ref{eq: aUnCorpo}), we are able to 
Hence we compute $\eta_1$ in $t=t_1$ and, inserting it in the first term of the right hand side of (\ref{eq: eqPERnu}), we obtain:
\begin{eqnarray}\label{eq: sorgPERnuRESULT}
&& \sum_{n\geq 0}(-1)^n\sum_{r=0}^{k}\ \sum_{\substack{ {\bf \underline{r}}_n:\ r_j>0\\\vert {\bf \underline{r}}_n\vert =k-r}}\ \left.\int_{0}^{t}\ud  t_2\int_{0}^{t_2}\ud t_3 \ud t_3\dots \int_{0}^{t_{n-1}}\ud t_n \eta_1\(z'_1;t,r,{\bf \underline{r}}_n,{\bf \underline{t}}_n,Z_N\)\right \vert_{t_1=t}=\nonumber\\
%&&=\sum_{n\geq 0}(-1)^n\sum_{r=0}^{k}\ \sum_{\substack{ {\bf \underline{r}}_n:\ r_j>0\\\vert {\bf \underline{r}}_n\vert =k-r}}\ \int_{ord}^{t}\ud  {\bf \underline{t}}_n (-1)^{r_1/2}c_{r_1}\int_{\R^3\times\R^3}\ud y\ \ud w\ D_x^{r_1 +1}\phi(x-y)\cdot D_{v}^{r_1 +1}\nonumber\\
%&&\qquad \qquad\mathcal{D}^{2r}T_N^{(r_n)}(t_n)T_N^{(r_{n-1})}(t_{n-1})\dots T_N^{(r_{2})}(t_{2}) \mu_{N,2}(x,v,y,w\vert Z_N(t))=\nonumber\\
&&=\sum_{ \substack{0<r_1\leq k\\r_1 \text{\ even}}}\ (-1)^{r_1/2}c_{r_1}\int
%_{\R^3\times\R^3}
\ud x'_2\ \ud v'_2\ D_{x'_1}^{r_1 +1}\phi(x'_1-x'_2)\cdot D_{v'_1}^{r_1 +1}\nu_2^{(k-r_1)}(x'_1,v'_1,x'_2,v'_2;t).\nonumber\\
&&
\end{eqnarray}
%where we made explicit that the sum over $r_1$ is over all $even$ integers as specified above (see Remark 7.0), because this feature determines which are the two-particles functions at orders lower than $k$ that are involved in (\ref{eq: sorgPERnuRESULT}).\\

Let us come back now to equation (\ref{eq: eqPERnu}). It is useful to observe that:
\begin{eqnarray}\label{eq: integraliINtempo}
\int_{ord}^t \ud{\bf \underline{t}}_{n} \sum_{\substack{I\subset I_n:\\|I|=m}}=
\int_{ord}^t \ud{\bf \underline{t}}_I \int_{ord}^t \ud{\bf \underline{t}}_{I_n\setminus I}.
\end{eqnarray}
Then, putting together (\ref{eq: eqPERnu}), (\ref{eq: sorgPERnuRESULT}), (\ref{eq: eqPEReta}) and (\ref{eq: integraliINtempo}), we obtain the following equation for $\nu_1^{(k)}$:
\begin{eqnarray}\label{eq: eqPERnuDEF}
&&\left(\pa_t+v'_1\cdot \nabla_{x'_1}\right)\nu_1^{(k)}(x'_1,v'_1;t)=L(\mu_N(t))\nu_1^{(k)}(x'_1,v'_1;t)+\nonumber\\
&&\qquad +\sum_{\substack{ 0<r_1\leq k\\r_1 \text{\ even}}}(-1)^{r_1/2}c_{r_1}\int
%_{\R^3\times\R^3}
\ud x'_2\ \ud v'_2\ D_{x'_1}^{r_1 +1}\phi(x'_1-x'_2)\cdot D_{v'_1}^{r_1 +1}\nu_2^{(k-r_1)}(x'_1,v'_1,x'_2,v'_2;t)+\nonumber\\
&&\qquad+\sum_{0<\ell<k}\(\nabla_{x'_{1}}\phi\ast\nu_1^{(\ell)}(t)\)\cdot\nabla_{v'_{1}}\nu_1^{(k-\ell)}(t)+ E_N^2,
\end{eqnarray}
with initial datum given by:
\begin{eqnarray}\label{eq: eqPERnuINDAT}
\nu_1^{(k)}(x'_1,v'_1;t)\vert_{t=0}=\eta_1((z'_1;0,k,{\bf \underline{r}}_0,{\bf \underline{t}}_0,Z_N)=\mathcal{D}^{2k}\mu_{N}(z'_1\vert Z_N).
\end{eqnarray}
Here $E_N^2$ arises from $E_N^1$ (see (\ref{eq: eqPEReta})).
Now, we want to prove that:
%, for any fixed configuration $Z_N$ which is typical with respect to $W_{N,0}^{(0)}$, we have:
\begin{eqnarray}\label{eq:convergenzaQO}
&&\nu_{1}^{(k)}(t)\rightarrow f^{(k)}(t),\ \text{as}\ \ N\to\infty,\ \ \mathcal{C}_b^{\infty}-\text{weakly},
%%%&&\text{for any fixed configuration $Z_N$ which is typical with respect to $W_{N,0}^{(0)}$}\nonumber
%&& \text{a.e with respect to $W_{N,0}^{(0)}$},
\end{eqnarray}
and
\begin{eqnarray}\label{eq:convergenzaQOadue}
&&\nu_{2}^{(k)}(t)\rightarrow f_2^{(k)}(t),\ \text{as}\ \ N\to\infty,\ \ \mathcal{C}_b^{\infty}-\text{weakly},
%%%&&\text{for any fixed configuration $Z_N$ which is typical with respect to $W_{N,0}^{(0)}$}.\nonumber
%&& \text{a.e with respect to $W_{N,0}^{(0)}$},
\end{eqnarray}
for any configuration $Z_N$ such that $\mu_N\to f_0^{(0)}$ in the weak sense of probability measure (namely, for any $Z_N$ typical with respect to $f_0^{(0)}$).
%in the sequel such configurations will be called ''typical'' with respect to $W_{N,0}^{(0)}$ ). 
As a consequence, reminding that $\nu_{1}^{(k)}(t)$ and $\nu_{2}^{(k)}(t)$ are equal to $\omega_{N,1}^{(k)}(t)$ and $\omega_{N,2}^{(k)}(t)$ respectively, a.e. with respect to $W_{N,0}^{(0)}$, 
%(except for an error going to zero in the limit), 
(\ref{eq:convergenzaQO}) and (\ref{eq:convergenzaQOadue}) are equivalent to:
%%These convergences can be rewritten as:
%%\begin{eqnarray}\label{eq:convergenzaQOBIS}
%%&&\mathbb{E}_N[\nu_{1}^{(k)}(t)]\stackrel{N\to\infty}{\rightarrow}f^{(k)}(t),\ \ \mathcal{C}_b^{\infty}-\text{weakly},
%&& \text{a.e with respect to $W_{N,0}^{(0)}$},
%%\end{eqnarray}
%%and
%%\begin{eqnarray}\label{eq:convergenzaQOBISadue}
%%&&\mathbb{E}_N[\nu_{2}^{(k)}(t)]\stackrel{N\to\infty}{\rightarrow}f_2^{(k)}(t),\ \ \mathcal{C}_b^{\infty}-\text{weakly},
%&& \text{a.e with respect to $W_{N,0}^{(0)}$},
%%\end{eqnarray}
%%which are equivalent to:
\begin{eqnarray}\label{eq:convergenza}
&&\omega_{N,1}^{(k)}(t)\rightarrow f^{(k)}(t),\ \text{as}\ \ N\to\infty,\ \ \mathcal{C}_b^{\infty}-\text{weakly},
\end{eqnarray}
and 
\begin{eqnarray}\label{eq:convergenzaadue}
&&\omega_{N,2}^{(k)}(t)\rightarrow f_2^{(k)}(t),\ \text{as}\ \ N\to\infty,\ \ \mathcal{C}_b^{\infty}-\text{weakly}.
\end{eqnarray}

As we already remarked, the $\mathcal{C}_b^{\infty}$-weak convergence implies the convergence in $\mathcal{S}'$, therefore, (\ref{eq:convergenza}) and (\ref{eq:convergenzaadue}) imply the convergence of $\omega_{N,1}^{(k)}(t)$ to $f^{(k)}(t)$ in $\mathcal{S}'(\R^3\times\R^3)$ and of $\omega_{N,2}^{(k)}(t)$ to $f_2^{(k)}(t)$ in $\mathcal{S}'(\R^{6}\times\R^6)$. 
%We will see at the end of this Section that  which is properly what is established by (\ref{eq: Th6.1}) in the cases $j=1$ and $j=2$. 
%and, in this way, to obtain again the empirical measure $\mu_N$these derivatives of $\mu_N$ ''versus'' bounded differentiable functions (thus it will be possible to integrate by parts).\\\\ 
%%%{\bf Remark 7.2}:\\
%%%From (\ref{eq: eqPERnuDEF}) we see that, in order to prove (\ref{eq:convergenzaQO}), we will need inductive assumptions on the convergence of %%%$\tilde{w}_{N,1}^{(h)}(t)$ and $\tilde{w}_{N,2}^{(h)}(t)$ for each $h<k$. 
\\\\
$\mathbf{7.1.}$ {\bf One and two-particle convergence.}\  In evaluating the behavior of $\nu_1^k(t)$ when $N\to\infty$, we note that it solves the initial value problem (\ref{eq: eqPERnuDEF})-(\ref{eq: eqPERnuINDAT}) for which we want to use Proposition 6.2. First, however, we have to verify the assumptions. The first one, namely {\bf i)}, is verified as follows by the considerations developed in Section 5.

Now, we have to check that assumption {\bf ii)} is satisfied, namely, we have to prove that
\begin{eqnarray}\label{eq:uniformBOUNDNESS}
&&\forall\ u_1,\ u_2\ \text{in\ }\ \mathcal{C}_b^{\infty}(\R^3\times\R^3),\nonumber\\
&& \text{there exists a constant $C=C(u_1,u_2)>0$, independent of $N$, such that:}\nonumber\\
&&\left\|  u_1 \ast \ \(u_2\ \nu_{1}^{(k)}(t)\) \right\|_{L^{\infty}\(\R^3\times\R^3\)}<C\ \ \ \text{for any} \  t.
\end{eqnarray}
%&& \text{sup}_{x,v}\left\vert\left( u_1* \ v\ \tilde{w}_{N,1}^{(k)}(t)\right)(x,v)\right\vert<C,\ \ \ \forall\ u,\ v\ \text{in\ }\ \mathcal{C}_b^{\infty}(\R^d\times\R^d).
%\end{eqnarray}
%%The property above is simply verified thanks to the structure of $\nu_{1}^{(k)}(t)$ and, more precisely, to the expression of $\eta_1$. In fact, w
We have:
\begin{eqnarray}\label{eq:uniformBOUNDNESSproof}
&& \left\|  u_1 \ast \ \(u_2\ \nu_{1}^{(k)}(t)\) \right\|_{L^{\infty}\(\R^3\times\R^3\)}
%=\text{sup}_{x,v}\left\vert\left( u_1\ast \ u_2\ \nu_{1}^{(k)}(t)\right)(x,v)\right\vert
=\sup_{x'_1,v'_1}\left\vert\int \ud y\ud w\ u_1(x'_1-y, v'_1-w)u_2(y,w)\nu_{1}^{(k)}(y, w; t)\right\vert\leq\nonumber\\
&&\leq\sum_{n\geq 0}\sum_{r=0}^{k}\ \sum_{\substack{ {\bf \underline{r}}_n:\ r_j>0\\\vert {\bf \underline{r}}_n\vert =k-r}}\ \int_{ord}^{t}\ud  {\bf \underline{t}}_n\nonumber\\
&&\ \ \ \sup_{x'_1,v'_1}\left\vert\int \ud y\ud w\ u_1(x'_1-y, v'_1-w)u_2(y,w)\eta_{1}(y, w; t;r,{\bf \underline{r}}_n ,{\bf \underline{t}}_n ,Z_N)\right\vert=\nonumber\\
&& =\sum_{n\geq 0}\sum_{r=0}^{k}\ \sum_{\substack{{\bf \underline{r}}_n:\ r_j>0\\\vert {\bf \underline{r}}_n\vert =k-r}}\ \int_{ord}^{t}\ud  {\bf \underline{t}}_n\nonumber\\
&&\ \ \ \sup_{x'_1,v'_1}\left\vert\int \ud y\ud w\ \left(u_1(x'_1-y, v'_1-w)u_2(y,w)\right)\mathcal{D}^{2r}T_N^{(r_n)}(t_n)\dots T_N^{(r_1)}(t_1)\mu_N(y, w\vert Z_N(t))\right\vert=\nonumber\\
&& =\sum_{n\geq 0}\sum_{r=0}^{k}\ \sum_{\substack{{\bf \underline{r}}_n:\ r_j>0\\\vert {\bf \underline{r}}_n\vert =k-r}}\ \int_{ord}^{t}\ud  {\bf \underline{t}}_n\nonumber\\
&&\ \ \ \sup_{x'_1,v'_1}\left\vert\int \ud y\ud w\ g(x'_1,v'_1,y,w)\mathcal{D}^{2r}T_N^{(r_n)}(t_n)\dots T_N^{(r_1)}(t_1)\mu_N(y, w\vert Z_N(t))\right\vert,
%&&=\sum_{n\geq 0}\sum_{r=0}^{k}\ \sum_{\substack{{\bf \underline{r}}_n:\ r_j>0\\\vert {\bf \underline{r}}_n\vert =k-r}}\ \int_{ord}^{t}\ud  {\bf \underline{t}}_n\nonumber\\
%&&\ \ \ \text{sup}_{x,v}\left\vert\frac{1}{N}\sum_{k=1}^{N}\left(\mathcal{D}^{2r}T_N^{(r_n)}(t_n)\dots T_N^{(r_1)}(t_1)\right)^{*}\left(u_1\(x-x_k(t),v-v_k(t)\)u_2\(x_k(t), v_k(t)\)\right)\right\vert,\nonumber\\
\end{eqnarray}
where we used the notation $g(x'_1,v'_1,y,w):=u_1(x'_1-y, v'_1-w)u_2(y,w)$ and, 
%in order to emphasize that, in this context, $x$ and $v$ play the role of parameters and the variables are indeed $y$ and $w$. 
clearly, we have $g(x'_1,v'_1,\cdot,\cdot)\in \mathcal{C}^{\infty}_b(\R^3\times\R^3)$ for any $x'_1$ and $v'_1$ and $g(\cdot,\cdot,y,w)\in \mathcal{C}^{\infty}_b(\R^3\times\R^3)$ for any $y$ and $w$.
%where $\left(\mathcal{D}^{2r}T_N^{(r_n)}(t_n)\dots T_N^{(r_1)}(t_1)\right)^{*}$ is the adjoint of the operator $\mathcal{D}^{2r}T_N^{(r_n)}(t_n)\dots T_N^{(r_1)}(t_1)$. 
By some estimates which will be proven in Appendix C (see Lemma C.2), we are guaranteed that, applying the operator $\mathcal{D}^{2r}T_N^{(r_n)}(t_n)\dots T_N^{(r_1)}(t_1)$ on the empirical measure $\mu_N(t)$ and integrating versus a function in $\mathcal{C}^{\infty}_b(\R^3\times\R^3)$ we obtain a quantity uniformly bounded in $N$. This feature, by virtue of the good properties of the function $g$
%, together with the fact that $\mu_N$ has bounded variation, 
ensures that 
%the quantity we obtained in 
(\ref{eq:uniformBOUNDNESSproof}) is finite. 

Let us now look at the initial datum for $\nu_1^{(k)}(t)$, in order to verify assumption {\bf iii)}. 

From (\ref{eq: eqPERnuINDAT}) we know that $\nu_1^{(k)}(0)=\mathcal{D}^{2k}\mu_N\in \mathcal{S}'(\R^{3}\times\R^{3})$. As regard to its limiting behavior, we find that:
\begin{eqnarray}\label{eq: convINDATstruttura}
&&\left. \nu_{1}^{(k)}(t)\right\vert_{t=0}=\mathcal{D}^{2k}\mu_N=
% \sum_{\substack{s_1,\dots,s_N\\0\leq s_j\leq k\\\sum_{j} s_j=k}}\prod_{j=1}^{N}D_{G,j}^{2 s_j}\mu=\nonumber\\
\sum_{n=1}^{N}\ \sum_{\substack{I\subset I_N\\ \vert I\vert=n}}\ \sum_{\substack{s_j: j\in I\\1\leq s_j\leq k\\\sum_{j} s_j=k}}\prod_{j\in I}D_{G,j}^{2 s_j}\ \mu_N,
\end{eqnarray}
where $I_N=\{1,\dots,N\}$.
For our convenience, we have written the action of the operator $\mathcal{D}^{2k}$ in a equivalent and slightly different way from that we used in (\ref{eq: opDstortoGEN}).
%, but with some calculation it is not difficult to see that they are perfectly equivalent. \\
%Looking at the definition of the operator $\mathcal{D}^{2r}$ given in (\ref{eq: opDstortoGEN}), it is not difficult to see that its action can be written as in (\ref{eq: convINDATstruttura}).
%The operator $\mathcal{D}$
%where $n$ is the number of initial one-particle functions that are different from the zero order one. 

We realize that the only surviving term in the sum (\ref{eq: convINDATstruttura}) is that with $n=1$.
%, because the others involve products of derivatives with respect to different variables, which, applied to $\mu_N$, are vanishing. 
Hence:
%Then, from (\ref{eq: convINDATstruttura}), it follows that:
%\begin{eqnarray}\label{eq: convINDATstruttura2}
%&&\nu_{1}^{(k)}(t)\vert_{t=0}=N D_{G,1}^{2 k}\mu_N+\sum_{n=2}^{N}\frac{N!}{n!(N-n)!}\sum_{\substack{s_1,\dots,s_n\\1\leq s_j\leq k\\\sum_{j} s_j=k}}\prod_{j=1}^{n}D_{G,j}^{2 s_j}\mu_N, \nonumber\\
%&&
%\end{eqnarray}
%where the first term corresponds to the case $n=1$. We observe that the second term in the right hand side of (\ref{eq: convINDATstruttura2}) is equal to zero. In fact, there are derivative with respect to the variables associated with at least two different particles and, because of the fact that all the variables in the game are at time $t=0$, this kind of derivatives applied to $\mu_N$ are equal to zero. Then, we have to deal only with the term $N D_{G,1}^{2 k}\mu$ and we have:
\begin{eqnarray}\label{eq: convINDATstruttura3}
&& \ \left. \nu_1^k(t)\right\vert_{t=0}=\sum_{j=1}^N  D_{G,j}^{2 k}\mu_N=\frac{1}{N}\sum_{j=1}^{N}D_{G,j}^{2 k}\delta(z'_1-z_j)=D_{G}^{2 k}\mu_N.
%=D_{G,1}^{2 k}\delta(z-z_1)=\nonumber\\
%&&=\frac{1}{N}\sum_{h=1}^{N}D_{G,h}^{2 k}\delta(z-z_h)=
%\sum_{h=1}^{N}D_{z_h}^{2 k}\mu=\frac{1}{N}\sum_{h=1}^{N}\sum_{j=1}^{N}D_{z_h}^{2 k}\delta(z-z_j)=\frac{1}{N}\sum_{h=1}^{N}D_{z_h}^{2 k}\delta(z-z_h)=\nonumber\\
\end{eqnarray}
Therefore we can conclude, by using the mean-field limit:
\begin{eqnarray}\label{eq: convINDAT}
\left(u, \nu_{1}^{(k)}(t)\vert_{t=0}\right)&&=\left(u,D_{G}^{2 k}\mu_N\right)=\nonumber\\
&&=\left( D_{G}^{2 k}u,\mu_N\right)\rightarrow \left(D_{G}^{2 k} u,f_0^{(0)}\right)=\(u,D_{G}^{2 k}f_0^{(0)}\right)=\(u,f_0^{(k)}\right),\ \text{as}\ \ N\to\infty,\ \ \nonumber\\
%&& \qquad\qquad\qquad\qquad\qquad= \left( u, f_0^{(k)}\right)\nonumber\\
&&\ \forall\ u\ \text{in}\ \ \mathcal{C}_{b}^{\infty}\left(\R^3\times\R^3\right).
\end{eqnarray}
Thus, $f_0^{(k)}$ plays the role of $\gamma_0$ in Proposition 6.2 and it is in $L^1\(\R^3\times \R^3\)$ because $f_0^{(0)}\in \mathcal{S}\(\R^3\times \R^3\)$.

We conclude the convergence proof (for the one and two-particle functions) by induction.
For $k=0$ we know that, for any configuration $Z_N$ which is typical with respect to $f_0^{(0)}$, 
%$W_{N,0}^{(0)}$, 
we have: 
\begin{eqnarray}\label{eq:convergenzaQOclas}
&&\nu_{1}^{(0)}(t)=\mu_N(t)\rightarrow f^{(0)}(t),\ \ \text{as}\ \ N\to\infty,
%&& \text{in the weak sense of probability measures, and, as a consequence,}\nonumber\\
%&&\text{ the convergence holds } \mathcal{C}_{b}^{\infty}-\text{weakly}.\nonumber\\
%&&
\end{eqnarray}
in the weak sense of probability measures, and, as a consequence,
the convergence holds $\mathcal{C}_{b}^{\infty}-\text{weakly}$. Moreover
\begin{eqnarray}\label{eq:convergenzaQOclasAdue}
&&\nu_{2}^{(0)}(t)=\mu_N(t)\otimes\mu_N(t)\rightarrow f_2^{(0)}(t)=f^{(0)}(t)\otimes f^{(0)}(t),\ \text{as}\ \ N\to\infty,
%&& \text{in the weak sense of probability measures, and, as a consequence,}\nonumber\\
%&&\text{ the convergence holds } \mathcal{C}_{b}^{\infty}-\text{weakly}.\nonumber\\
%&&
\end{eqnarray}
in the weak sense of probability measures, and, as a consequence,
the convergence holds $\mathcal{C}_{b}^{\infty}-\text{weakly}$.

We make the following inductive assumptions for all $h<k$:
\begin{eqnarray}\label{eq:convergenzaQOinduz}
&&\nu_{1}^{(h)}(t)\rightarrow f^{(h)}(t),\ \text{as}\ \ N\to\infty,\ \ \  \mathcal{C}_{b}^{\infty}-\text{weakly}, 
%\text{for any configuration $Z_N$ which is typical with respect to $W_{N,0}^{(0)}$, }
\end{eqnarray}
for any configuration $Z_N$ which is typical with respect to $f_0^{(0)}$,
%$W_{N,0}^{(0)}$, 
and
\begin{eqnarray}\label{eq:convergenzaQOinduzAdue}
&&\nu_{2}^{(h)}(t)\rightarrow f_2^{(h)}(t)=\sum_{0\leq q\leq h}f^{(q)}(t)f^{(h-q)}(t),\ \text{as}\ \ N\to\infty,\ \  \mathcal{C}_{b}^{\infty}-\text{weakly},
%&&\text{for any configuration $Z_N$ which is typical with respect to $W_{N,0}^{(0)}$, }\nonumber\\
\end{eqnarray}
for any configuration $Z_N$ which is typical with respect to $f_0^{(0)}$.
%$W_{N,0}^{(0)}$.\\ 
%(we note that in (\ref{eq:convergenzaQOinduzAdue}) we made use of the structure of %$f_2^{(h)}(t)$). 

Now we want to prove that (\ref{eq:convergenzaQOinduz}) and (\ref{eq:convergenzaQOinduzAdue})
%, assuming the convergences of the one-particle function and of the two-particles one for $k=0$ and for any $h<k$, they 
hold also for $h=k$.\\
Thanks to (\ref{eq:convergenzaQOinduz}), we can affirm that:
\begin{eqnarray}\label{eq:convergenzaQOinduzFINE1}
&&\sum_{0<\ell<k}\left(\nabla_{x'_1} \phi \ast \nu_{1}^{(\ell)}\right) \cdot \nabla_{v'_1} \nu_{1}^{(k-\ell)}\to\sum_{0<\ell<k}\left(\nabla_{x'_1} \phi \ast f^{(\ell)}\right) \cdot \nabla_{v'_1} f^{(k-\ell)}=\sum_{0<\ell<k}T^{(0)}_{\ell}f^{(k-\ell)},\nonumber\\
&&\nonumber\\
&&\mathcal{C}_{b}^{\infty}-\text{weakly},
\end{eqnarray}
and, thanks to (\ref{eq:convergenzaQOinduzAdue}), have:
\begin{eqnarray}\label{eq:convergenzaQOinduzFINE2}
&& \sum_{\substack{0<r_1 \leq k\\r_1\ \text{even}}}(-1)^{r_1/2}c_{r_1}\int \ud x'_2  \ud v'_2   D_{x'_1}^{r_1 +1}
% \sum_{\alpha :\vert \alpha\vert =r_1 +1} \frac{\pa^{r_1 +1}}{\pa_{x_1}^{\alpha_1}\dots \pa_{x_d}^{\alpha_d}} 
\phi (x'_1-x'_2)  \cdot D_{v'_1}^{r_1 +1}
%\frac{\pa^{r_1 +1}}{\pa_{v_1}^{\alpha_1}\dots \pa_{v_d}^{\alpha_d}} 
\nu_{2}^{(k-r_1)}(x'_1,v'_1,x'_2,v'_2;t)\nonumber\\
%\Theta_N^{(k)}(t)\nonumber\\
&&\nonumber\\
&&\downarrow\ \ \ \mathcal{C}_{b}^{\infty}-\text{weakly}\nonumber\\
&&\nonumber\\
&&  \sum_{\substack{0<r_1 \leq k\\r_1\ \text{even}}}(-1)^{r_1/2}c_{r_1}\int \ud x'_2  \ud v'_2   D_{x'_1}^{r_1 +1}
% \sum_{\alpha :\vert \alpha\vert =r_1 +1} \frac{\pa^{r_1 +1}}{\pa_{x_1}^{\alpha_1}\dots \pa_{x_d}^{\alpha_d}} 
\phi (x'_1-x'_2)  \cdot D_{v'_1}^{r_1 +1}
%\frac{\pa^{r_1 +1}}{\pa_{v_1}^{\alpha_1}\dots \pa_{v_d}^{\alpha_d}} 
f_{2}^{(k-r_1)}(x'_1,v'_1,x'_2,v'_2;t)=\nonumber\\
&&=\sum_{\substack{0<r_1 \leq k\\r_1\ \text{even}}}\ \sum_{0\leq q\leq k-r_{1}}(-1)^{r_1/2}c_{r_1}\int \ud x'_2 \ud v'_2  D_{x'_1}^{r_1 +1}
%\sum_{\alpha :\vert \alpha\vert =r_1 +1} \frac{\pa^{r_1 +1}}{\pa_{x_1}^{\alpha_1}\dots \pa_{x_d}^{\alpha_d}}  
\phi (x'_1-x'_2)f^{(k-r_{1})}(x'_2,v'_2;t) \cdot  D_{v'_1}^{r_1 +1}
%\frac{\pa^{r_1 +1}}{\pa_{v_1}^{\alpha_1}\dots \pa_{v_d}^{\alpha_d}} 
f^{(q)}(x'_1,v'_1;t)=\nonumber\\
&&=\sum_{\substack{0<r_1\leq k\\r_1\ \text{even}}}\ \sum_{0\leq q\leq k-r_{1}}T_{q}^{(r_1)}f^{(k-r_1-q)}(t).\nonumber\\
&& 
\end{eqnarray}

At the end, putting together (\ref{eq:convergenzaQOinduzFINE1}) and (\ref{eq:convergenzaQOinduzFINE2}), we find that the sum of the source terms in equation (\ref{eq: eqPERnuDEF}) converges $\mathcal{C}_{b}^{\infty}$-weakly to:
\begin{eqnarray}\label{eq:sorgenteNUOVA}
&&\sum_{\substack{0<\ell<k}}T^{(0)}_{\ell}f^{(k-\ell)}+\sum_{\substack{ 0<r_{1}\leq k\\ 0 \leq q\leq k-r_{1}}}T_{q}^{(r_1)}f^{(k-r_1-q)},
\end{eqnarray}
which plays the role of $\Theta$ in Proposition 6.2 and it is easy to check that it is in $\mathcal{C}^0\(L^1(\R^3\times\R^3), \mathbb{R}^+\)$.
%and it is properly the source term we find in the equation for $f^{(k)}$.\\\\
Therefore, we can apply Proposition 6.2 claiming that, for any typical configuration $Z_N$ with respect to $f_0^{(0)}$,
%$W_{N,0}^{(0)}$, 
$\nu_1^{(k)}(t)$ converges $\mathcal{C}_{b}^{\infty}$-weakly to the solution of the problem (\ref{eq: Prop6.1}).
%%, whose well-posedeness is established by Proposition 6.1, with $h$ replaced by $f^{(0)}$, initial datum given by $f_0^{(k)}$ and source terms given by (\ref{eq:sorgenteNUOVA}). At this point, l
Looking at (\ref{eq: WigLIOUseq}) and (\ref{eq: Teta}), we realize that we obtained the equation satisfied by $f^{(k)}(t)$.
% namely, the one-particle convergence is proven. 
%%%\\Furthermore, we have seen that Proposition 6.1, together with our choice of the initial datum and to the good properties of the Vlasov flow, guarantees the well-posedeness in $L^1(\R^3\times\R^3)$ of the initial value problems associated with the coefficients $f^{(k)}(t)$.\\

In order to ''close'' the recurrence procedure, it remains to show the two-particle convergence at order $k$. It follows from the one-particle analysis and from the following computation
% as regard to $\eta_2$ 
(see (\ref{eq: aDueCorpi})):
\begin{eqnarray}\label{eq: eta2BIS}
&&\eta_{2}(z'_1,z'_2;t,r,{\bf \underline{r}}_n,{\bf \underline{t}}_n,Z_N)=\nonumber\\
&&=\sum_{0\leq \ell\leq k}\sum_{0\leq m\leq n}\sum_{\substack{I:I\subseteq I_n\\ \vert I\vert=m}}\eta_{1}(z'_1;t,\ell,{\bf \underline{r}}_I,{\bf \underline{t}}_I,Z_N)\eta_{1}(z'_2;t, k-\ell,{\bf \underline{r}}_{I_n\setminus I},{\bf \underline{t}}_{I_n\setminus I},Z_N)+R_N^2,
% \ o(1) \ \ \ \text{in the limit $N\to\infty$,}
%O\left(\frac{1}{N}\right)
\end{eqnarray}
where $R_N^2$ is a remainder arising from the action of the operator $\mathcal{D}^{2r}T_N^{(r_n)}(t_n)\dots T_N^{(r_1)}(t_1)$ on a product of two empirical measures $\mu_N(t)$. In Appendix C we will see that it is vanishing in the limit.
%thanks to the mean-field estimates (see Proposition 5.1) and to the considerations done previously as regard to the action of the operators $T_N^{(r_j)}(t_j)$ and $\mathcal{D}^{2r}$. 
As a consequence, $\nu_{2}^{(k)}$ (see (\ref{eq: senza MEDIA}) for $j=2$) is such that:
\begin{eqnarray}\label{eq:kORDdue2TRIS}
&&\nu_{2}^{(k)}(t)= \sum_{0\leq q\leq k}\nu_{1}^{(q)}(t)\nu_{1}^{(k-q)}(t)+ o(1),
%O\left(\frac{1}{N}\right).\nonumber\\
%&&\text{in the limit $N\to\infty$.}
\end{eqnarray}
in the limit $N\to\infty$. Therefore, from the inductive assumption (\ref{eq:convergenzaQOinduz}) and from the one-particle convergence at order $k$, we conclude that:
\begin{eqnarray}\label{eq:convAdueFINALE}
&&\nu_{2}^{(k)}(t)\rightarrow \sum_{0\leq q\leq k}f^{(q)}(t)f^{(k-q)}(t)=f_2^{(k)}(t),\ \text{as}\ \ N\to\infty,\  \mathcal{C}_{b}^{\infty}-\text{weakly},
\end{eqnarray}
for any configuration $Z_N$ which is typical with respect to $f_{0}^{(0)}$.
%$W_{N,0}^{(0)}$. \\
Thus, we have just proven the convergence of $\omega_{N,j}^{(k)}$ in the cases $j=1$, $j=2$.
\\\\
$\mathbf{7.2.}$ {\bf $j$-particle convergence.}\  As for $j=2$, the $j$-particle convergence can be reduced by the one-particle control. Indeed by (\ref{eq: senza MEDIA}) and (\ref{eq: eta_j}) we have:
\begin{eqnarray}\label{eq:kORDj}
&&\nu_{j}^{(k)}(t)= \sum_{\substack{s_1\dots s_j\\0\leq s_m\leq k\\\sum_m s_m=k}}\prod_{m=1}^{j}\nu_{1}^{(s_m)}(t)+ R_N^j,\\
%O\left(\frac{1}{N}\right).\nonumber\\
&&\text{ with $R_N^j\to 0$ when $N\to\infty$.}\nonumber
\end{eqnarray}
Again the error term $R_N^j$ arises from the presence of products of derivatives with respect to the same variable.
In conclusion, the result we proved for $\nu_1^{(k)}(t)$, together with the estimates proven in Appendix C, is sufficient to guarantee the $\mathcal{C}^{\infty}_b$-weak convergence of $\nu_{j}^{(k)}(t)$ to $f_j^{(k)}(t)$ for any $j$ (for any typical configuration $Z_N$ with respect to %$W_{N,0}^{(0)}$
$f_0^{(0)}$), and, as a consequence, the $\mathcal{C}^{\infty}_b$-weak convergence of $\omega_{N,j}^{(k)}(t)$ is $f_j^{(k)}(t)$, for any $j$. 

The final step is to realize that this convergence does imply that for the coefficients $W_{N,j}^{(k)}(t)$, namely what is established by Theorem 6.1.

First of all, we observe that, for any test function $u$ we have:
\begin{eqnarray}\label{eq: W_N,1sufunztestSEC7}
\(u,W_{N,1}^{(k)}(t)\)&&=\int_{\R^6}\ud z_1 W_{N,1}^{(k)}(z_1;t)u(z_1)=\nonumber\\
%&&=\int_{\R^6}\ud z_1\int_{\R^{3(N-1)}\times \R^{3(N-1)}}\ud Z_{N-1}W_{N}^{(k)}(Z_N;t)u(z_1)=\nonumber\\
&&=\int_{\R^{3N}\times \R^{3N}}\ud Z_{N}W_{N}^{(k)}(Z_N;t)u(z_1)=\nonumber\\
&&=\int_{\R^{3N}\times \R^{3N}}\ud Z_{N}W_{N}^{(k)}(Z_N;t)\frac{1}{N}\sum_{l=1}^{N}u(z_l)=\nonumber\\
&&=\int_{\R^{3N}\times \R^{3N}}\ud Z_{N}W_{N}^{(k)}(Z_N;t)\(u,\mu_N\)=\(u,\omega_{N,1}^{(k)}(t)\),\nonumber\\
&&
\end{eqnarray}
where we made use of the symmetry of the coefficient $W_{N}^{(k)}(Z_N;t)$ with respect to any permutation of the variables (the computation is the same we did in Section 5 for $W_{N,1}^{(1)}(t)$).
% see (\ref{eq: W_N,1sufunztest})). 
From (\ref{eq: W_N,1sufunztestSEC7}), we can see that $W_{N,1}^{(k)}(t)$ and $\omega_{N,1}^{(k)}(t)$ are equal as distributions in $\mathcal{S}'\(\R^3\times\R^3\)$ (in particular, we can choose test functions belonging to $\mathcal{C}^{\infty}_b\(\R^3\times\R^3\)$), then the convergence of $W_{N,1}^{(k)}(t)$ is proven. Moreover, for $j\geq 2$, a straightforward computation shows that, by fixing an index $\overline{j}$, we have 
\begin{eqnarray}\label{eq: ERR4}
%&&\int_{\R^{6N}}\ud Z_N\ W_{N}^{(k)}(Z_N;t)\int_{\R^{6j}}\ud Z_j\mu_N(z_1\vert Z_N)\dots \mu_N(z_j\vert Z_N)\varphi_j(Z_j)=\nonumber\\
&&\(u_{\overline{j}},\omega_{N,\overline{j}}^{(k)}(t)\)=
%%%%%&&=\frac{1}{N^{j-1}}<W_{N,1}^{(k)}(t),u_j\vert_{Z_{j-1}\equiv z_1}>
%\int_{\R^{6}}\ud Z_{j}W_{N,1}^{(k)}(z_1;t)
%\sum_{\substack{1\leq l_1\leq N}}
%\varphi_j(z_{1},\dots,z_{1})
%%%%%+\nonumber\\
%%%%%&&+\frac{N(N-1)}{N^j}<W_{N,2}^{(k)}(t),u_j\vert_{Z_{j-2}\equiv z_1}>
%\int_{\R^{6}\times\R^6}\ud Z_{2}W_{N,2}^{(k)}(Z_2;t)
%\sum_{\substack{l_1\neq l_2\\1\leq l_1,l_2\leq N}}
%\varphi_j(z_{1},z_{2},z_{1},\dots,z_{1})
%%%%%%%%\frac{N(N-1)\dots (N-j+1)}{N^j}<W_{N,j-1}^{(k)}(t),u_j\vert_{z_j\equiv z_1}>
%\int_{\R^{6{j-1}}}\ud Z_{j-1}W_{N,j-1}^{(k)}(Z_{j-1};t)
%\sum_{\substack{l_1\dots l_{j-1}\\l_i\neq l_h\ \text{if}\ i\neq h\\1\leq l_i\leq N}}
%\varphi_j(z_{1},z_{2},\dots,z_{j-1},z_{1})
\frac{N(N-1)\dots (N-\overline{j}+1)}{N^{\overline{j}}}\(u_{\overline{j}},W_{N,\overline{j}}^{(k)}(t)\)+\frac{C_{j<\overline{j}}}{N},
%\int_{\R^{6j}}\ud Z_{j}W_{N,j}^{(k)}(Z_j;t)
%\sum_{\substack{l_1\dots l_j:\\l_i\neq l_h,\ \text{if}\ i\neq h\\1\leq l_i\leq N}}
%\varphi_j(z_{1},\dots,z_{j}),\nonumber\\
%&&
\end{eqnarray}
where $C_{j<\overline{j}}<\infty$ provided that $\(u_j,W_{N,j}^{(k)}(t)\)$ is uniformly bounded for each $j<\overline{j}$.
Then, to conclude the proof of Theorem 6.1, it is enough to use a recurrence argument.
% in fact, the convergence for $j=1$ has been proven and (\ref{eq: ERR4}), together with the convergence of $\omega_{N,j}^{(k)}(t)$ for any $j$, implies that the convergence for $j<\overline{j}$ implies the convergence for $j=\overline{j}$. Therefore, the proof of Theorem 6.1 is done.

%this fact could mean that in the proof of the convergence of $\nu_3^{(k)}$ we should do inductive assumptions on $\nu_4^{(h)}$ for any $h<k$ and, step by step, in order to prove the convergence of the $j$-particle function at order $k$, we should do inductive assumptions on the $j+1$-particles function at any order $h<k$. It would be extremely complicated and it would mean that the proof should become more and more involved going to higher order in $\var$ but, fortunately, the structure of the problem is much more simpler. In fact, if we want to prove the convergence of $\nu_j^{(k)}(t)$, $k$ fixed, we need to use the equation for the correlation function only in the case $j=1$. Thus, the two-particles function $\nu_2^{(h)}(t)$ for $h<k$ is necessarily involved and we have to control it. But for any $j\geq 2$ we do not need the equation anymore and we can obtain the convergence as a consequence of the particular structure of correlations functions.

%\addcontentsline{toc}{section}{Appendix A}
\newpage
\section*{Appendix A}
\setcounter{equation}{0}    
\def\theequation{A.\arabic{equation}}
{\bf Proof of Proposition 5.1}\\
 
To avoid inessential notational complications, we deal with the one-dimensional case. \\
By the Newton equations, we have: 
\begin{eqnarray}\label{eq:est7v}
&&\frac{\pa x_{i}(t)}{\pa v_r}=\delta_{ir}t+\int_{0}^{t}\ud s (t-s)\frac{1}{N}\sum_{ j\neq i}^N \pa_{x}F\left(x_{i}(s)-x_{j}(s)\right)\left(\frac{\pa x_{i}(s)}{\pa v_{r}}-\frac{\pa x_{j}(s)}{\pa v_{r}}\right),\\
&&\nonumber\\
&&\frac{\pa v_{i}(t)}{\pa v_{r}}=\delta_{ir} + \int_{0}^{t} \ud s\frac{1}{N}\sum_{j\neq i}^{N}\pa_{x} F\left(x_{i}(s)-x_{j}(s)\right)\left(\frac{\pa x_{i}(s)}{\pa v_r}-\frac{\pa x_{j}(s)}{\pa v_{r}}\right),\label{eq:est7vBIS}
\end{eqnarray}
where:
\begin{eqnarray}\label{eq:est1}
F=-\nabla_{x}\phi,
\end{eqnarray}
is the force associated with the potential $\phi$.

Let us analyze in detail the derivative of $x_{i}(t)$. From (\ref{eq:est7v}), we get:
\begin{eqnarray}\label{eq:estAppA}
&&
%\text{sup}_{t >0}
\max_{\substack{i,r\\t\leq T}}\left\vert \frac{\pa x_{i}(t)}{\pa v_r}\right\vert\leq C.
\end{eqnarray}
%$C$ depending on $F$ and $t$.
Inserting this estimate again in (\ref{eq:est7v}), we realize that we can obtain a better bound for $\frac{\pa v_{i}(t)}{\pa v_{r}}$ in the case $r\neq i$ (see \cite{Graffi}), namely:
\begin{eqnarray}\label{eq:estAppA1}
\left\vert \frac{\pa x_{i}(t)}{\pa v_r}\right\vert &&\leq C \int_{0}^{t}\ud s (t-s) \left\vert\frac{\pa x_{i}(s)}{\pa v_{r}}\right\vert+\\
&&+C \int_{0}^{t}\ud s (t-s)\frac{1}{N}\left\vert \frac{\pa x_{r}(s)}{\pa v_{r}}\right\vert +\nonumber\\
&&+C \int_{0}^{t}\ud s (t-s)\frac{1}{N}\sum_{ \substack{j\neq i\\j\neq r}}^N \pa_{x}F\left(x_{i}(s)-x_{j}(s)\right)\left\vert \frac{\pa x_{j}(s)}{\pa v_{r}}\right\vert.\nonumber\\
&&
\end{eqnarray}
Hence, by virtue of the Gronwall lemma, we find:
\begin{eqnarray}\label{eq:estAppA2}
&&
%\text{sup}_{t >0}
\max_{\substack{i\neq r\\ t\leq T}}\left\vert \frac{\pa x_{i}(t)}{\pa v_r}\right\vert\leq \frac{C}{N}.
\end{eqnarray}
%By (\ref{eq:estAppA}) and (\ref{eq:estAppA2}), we see that the derivative of $x_{i}(t)$ with respect to $ v_r$ is $O(1)$ if and only if $r=i$, otherwise it is $O(1/N)$. Then, we have:
%%%%Thus:
%%%%\begin{eqnarray}\label{eq:estAppA2BIS}
%%%%&&
%\text{sup}_{t >0}
%%%%\frac{1}{N}\sum_{i=1}^{N}\left\vert \frac{\pa x_{i}(t)}{\pa v_r}\right\vert\leq 
%\text{sup}_{t >0}
%%%%\(\frac{1}{N}\left\vert \frac{\pa x_{r}(t)}{\pa v_r}\right\vert
%%%%+\frac{1}{N}\sum_{i\neq r}^{N}\left\vert \frac{\pa x_{i}(t)}{\pa v_r}\right\vert\)\leq \frac{C}{N}. 
%%%%\end{eqnarray}
By (\ref{eq:est7vBIS}), we find that the same estimate holds for the derivative of $v_i(t)$ with respect to $v_r$. 
%Thus, looking at the derivative of the time evolved empirical measure tested versus a smooth function $u$, we have:
%\begin{eqnarray}\label{eq:estAppA2TRIS}
%\text{sup}_{t >0}\left\vert \(u,\frac{\pa }{\pa v_r}\mu_N(t)\)\right\vert&&=\text{sup}_{t >0}\left\vert\int \ud x\ud v\ u(x,v) \ 
%\frac{1}{N}\sum_{i=1}^{N}\frac{\pa }{\pa v_r}\(\delta(x-x_i(t))\delta(v-v_i(t))\)\right\vert= \nonumber\\
%&&= \text{sup}_{t >0}\left\vert\ \frac{1}{N}\sum_{i=1}^{N}\ \frac{\pa }{\pa v_r}u(x_i(t),v_i(t))\right\vert\leq \nonumber\\
%&&\leq \text{sup}_{t >0}\frac{1}{N}\sum_{i=1}^{N}\left\vert\frac{\pa }{\pa x}u(x,v)\vert_{\substack{x=x_i(t),\\ v=v_i(t)}}\frac{\pa x_{i}(t)}{\pa v_r}+\frac{\pa }{\pa v}u(x,v)\vert_{\substack{x=x_i(t),\\ v=v_i(t)}}\frac{\pa v_{i}(t)}{\pa v_r}\right\vert\leq\nonumber\\
%&&\leq  \text{sup}_{t >0}\frac{1}{N}\sum_{i=1}^{N}C_{u,t}\(\left\vert\frac{\pa x_{i}(t)}{\pa v_r}\right\vert+\left\vert\frac{\pa v_{i}(t)}{\pa v_r}\right\vert\)\leq \frac{C}{N}. 
%\end{eqnarray}
Analogous estimates hold for the derivatives with respect to the initial positions (see also \cite{Graffi}).\\ 
Therefore the claim of Proposition 5.1 is proven for derivatives of order one.

Now, let us consider a sequence $I:=(j_1,\dots, j_k)$ of possibly repeated indices. We show that:
\begin{eqnarray}\label{eq:estAppA3}
%\text{sup}_{t >0}
\frac{1}{N}\sum_{i=1}^N
\left\vert \frac{\pa^k x_{i}(t)}{\pa v_{j_1}\dots \pa v_{j_k}}\right\vert \leq \frac{C}{N^{d_k}},
\end{eqnarray}
where $d_k$ is the number of different indices in the sequence $j_1,\dots, j_k$.
%which are different from $i$ (counted without their multiplicity). 
We know that (\ref{eq:estAppA3}) is verified for $k=1$ (it follows directly by (\ref{eq:estAppA}) and (\ref{eq:estAppA2})), thus we prove (\ref{eq:estAppA3}) by induction on $k$. 
Denoting by:
\begin{eqnarray}\label{eq:estAppA4}
D(I):= \frac{\pa^k }{\pa v_{j_1}\dots \pa v_{j_k}},
\end{eqnarray}
estimate (\ref{eq:estAppA3}) can be rewritten as:
\begin{eqnarray}\label{eq:estAppA3bis}
%%\text{sup}_{t >0}
%\max_{i,I:\vert I\vert=k}
 \frac{1}{N}\sum_{i=1}^{N}\left\vert D(I)x_i(t)\right\vert \leq \frac{C}{N^{d_k}}.
\end{eqnarray}
By (\ref{eq:est7v}) we derive the following estimate for $ D(I)x_i(t)$:
\begin{eqnarray}\label{eq:estAppA5prima}
&&
%\frac{1}{N}\sum_{i=1}^{N}
\left\vert D(I) x_{i}(t)\right\vert \leq  \int_{0}^{t} \ud s (t-s)
%\frac{1}{N^2}
\frac{C}{N}
%\sum_{i=1}^{N}
\sum_{ j\neq i}^N 
%\pa_{x}F\left(x_{i}(s)-x_{j}(s)\right)
\left\vert D(I)\left( x_{i}(s)-x_{j}(s)\right)\right\vert +M_i(t),\nonumber\\
&&
\end{eqnarray}
where the term $M_i(t)$ can be computed from (\ref{eq:est7v}) according to the Leibniz rule. Let $\mathcal{P}_n:=\{I_1,\dots, I_n\}$ be a partition of the set $I$ of cardinality $n$, with $2\leq n\leq k$, then we have:
\begin{eqnarray}\label{eq:estAppA6prima}
&& M_i(t) \leq\int_{0}^{t} \ud s (t-s)
%\frac{1}{N^2}
\frac{1}{N}
\sum_{ j\neq i}^N \sum_{n=2}^{k}\sum_{\mathcal{P}_n}C(\mathcal{P}_n)\left\vert\prod_{H\in \mathcal{P}_n} \[D(H) \(x_{i}(s)-x_{j}(s)\right)\]\right\vert\leq\nonumber\\
&&\leq\int_{0}^{t} \ud s (t-s)\sum_{n=2}^{k}\sum_{\mathcal{P}_n}C(\mathcal{P}_n)\frac{1}{N}
\sum_{ j=1}^N \left\vert\prod_{H\in \mathcal{P}_n} \[D(H) \(x_{i}(s)-x_{j}(s)\right)\]\right\vert,
\end{eqnarray}
where $D(H):= \prod_{h\in H}\frac{\pa}{\pa v_{h}}$ and $C(\mathcal{P}_n)$ are coefficients depending on the partition $\mathcal{P}_n$ and on suitable derivatives of $F$.
By 
%(\ref{eq:estAppA6prima}) and 
(\ref{eq:estAppA5prima}), it follows that:
\begin{eqnarray}\label{eq:estAppA5}
&&
\frac{1}{N}\sum_{i=1}^{N}
\left\vert D(I) x_{i}(t)\right\vert \leq \int_{0}^{t} \ud s (t-s) 
\frac{C}{N}
\sum_{i=1}^{N} \left\vert D(I)x_{i}(s)\right\vert
%+\frac{1}{N}\sum_{j=1}^{N}\left\vert D(I)x_{j}(s)\right\vert 
+M(t),
\end{eqnarray}
where $M(t)=\frac{1}{N}\sum_{i=1}^{N}M_i(t)$ and, by (\ref{eq:estAppA6prima}), we have:
\begin{eqnarray}\label{eq:estAppA6}
&& M(t) \leq\int_{0}^{t} \ud s (t-s)\sum_{n=2}^{k}\sum_{\mathcal{P}_n}C(\mathcal{P}_n)\frac{1}{N^2}
%\frac{1}{N}
\sum_{i=1}^{N}\sum_{ j=1}^N \left\vert\prod_{H\in \mathcal{P}_n} \[D(H) \(x_{i}(s)-x_{j}(s)\right)\]\right\vert,\nonumber\\
&&
\end{eqnarray}
We observe that:
\begin{eqnarray}\label{eq:estAppA6bis}
&&\frac{1}{N^2}
%\frac{1}{N}
\sum_{i,j=1}^{N}
%\sum_{ j=1}^N 
\left\vert\prod_{H\in \mathcal{P}_n}\left[D(H) \(x_{i}(s)-x_{j}(s)\right)\]\right\vert\leq\frac{1}{N}
\sum_{i=1}^{N}\prod_{H\in\mathcal{P}_n}\left\vert D(H)x_{i}(s)\right\vert+ \nonumber\\
&&+
%(-1)^n
\frac{1}{N}\sum_{j=1}^{N}\prod_{H\in\mathcal{P}_n}\left\vert D(H)x_{j}(s)\right\vert+\nonumber\\
&&+\sum_{\mathcal{Q}\subset \mathcal{P}_n}C(\mathcal{Q})\(\frac{1}{N}
\sum_{i=1}^{N}\prod_{Q\in\mathcal{Q}}\left\vert D(Q)x_{i}(s)\right\vert\)\(\frac{1}{N}
\sum_{j=1}^{N} \prod_{J\in\mathcal{P}_n\setminus\mathcal{Q} }\left\vert D(J)x_{j}(s)\right\vert\),\nonumber\\
&&
%&&+\sum_{J,H:J\sqcup H=I }C(J,H)D(J)x_{i}(s)D(H)x_{i}(s).
\end{eqnarray}
where $\mathcal{Q}$ is any subpartition of $\mathcal{P}_n$ and $C(\mathcal{Q})$ are coefficients depending on $\mathcal{Q}$.\\
We assume that the estimate (\ref{eq:estAppA3bis}) holds for any $m\leq k-1$, namely:
\begin{eqnarray}\label{eq:estAppA7}
%%\text{sup}_{t >0}
%\max_{\substack{i,\ M\subset I:\\\vert M\vert=m}}
 \frac{1}{N}\sum_{i=1}^{N}\left\vert D(M) x_{i}(t)\right\vert \leq \frac{C}{N^{d_m}},\ \ \text{for any } M\subset I\ \text{s.t.\ } \vert M\vert=m\leq k-1,
\end{eqnarray}
where $d_m$ is the number of different indices in the sequence $M$.\\ 
%which are different from $i$ (counted without their multiplicity).\\
Indeed, if we consider a partition $\mathcal{P}_n$ of cardinality $n\geq 2$, we are guaranteed 
%by (\ref{eq:estAppA7}) 
that $\vert M\vert\leq k-1$ for each $M\in\mathcal{P}_n$.  Then, by noting that:
\begin{eqnarray}\label{eq:estAppAnuova}
 \frac{1}{N}\sum_{i=1}^{N}\prod_{H\in\mathcal{H}}\left\vert D(H) x_{i}(t)\right\vert \leq\prod_{H\in\mathcal{H}} \frac{1}{N}\sum_{i=1}^{N}\left\vert D(H) x_{i}(t)\right\vert,\ \forall\ \text{subpartition} \ \mathcal{H}\subseteq \mathcal{P}_n,
\end{eqnarray}
we can apply the inductive hypotheses (\ref{eq:estAppA7}) to estimate the derivatives of $x_i(s)$ and $x_j(s)$ appearing in (\ref{eq:estAppA6bis}). Thus, we obtain:
%in estimating (\ref{eq:estApp6bis}) and we get:
\begin{eqnarray}\label{eq:estAppA8}
%\prod_{H\in \mathcal{P}_n}\left\vert D(H) \(x_{i}(s)-x_{j}(s)\right)\right\vert &&\leq \prod_{H\in\mathcal{P}_n}\frac{C}{N^{d_h}}+ \prod_{H\in\mathcal{P}_n}\frac{C}{N^{d_h}}
 \frac{1}{N}\sum_{i=1}^{N}\prod_{H\in \mathcal{P}_n}\left\vert D(H)x_{i}(s)\right\vert&&\leq \prod_{H\in \mathcal{P}_n} \frac{1}{N}\sum_{i=1}^{N}\left\vert D(H)x_{i}(s)\right\vert\leq\nonumber\\
&& \leq\prod_{H\in \mathcal{P}_n}\frac{C}{N^{d_h}}=\frac{C}{N^{\sum d_h}}\leq \frac{C}{N^{d_k}} ,
%%+\nonumber\\
%%&&+\sum_{\mathcal{Q}\subset \mathcal{P}_n}C(\mathcal{Q})\prod_{Q\in\mathcal{Q}}\frac{C}{N^{d_q}}\prod_{J\in\mathcal{P}_n\setminus\mathcal{Q} }\frac{C}{N^{d_j}}.\nonumber\\
%&&+\sum_{J,H:J\sqcup H=I }C(J,H)D(J)x_{i}(s)D(H)x_{i}(s).
\end{eqnarray}
where $d_h$ is the number of different indices in the sequence $H$ and we used that $\sum_{H\in \mathcal{P}_n}d_h\geq d_k$.\\
% because $H\subset I$.\\
In a similar way, we find
\begin{eqnarray}\label{eq:estAppA9}
\frac{1}{N}\sum_{i=1}^{N}\prod_{Q\in \mathcal{Q}}\left\vert D(Q)x_{i}(s)\right\vert\leq \prod_{Q\in \mathcal{Q}}\frac{C}{N^{d_q}},
%\leq \frac{C}{N^{d}},
\end{eqnarray}
where $d_q$ is the number of different indices in the sequence $Q$.\\Moreover, we have:
\begin{eqnarray}\label{eq:estAppA10}
&&\frac{1}{N}\sum_{j=1}^{N}\prod_{H\in \mathcal{P}_n}\left\vert D(H)x_{j}(s)\right\vert\leq \prod_{H\in \mathcal{P}_n}\frac{C}{N^{d_h}}=\frac{C}{N^{\sum d_h}}\leq \frac{C}{N^{d_k}},
%%+
%\delta_j(I)\prod_{H\in \mathcal{P}_n}\frac{C}{N^{d_h -1}}+(1-\delta_j(I))\prod_{H\in \mathcal{P}_n}\frac{C}{N^{d_h }}\leq\nonumber\\
%&&\leq\delta_j(I)\frac{C}{N^{d-1}}+(1-\delta_j(I))\frac{C}{N^{d}},
\end{eqnarray}
and
\begin{eqnarray}\label{eq:estAppA11}
&&\frac{1}{N}\sum_{j=1}^{N}\prod_{J\in \mathcal{P}_n\setminus\mathcal{Q}}\left\vert D(J)x_{j}(s)\right\vert
%&&\leq \prod_{J\in \mathcal{P}_n}\left\vert D(J)x_{j}(s)\right\vert\leq\nonumber\\
\leq \prod_{J\in \mathcal{P}_n\setminus\mathcal{Q}} \frac{C}{N^{d_j}},
%%%%\delta_j(K)\prod_{J\in \mathcal{P}_n\setminus\mathcal{Q}} \frac{C}{N^{d_j -1}}+(1-\delta_j(K))\prod_{J\in \mathcal{P}_n\setminus\mathcal{Q}}\frac{C}{N^{d_j }},\nonumber\\
%&&\leq \delta_j(I)\frac{C}{N^{d-1}}+(1-\delta_j(I))\frac{C}{N^{d}},
\end{eqnarray}
where $d_j$ is the number of different indices in the sequence $J$.
%$K$ is the subset of $I$ for which $\mathcal{P}_n\setminus \mathcal{Q}$ is a partition and $\delta_j(\cdot)$ is the Dirac measure associated with $j$, namely $\delta_j(I)=1$ if $j\in I$ and $\delta_j(I)=0$ if $j\notin I$. 
Then, putting together (\ref{eq:estAppA9}) and (\ref{eq:estAppA11}), we find:
\begin{eqnarray}\label{eq:estAppA12}
&&\sum_{\mathcal{Q}\subset \mathcal{P}_n}C(\mathcal{Q})\(\frac{1}{N}\sum_{i=1}^{N}\prod_{Q\in\mathcal{Q}}\left\vert D(Q)x_{i}(s)\right\vert\)\(\frac{1}{N}\sum_{j=1}^{N}\prod_{J\in\mathcal{P}_n\setminus\mathcal{Q} }\left\vert D(J)x_{j}(s)\right\vert\)\leq\nonumber\\
&&\sum_{\mathcal{Q}\subset \mathcal{P}_n}C(\mathcal{Q})\prod_{Q\in\mathcal{Q}}\prod_{J\in\mathcal{P}_n\setminus\mathcal{Q} }\frac{C}{N^{d_{q}+d_j}}\leq\nonumber\\
&&\leq \sum_{\mathcal{Q}\subset \mathcal{P}_n}C(\mathcal{Q})\prod_{Q\in\mathcal{Q}}\prod_{J\in\mathcal{P}_n\setminus\mathcal{Q} }\frac{C}{N^{d_k}}\leq \frac{C}{N^{d_k}}.
%%
%%\leq \delta_j(I)\frac{C}{N^{d-1}}+(1-\delta_j(I))\frac{C}{N^{d}}.
\end{eqnarray}
In the end, we have just proven that each term in (\ref{eq:estAppA6bis}) is bounded by $\frac{C}{N^{d_k}}$. Therefore, by using this estimate in (\ref{eq:estAppA6}), we find:
\begin{eqnarray}\label{eq:estAppA13}
&& M(t) \leq\frac{C}{N^{d_k}}.
\end{eqnarray}
By (\ref{eq:estAppA13}) and (\ref{eq:estAppA5}), it follows that:
\begin{eqnarray}\label{eq:estAppA14}
&&\frac{1}{N}\sum_{i=1}^{N}\left\vert D(I) x_{i}(t)\right\vert \leq \int_{0}^{t} \ud s (t-s)
%\frac{1}{N}
\frac{C}{N}\sum_{i=1}^{N}\left\vert D(I)x_{i}(s)\right\vert
+\frac{C}{N^{d_k}}.\nonumber\\
&&
\end{eqnarray}
Therefore, by using the Gronwall lemma, we find:
\begin{eqnarray}\label{eq:estAppA15}
&&\frac{1}{N}\sum_{i=1}^{N}\left\vert D(I) x_{i}(t)\right\vert \leq\frac{C}{N^{d_k}}.
\end{eqnarray}

As regard to the derivatives of $v_i(t)$ with respect to some initial velocities $v_{j_1},\dots,v_{j_k}$, an analogous estimate holds and the proof  works in the same way.
% by computing the derivative $D(I)$ of $v_i(t)$ and by doing estimates through an induction procedure. 
Furthermore, this strategy leads to the same estimate for the derivatives of the function $\frac{1}{N}\sum_{i=1}^{N}z_i(t)$ with respect to some initial positions $x_{j_1},\dots,x_{j_k}$.

Now, thanks to the estimate we have just proven for the derivatives of the function $\frac{1}{N}\sum_{i=1}^{N}z_i(t)$, we are able to prove the claim of Proposition 5.1. In fact, we have: 
%By Proposition 5.1 we know that:
%\begin{eqnarray}\label{eq: specifico}
%&&\frac{1}{N}\sum_{i=1}^{N}\left\vert\frac{\pa^{m} z_i(t)}{\pa^{m} z_j} \right\vert\leq \frac{C}{N},
%\end{eqnarray}
%namely
%\begin{eqnarray}\label{eq: specificoII}
%&&\frac{1}{N}\sum_{i\neq j}^{N}\left\vert\frac{\pa^{m} z_i(t)}{\pa^{m} z_j} \right\vert+\frac{1}{N}\left\vert\frac{\pa^{m} z_j(t)}{\pa^{m} z_j} \right\vert\leq \frac{C}{N}.
%\end{eqnarray}
%Therefore, as for the case of derivatives of order one, we obtain (see also \cite{Graffi}):
%\begin{eqnarray}\label{eq: specificoIII}
%&&\left\vert\frac{\pa^{m} z_i(t)}{\pa^{m} z_j} \right\vert\leq C\(\delta_{ij}+\frac{1}{N}\),\ \ \text{for any}\ m
%\end{eqnarray}
%and, in general, Proposition 5.1 implies that given a sequence $(j_1\dots j_m)$ of possibly repeated indices we have:
\begin{eqnarray}\label{eq: specificoIV}
&&\frac{1}{N}\sum_{i=1}^{N}\left\vert D(I)z_i(t) \right\vert=\frac{1}{N}\sum_{\substack{i=1\\i\in D}}^{N}\left\vert D(I) z_{i}(t)\right\vert+\frac{1}{N}\sum_{\substack{i=1\\i\notin D}}^{N}\left\vert D(I)z_i(t) \right\vert\leq \frac{C}{N^{d_k}},
%$\nonumber\\
%&&\leq C\(\frac{\sum_{\ell=1}^{k}\delta_{i j_\ell}}{N^{d_m -1}}+\frac{1}{N^{d_m}}\),
\end{eqnarray}
where $D\subset I$ contains the different indices appearing in the sequence $I$. Thus, according to our previous notation, $\vert D\vert =d_k$ and we denote the elements of $D$ by $\tilde{j}_1,\dots,\tilde{j}_{d_k}$. Then by (\ref{eq: specificoIV}) we find:
\begin{eqnarray}\label{eq: correzME1}
\frac{1}{N}\sum_{i=1}^{N}\left\vert D(I)z_i(t) \right\vert&&=\frac{1}{N}
%\sum_{\substack{i\in \(\tilde{j}_1,\dots,\tilde{j}_{d_k}}\)
\left\vert D(I)z_{\tilde{j}_1}(t) \right\vert+\dots + \frac{1}{N}\left\vert D(I)z_{\tilde{j}_{d_k}}(t) \right\vert+\nonumber\\
&&+\frac{1}{N}\sum_{\substack{i=1\\i\notin D}}^{N}\left\vert D(I)z_i(t) \right\vert\leq \frac{C}{N^{d_k}},
%$\nonumber\\
%&&\leq C\(\frac{\sum_{\ell=1}^{k}\delta_{i j_\ell}}{N^{d_m -1}}+\frac{1}{N^{d_m}}\),
\end{eqnarray}
which implies
\begin{eqnarray}\label{eq: correzME2}
&&\left\vert D(I)z_i(t) \right\vert\leq  C\(\frac{\sum_{\ell=1}^{d_k}\delta_{i \tilde{j}_\ell}}{N^{d_k -1}}+\frac{1}{N^{d_k}}\),
\end{eqnarray}
or
\begin{eqnarray}\label{eq: correzME3}
&&\left\vert D(I)z_i(t) \right\vert\leq  \frac{C}{N^{d_k^{(i)}}},
\end{eqnarray}
where $d_k^{(i)}$ is the number of different indices in the sequence $I$ which are also different from $i$.
%So that, the claim of Proposition 5.1 is proven for any $k$.
\\ \spaz \hspace{1cm} \hfill $\square$ \newline
% by virtue of (\ref{eq:estAppA8}); on the contrary, for the terms involving derivatives of $x_j(s)$, by (\ref{eq:estAppA10}) and (\ref{eq:estAppA12}) we have:
%\begin{eqnarray}\label{eq:estApp13}
%&&\frac{1}{N}\sum_{j\neq i}^N \(\delta_j(I)\frac{C}{N^{d-1}}+(1-\delta_j(I))\frac{C}{N^{d}}\)=\frac{k}{N}\frac{C}{N^{d-1}}+\frac{C}{N^{d}}\leq \frac{C}{N^{d}}.\nonumber\\
%&&
%\end{eqnarray}
%CON QUESTA INDUZIONE NON CE LA FACCIO A STIMARE IL PEZZO CON $D(I)x_j(s)$ se $j\in I$!!!! QUELLO CHE PER LE DERIVATE DI ORDINE MAX GIOCA IL RUOLO DELLA STIMA UNIFORME DI ORDINE UNO NEL CASO DI DERIVATA PRIMA, E' L'ipotesi induttiva, ma sul numero di variabili diverse va fatta!!!!
%\addcontentsline{toc}{section}{Appendix B}
\section*{Appendix B}
\setcounter{equation}{0}    
\def\theequation{B.\arabic{equation}}
{\bf Proof of Proposition 6.1}\\

Let $U_h(t,s)$ be the two parameters semigroup solution of the linear problem:
\begin{equation}\label{eq: AppB1}
\left\{
\begin{aligned}
&\left(\pa_t+v\cdot \nabla_x\right)U_h(t,s)\gamma_0=\left(\nabla \phi\ast h\right)\ast \nabla_v U_h(t,s)\gamma_0,\\
&U_h(s,s)\gamma_0=\gamma_0.
\end{aligned}
\right.
\end{equation}
The solution of (\ref{eq: AppB1}) is obtained by carrying the initial datum $\gamma_0$ along the characteristic flow
\begin{equation}\label{eq: AppB2}
\left\{
\begin{aligned}
&\dot{x}=v,\\
&\dot{v}=-\nabla \phi \ast h.
\end{aligned}
\right.
\end{equation}

Next, we consider the problem
\begin{equation}\label{eq: AppB3}
\left\{
\begin{aligned}
&\left(\pa_t+v\cdot \nabla_x\right)\tilde{\gamma}=L(h)\tilde{\gamma},\\
&\tilde{\gamma}\vert_{t=0}=\gamma_0.
\end{aligned}
\right.
\end{equation}
which can be reformulated in integral form:
\begin{equation}\label{eq: AppB4}
\tilde{\gamma}(t)=U_h(t,0)\gamma_0+\int_0^t\ud s \ U_h(t,s)\left[\left(\nabla \phi\ast \tilde{\gamma}(s)\right)\cdot \nabla_v h(s)\right].
\end{equation}
The above formula can be iterated to yield the formal solution
\begin{eqnarray}\label{eq: expAppB5}
\tilde{\gamma}(x,v;t)=&& U_h(t,0)\gamma_0(x,v)+\sum_{n\geq 1}\int_0^{t}\ud t_1\int_{0}^{t_1}\ud t_2\dots \int_{0}^{t_{n-1}}\ud t_n\int \ud x_1\int \ud v_1\dots \int \ud x_n\int \ud v_n\nonumber\\
&&U_h(t,t_1)\left[\nabla_v h(x,v;t_1)\cdot \nabla_x\phi(x-x_1)\right]\nonumber\\
&&U_h(t_1,t_2)\left[\nabla_{v_1} h(x_1,v_1;t_2)\cdot \nabla_{x_1}\phi(x_1-x_2)\right]\nonumber\\
&&\dots\nonumber\\
&&U_h(t_{n-1},t_n)\left[\nabla_{v_{n-1}} h(x_{n-1},v_{n-1};t_n)\cdot \nabla_{x_{n-1}}\phi(x_{n-1}-x_n)\right]\nonumber\\
&&U_h(t_{n},0)\gamma_0(x_n, v_n).
\end{eqnarray}
%with the convention $t_0=t$. 
We remark that $U_h(t_k,t_{k+1})$ acts on the variables $x_k, v_k$ with the convention that $(x_0,v_0)=(x,v)$ and, furthermore, $U_h$ is multiplicative and preserves the $L^p(\R^3\times\R^3)$ norms $(p=1,2,\dots,\infty)$.\\
Under the assumptions of Proposition 6.1, the above series is bounded in $L^1(\R^3\times\R^3)$ by:
\begin{eqnarray}\label{eq: expUNIFBOUNDAppB6}
&&\sum_{n\geq 0}\frac{t^n}{n!}\left(\text{sup}_{\tau \in [0,t]}\left\|\nabla_v h(\tau)\right\|_{L^1(\R^3\times\R^3)}\right)^n \left\|\nabla_x \phi\right\|_{L^{\infty}(\R^3)}^n \left\|\gamma_0\right\|_{L^{1}(\R^3\times\R^3)},\nonumber\\
&&
\end{eqnarray}
which is converging for each $t$. Now, we denote by $\Sigma_h(t,s):L^1(\R^3\times\R^3)\rightarrow L^1(\R^3\times\R^3)$, the two parameters semigroup given by the series (\ref{eq: expAppB5}). Then, the solution $\gamma$ to the problem (\ref{eq: Prop6.1}) is given by:
\begin{equation}\label{eq: AppB7}
\gamma(t)=\Sigma_h(t,0)\gamma_0+\int_0^t\ud s \ \Sigma_h(t,s)\Theta(s),
\end{equation}
and, thanks to the assumption we made on $\Theta$ and to the fact that the above series (\ref{eq: expAppB5}) is converging for any $t$, we are guaranteed that $\gamma\in \mathcal{C}^0\(L^1(\R^3\times\R^3),\R^+\)$.

The $\mathcal{C}^k$ regularity of $\tilde{\gamma}(t)=\Sigma_h(t,0)\gamma_0$ follows by (\ref{eq: expAppB5}) and
%the $\mathcal{C}^k$ regularity and 
the fact that $U_h(t,t_1)$ propagates the $\mathcal{C}^k$ regularity.
\newline \spaz \hspace{1cm} \hfill $\square$ \newline

{\bf Proof of Proposition 6.2}\\

The proof consists of two steps.\\\\
\emph{Step 1)}:

Let $\gamma_N$ be as in Proposition 6.2. Then, we show that $\gamma_N$ solves the problem:
\begin{equation}\label{eq: AppB8}
\left\{
\begin{aligned}
&\left(\pa_t+v\cdot \nabla_x\right)\gamma_N=L(h)\gamma_N+\Theta_{N}',\\
&\gamma_N\vert_{t=0}=\gamma_{N,0},
\end{aligned}
\right.
\end{equation}
with
\begin{equation}\label{eq: AppB9}
\Theta_{N}'=\Theta_N +R_N,
\end{equation}
and $R_N$ is such that:
\begin{equation}\label{eq: AppB10}
R_N\rightarrow 0 ,\ \ \ \mathcal{C}_{b}^{\infty}-\text{weakly}.
\end{equation}
In proving (\ref{eq: AppB10}), the assumption {\bf ii)} on $\gamma_N$ is crucial.\\\\
\emph{Step 2)}:

By virtue of Step 1), the hypotheses we made on $\nabla_v h$ and Proposition 6.1, we find that:
\begin{equation}\label{eq: AppB11}
\gamma_N(t)=\Sigma_{h}(t,0)\gamma_{N,0}+\int_0^t\ud s \ \Sigma_{h}(t,s)\Theta_{N}'(s).
\end{equation}
Then, reminding that:\\
$\circ$ $h(t)\in \mathcal{C}_b^{\infty}(\R^3\times\R^3)$ for any $t$,\\
$\circ$ the flow $\Sigma_{h}$ propagates the $\mathcal{C}^k$ regularity,\\
%preserves the regularity features common to $\nabla_v h(t)$ (for any $t$) and to the initial datum,\\
$\circ$ $R_N\rightarrow 0 ,\ \ \ \mathcal{C}_{b}^{\infty}-\text{weakly}$,\\
and by virtue of
%(ensured by the expansion (\ref{eq: expAppB5}) and by the assumption on $h$ AGGIUNGERE), 
%for each $t$ and $s$, 
the assumptions on $\gamma_{N,0}$ and $\Theta_N$,
% and to what has been proven in Step 1) as regard to $R_N$, 
we can easily show that:
\begin{equation}\label{eq: AppB12}
\gamma_N\rightarrow\gamma,\ \text{as}\ \ N\to\infty,\ \ \mathcal{C}_{b}^{\infty}-\text{weakly},
\end{equation}
where 
\begin{equation}\label{eq: AppB13}
\gamma(t)=\Sigma_{h}(t,0)\gamma_{0}+\int_0^t\ud s \ \Sigma_{h}(t,s)\Theta(s).
\end{equation}
Therefore, we recognize that $\gamma$ solves the problem (\ref{eq: Prop6.1})
and, by virtue of Proposition 6.1, it is uniquely determined by (\ref{eq: AppB13}) and hence it is in $\mathcal{C}^0\(L^1(\R^3\times\R^3), \R^+\)$.\\\\
\emph{Proof of Step 1)}:

We have:
%We know that $\gamma_N$ solves the problem (\ref{eq: Prop6.2}). Then, it follows that:
\begin{equation}\label{eq: Prop6.2APP}
\left\{
\begin{aligned}
&\left(\pa_t+v\cdot \nabla_x\right)\gamma_N=L(h)\gamma_N+\Theta_N+L(h_N - h)\gamma_N\\
&\left. \gamma_N(x,v;t)\right\vert_{t=0}=\gamma_{N,0}(x,v),
\end{aligned}
\right.
\end{equation}
where
\begin{equation}\label{eq: AppB15}
R_N=R_N(x,v;t):=L(h_N - h)\gamma_N.
\end{equation}
We want to show that $R_N\to 0$,\  \ $\mathcal{C}_b^{\infty}$-weakly. According to the definition of the operator $L$, we have:
\begin{equation}\label{eq: AppB16}
R_N=\(\nabla_x\phi\ast (h_N -h)\)\nabla_v\gamma_N+\(\nabla_x\phi\ast \gamma_N\)\nabla_v(h_N -h),
\end{equation}
thus, we have to show that
\begin{equation}\label{eq: AppB17}
\(u,\(\nabla_x\phi\ast (h_N -h)\)\nabla_v\gamma_N\)\rightarrow 0,\ \text{as}\ \ N\to\infty,\ \ \forall\  u\in \mathcal{C}_b^{\infty}(\R^3\times\R^3),
\end{equation}
and
\begin{equation}\label{eq: AppB18}
\(u,\(\nabla_x\phi\ast \gamma_N\)\nabla_v(h_N -h)\)\rightarrow 0,\ \text{as}\ \ N\to\infty,\ \ \forall\  u\in \mathcal{C}_b^{\infty}(\R^3\times\R^3).
\end{equation}
%The proofs of (\ref{eq: AppB17}) and (\ref{eq: AppB18}) are quite similar, so we do in detail 
We show only (\ref{eq: AppB18}) in detail because (\ref{eq: AppB17}) will follow the same line.
We have:
\begin{eqnarray}\label{eq: AppB19}
&&\(u,\(\nabla_x\phi\ast \gamma_N\)\nabla_v(h_N -h)\)=\int \ud x \ud v\int \ud y\ud w\ u(x,v)\nabla_x\phi(x-y)\gamma_N(y,w;t)\cdot\nonumber\\
&&\qquad\qquad\qquad\qquad \qquad\qquad\qquad\qquad\qquad\nabla_v(h_N(x,v;t) -h(x,v;t))=\nonumber\\
&&\qquad\qquad\qquad\qquad=-\int \ud x \ud v\int \ud y\ud w\ \nabla_v u(x,v)\nabla_x\phi(x-y)\gamma_N(y,w;t)\cdot\nonumber\\
&&\qquad\qquad\qquad\qquad\qquad\qquad\qquad\qquad\qquad (h_N(x,v;t) -h(x,v;t))=\nonumber\\
&&\qquad\qquad\qquad\qquad=\int \ud x \ud v\int \ud y\ud w\ \nabla_v u(x,v)\(\nabla_x\phi\ast\gamma_N\)(x,v;t)(h-h_N)(x,v;t).\nonumber\\
&&
\end{eqnarray}

Setting
\begin{eqnarray}\label{eq: AppB20}
\zeta_N(x,v):=\nabla_v u(x,v)\int \ud y\ud w\nabla_x\phi(x-y)\gamma_N(y,w;t),
\end{eqnarray}
we can write (\ref{eq: AppB19}) as:
\begin{eqnarray}\label{eq: AppB21}
&&\(u,\(\nabla_x\phi\ast \gamma_N\)\nabla_v(h_N -h)\)=\int \ud x\ud v\ \zeta_N(x,v)(h(x,v;t) -h_N(x,v;t))=\nonumber\\
&&\qquad\qquad\qquad\quad=\int \ud x\ud v\int \ud x'\ud v'\(\zeta_N(x,v)-\zeta_N(x',v')\)P_N(x,v;x',v';t),\nonumber\\
&&
\end{eqnarray}
where $P_N$ is a coupling of $h$ and $h_N$, namely a probability density in $\R^6\times \R^6$ with marginals given by $h$ and $h_N$. Now we observe that:
\begin{eqnarray}\label{eq: AppB22}
\nabla_{x,v}\zeta_N(x,v):=\int \ud y\ud w\nabla_{x,v}\left[\nabla_v u(x,v)\nabla_x\phi(x-y)\right]\gamma_N(y,w;t),\nonumber\\
&&
\end{eqnarray}
and, thanks to the assumption {\bf ii)} we made on $\gamma_N$, we know that there exists a constant $C=C(u,\phi)>0$ such that:
\begin{eqnarray}\label{eq: AppB23}
\sup_{x,v}\left\vert \nabla_{x,v}\zeta_N(x,v)\right\vert =\left\|\nabla\zeta_N\right\|_{L^{\infty}(\R^3\times\R^3)}<C<+\infty.\nonumber\\
&&
\end{eqnarray}
Therefore, coming back to (\ref{eq: AppB21}), we find:
\begin{eqnarray}\label{eq: AppB24}
\left\vert \(u,\(\nabla_x\phi\ast \gamma_N\)\nabla_v(h_N -h)\)\right\vert&&\leq \int \ud z\int \ud z'\left\vert \zeta_N(z)-\zeta_N(z')\right\vert P_N(z;z';t)\nonumber\\
&&\leq \int \ud z\int \ud z'C\left\vert z-z'\right\vert P_N(z;z';t).\nonumber\\
&&
\end{eqnarray}
where we used the standard notation  $z=(x,v)$ and $z'=(x',v')$. Then, taking in (\ref{eq: AppB24}) the infimum over all couplings between $h$ and $h_N$, we obtain that:
\begin{eqnarray}\label{eq: AppB25}
\left\vert \(u,\(\nabla_x\phi\ast \gamma_N\)\nabla_v(h_N -h)\)\right\vert\leq C \mathcal{W}(h_N,h),
\end{eqnarray}
where, as in Section 5, $\mathcal{W}$ denotes the Wasserstein distance. 
%(see \ref{DOB}). 
But we know that the right hand side of (\ref{eq: AppB25}) goes to zero because of the assumption {\bf i)}, then we have just proven that:
\begin{eqnarray}\label{eq: AppB26}
\left\vert \(u,\(\nabla_x\phi\ast \gamma_N\)\nabla_v(h_N -h)\)\right\vert\rightarrow 0,\ \ \forall\ u\in\mathcal{C}_b^{\infty}(\R^3\times\R^3).\nonumber\\
&&
\end{eqnarray}

Analogously, we can prove that
\begin{eqnarray}\label{eq: AppB27}
\left\vert \(u,\(\nabla_x\phi\ast (h_N -h)\)\nabla_v\gamma_N\)\right\vert\rightarrow 0,\ \ \forall\ u\in\mathcal{C}_b^{\infty}(\R^3\times\R^3).\nonumber\\
&&
\end{eqnarray}
Therefore we have just proven that $R_N$ goes to zero in the $\mathcal{C}_b^{\infty}$-weak sense and the proof of Step 1) is done.\\\\
\emph{Proof of Step 2)}:

Thanks to Step 1) and to the assumption on $\nabla_v h$, we know that $\gamma_N(t)$ can be written as in (\ref{eq: AppB11}). Then, for any function $u$ in $\mathcal{C}_b^{\infty}(\R^3\times\R^3)$, we have that:
\begin{equation}\label{eq: AppB28}
\(u,\gamma_N(t)\)=\(u,\Sigma_{h}(t,0)\gamma_{N,0}\)+\int_0^t\ud s \ \(u,\Sigma_{h}(t,s)\Theta_{N}'(s)\),
\end{equation}
namely
\begin{equation}\label{eq: AppB29}
\(u,\gamma_N(t)\)=\(\left(\Sigma_{h}(t,0)\right)^* u,\gamma_{N,0}\)+\int_0^t\ud s \ \(\(\Sigma_{h}(t,s)\)^*u,\Theta_{N}'(s)\),
\end{equation}
where $\Sigma_{h}^*$ is the adjoint of $\Sigma_{h}$. We remind that the two-parameters semigroup $\Sigma_{h}(t,s)$ propagates the $\mathcal{C}^k$ regularity, provided that $\nabla_v h\in \mathcal{C}^k\(\R^3\times\R^3\)$. In particular, if $\Sigma_h$ acts on a function $u$ which is in $\mathcal{C}_b^{\infty}(\R^3\times\R^3)$ and the function $h(t)$ is supposed to be in $\mathcal{C}_b^{\infty}(\R^3\times\R^3)$ for any $t$, as it is in the assumptions of Proposition 6.2, we are clearly guaranteed that $\nabla_v h(t)$ is in $\mathcal{C}_b^{\infty}(\R^3\times\R^3)$ for any $t$, and then, $u(t):=\Sigma_h(t,0)u(x,v)$ is also in $\mathcal{C}_b^{\infty}(\R^3\times\R^3)$ for any $t$. Obviously, the same holds for $\Sigma_{h}^*$. Thus, the functions $\left(\Sigma_{h}(t,0)\right)^* u$ and $\(\Sigma_{h}(t,s)\)^*u$ appearing in (\ref{eq: AppB29}) are in $\mathcal{C}_b^{\infty}(\R^3\times\R^3)$ for any $t$. Therefore, thanks to the assumptions we made on $\gamma_{N,0}$ and $\Theta_N$, and of what we know about $R_N$, we find that:
\begin{eqnarray}\label{eq: AppB29BIS}
&&\(\left(\Sigma_{h}(t,0)\right)^* u,\gamma_{N,0}\)+\int_0^t\ud s \ \(\(\Sigma_{h}(t,s)\)^*u,\Theta_{N}'(s)\)\nonumber\\
&&\nonumber\\
&&\qquad\qquad\qquad \downarrow\ \ \ \ \ \ N\to\infty\nonumber\\
&&\nonumber\\
&&\(\left(\Sigma_{h}(t,0)\right)^* u,\gamma_{0}\)+\int_0^t\ud s \ \(\(\Sigma_{h}(t,s)\)^*u,\Theta(s)\)=\nonumber\\
&&=\( u,\Sigma_{h}(t,0)\gamma_{0}\)+\int_0^t\ud s \ \(u,\Sigma_{h}(t,s)\Theta(s)\).
\end{eqnarray}
Finally, by Proposition 6.1, we know that the expression (\ref{eq: AppB29BIS}) identifies properly the unique solution of the problem (\ref{eq: Prop6.1}) in $\mathcal{C}^0\(L^1\(\R^3\times\R^3\),\R^+\)$ and Proposition 6.2 is proven.
\newline \spaz \hspace{1cm} \hfill $\square$ \newline

%\addcontentsline{toc}{section}{Appendix C}
\section*{Appendix C}
\setcounter{equation}{0}    
\def\theequation{C.\arabic{equation}}
%%{\bf Estimate of the error term $\mathbf{E_N^1}$}\\\\
{\bf Lemma C.1:} \emph{For each time $\tau>0$, let us define the operator $\hat{T}_N^{(n)}(\tau)$ as follows:}
$$
\hat{T}_N^{(n)}(\tau):=S_N(-\tau)\hat{T}_N^{(n)}S_N(\tau).
$$

\emph{Then, for each $m\geq 0$ and for each $u\in\mathcal{C}_b^{\infty}(\R^3\times\R^3)$, there exists a constant $C>0$, not }

\emph{depending on $N$, such that:}
\begin{eqnarray}
&&\text{{\bf i)}}\left\vert \left(u,\hat{T}_N^{(r_m)}(t_m)\dots \hat{T}_{N}^{(r_1)}(t_1)\mu_N(t)\right)\right\vert<C.\ \ \label{eq: streak2O1N}
\end{eqnarray}

\emph{Moreover, we have:}
%there exists a positive constant $C$ such that:
\begin{eqnarray}
&&\text{{\bf ii)}}\left\vert \left(u,T_N^{(r_m)}(t_m)\dots T_{N}^{(r_1)}(t_1)\mu_N(t)\right)\right\vert\leq \left\vert \left(u,\hat{T}_N^{(r_m)}(t_m)\dots \hat{T}_{N}^{(r_1)}(t_1)\mu_N(t)\right)\right\vert+O\(\frac{1}{N}\).\nonumber\\
%&&\ \ \forall\ u\in\mathcal{C}_b^{\infty}(\R^3\times\R^3),\nonumber\\
&&\label{eq: streak2O1}
\end{eqnarray}
\\
{\bf Proof:}

We observe that: 
%the action of a streak of operators $\hat{T}_N^{(r_j)}(t_j)$ acting on $\mu_N(t)$ and tested versus a smooth function $u$ can be rewritten in this way:
\begin{eqnarray}\label{eq: rewrite}
&&\left(u,\hat{T}_N^{(r_m)}(t_m)\dots \hat{T}_{N}^{(r_1)}(t_1)\mu_N(t)\right)=\hat{T}_N^{(r_m)}(t_m)\hat{T}_N^{(r_{m-1})}(t_{m-1})\dots \hat{T}_N^{(r_{1})}(t_1)U(Z_N(t)),\nonumber\\
&&
%&&=S_N(-t_m)\hat{T}_N^{(r_m)}S_N(t_m-t_{m-1})\hat{T}_N^{(r_{m-1})}S_N(t_{m-1}-t_{m-2})\dots \hat{T}_N^{(r_{1})}S_N(t_{1})U(Z_N(t)),\nonumber\\
\end{eqnarray}
where:
% we denoted by $U(Z_N(t))$ the following function of the time evolved configuration $Z_N(t)$:
\begin{eqnarray}\label{eq: funzU}
U(Z_N(t)):=\(u,\mu_N(t)\)=\frac{1}{N}\sum_{\ell=1}^{N}u(z_\ell(t)).
\end{eqnarray}

We assume $m>0$ being the case $m=0$ obvious.
%%%We note that for $m=0$, we have to deal with the empirical measure only. Therefore, the claim of Lemma C.1 has already been established by Proposition 5.2. Then, let us look at the case $m>0$.\\
%Reminding the decomposition (\ref{eq: restoOPt}), we are going to show that all terms of the streak involving some operator $R_N^{(r_j)}$ are of order $1/N$, therefore, they are vanishing in the limit. On the contrary, the streak involving all operators $\hat{T}_N^{(r_j)}(t_j)$ is of order one and gives rise to the estimate ${\bf i)}$. In the same way, by differentiating the terms involving some operator $R_N^{(r_j)}$ with respect to $z_1,\dots, z_s$, we obtain quantities of size $1/N^{s+1}$, while the term involving all operators $\hat{T}_N^{(r_j)}(t_j)$ gives rise to a decay factor of order $1/N^s$. The estimate ${\bf ii)}$ follows properly from that term.\\
%The proof for the case $m=0$ follows straightforward by Proposition 5.1. On the contrary, the general case is technically more involved but the main tools are again the mean-field estimates.\\ 
\\
By using the notations:
\begin{eqnarray}\label{eq: notazSTRINGA}
&& S({\bf \underline{r}}_m,{\bf \underline{t}}_m):=T_N^{(r_m)}(t_m)\dots T_{N}^{(r_1)}(t_1)
\end{eqnarray}
and
\begin{eqnarray}\label{eq: notazSTRINGAdiag}
&&\hat{S}({\bf \underline{r}}_m,{\bf \underline{t}}_m):=\hat{T}_N^{(r_m)}(t_m)\dots \hat{T}_{N}^{(r_1)}(t_1),
\end{eqnarray}
we have (see the first term in the right hand side of (\ref{eq: opTtemposuG})):
%and reminding the definition for the operators $\hat{T}_N^{(r_j)}(t_j)$ (i.e. (\ref{eq: diagOPt})), 
%let us write the action of $\hat{S}({\bf \underline{r}}_m,{\bf \underline{t}}_m)$ on $U(Z_N(t))$:
\begin{eqnarray}\label{eq: azSTRINGAdiag}
\hat{S}({\bf \underline{r}}_m,{\bf \underline{t}}_m)U(Z_N(t))&&=\frac{C}{N^m}\sum_{j_1\dots j_m}\sum_{l_1\dots l_m}D_x^{r_m+1}\phi(x_{j_m}(t_m)-x_{l_m}(t_m))\cdot D_{v_{j_m}}^{r_m+1}(t_m)\nonumber\\
&&D_x^{r_{m-1}+1}\phi(x_{j_{m-1}}(t_{m-1})-x_{l_{m-1}}(t_{m-1}))\cdot D_{v_{j_{m-1}}}^{r_{m-1}+1}(t_{m-1})\nonumber\\
&&\dots\nonumber\\
&&D_x^{r_1+1}\phi(x_{j_1}(t_1)-x_{l_1}(t_1))\cdot D_{v_{j_1}}^{r_1+1}(t_1)U(Z_N(t)),
\end{eqnarray}
$C$ depending on ${\bf \underline{r}}_m$. By setting:
\begin{eqnarray}
&&\Phi_{j_n}(Z_N(t_n)):=\frac{1}{N}\sum_{l_n=1}^{N}D_x^{r_n+1}\phi(x_{j_n}(t_n)-x_{l_n}(t_n))\label{eq: PHI}\\
%&& \ \ \text{and}\ \ \nonumber\\
%&& \Phi'_n(Z_N(t_n)):=\frac{\pa^{r_n+1}}{\pa_{x_{l_n}}^{\vert k_1\vert}\pa_{x_{j_n}}^{\vert k_2\vert}}(t_n)\phi(x_{j_n}(t_n)-x_{l_n}(t_n)), \label{eq: PHIprimo}\\
&& \forall\ \ n=1,2,\dots, m\nonumber
\end{eqnarray}
(\ref{eq: azSTRINGAdiag}) can be rewritten as
\begin{eqnarray}\label{eq: azSTRINGAdiagII}
\hat{S}({\bf \underline{r}}_m,{\bf \underline{t}}_m)U(Z_N(t))&&=C\sum_{j_1\dots j_m}\Phi_{j_m}(Z_N(t_m))\cdot D_{v_{j_m}}^{r_m+1}(t_m)\nonumber\\
&&\Phi_{j_{m-1}}(Z_N(t_{m-1}))\cdot D_{v_{j_{m-1}}}^{r_{m-1}+1}(t_{m-1})\nonumber\\
&&\dots\nonumber\\
&&\Phi_{j_{1}}(Z_N(t_1))\cdot D_{v_{j_1}}^{r_1+1}(t_1)U(Z_N(t)).
\end{eqnarray}

We observe that, thanks to the smoothness of the potential $\phi$, $\Phi_{j_n}$ (for each $n$) is a uniformly bounded function of the configuration $Z_N$, together with its derivatives.\\
% namely, there exists a constant $C$ not depending on $N$, such that:
%\begin{eqnarray}
%&&\left\vert \Phi_n(Z_N(t_n))\right\vert<C \ \ \ \forall\ \ n=1,2,\dots, m,\label{eq: PHIstima}
%\end{eqnarray}
%moreover
%\begin{eqnarray}
%&&\left\vert D\Phi_n(Z_N(t_n))\right\vert<C \ \ \ \forall\ \ n=1,2,\dots, m,\label{eq: PHIstimaDER}
%\end{eqnarray}
%where $D$ stands for some derivative with respect to some variable $v_{j_k}$ or $v_{l_k}$.\\
Performing the derivatives in (\ref{eq: azSTRINGAdiagII}), we realize that $\hat{S}({\bf \underline{r}}_m,{\bf \underline{t}}_m)U(Z_N(t))$ is a linear combination of terms of the following type:
\begin{eqnarray}\label{eq: azSTRINGAdiagIII}
&&\sum_{j_1\dots j_m}\Phi_{j_m}(Z_N(t_m))\cdot D_{v_{j_m}}^{a_{m,1}}(t_m)\dots D_{v_{j_2}}^{a_{2,1}}(t_2)D_{v_{j_1}}^{a_{1,1}}(t_1)U(Z_N(t))\nonumber\\
&& \qquad \quad D_{v_{j_m}}^{a_{m,2}}(t_m)\dots D_{v_{j_2}}^{a_{2,2}}(t_2)\Phi_{j_1}(Z_N(t_1))\nonumber\\
&&\qquad \quad\dots\nonumber\\
&&\qquad  \quad D_{v_{j_m}}^{a_{m,m-1}}(t_{m})D_{v_{j_{m-1}}}^{a_{m-1,m-1}}(t_{m-1})\Phi_{j_{m-2}}(Z_N(t_{m-2}))\nonumber\\
&&\qquad \quad D_{v_{j_m}}^{a_{m,m}}(t_m)\Phi_{j_{m-1}}(Z_N(t_{m-1})),
\end{eqnarray}
with the constraint
\begin{equation}\label{eq: constraint}
\left\{
\begin{aligned}
& a_{1,1}=r_1 +1\\
& a_{2,1}+a_{2,2}=r_2+1\\
&\dots\\
& a_{m,1}+a_{m,2}+\dots +a_{m,m}=r_m+1.
\end{aligned}
\right.
\end{equation}

For a fixed sequence $a_{\ell,s}$, we have to compensate the divergence arising from the sum $\sum_{j_1\dots j_m}$, which is $O\(N^m\)$, by the decay of the derivatives as given by Proposition 5.1 and Proposition 5.2. Indeed we have:
\begin{eqnarray}\label{eq: correzMARIO1}
&&\left\vert D_{v_{j_m}}^{a_{m,1}}(t_m)\dots D_{v_{j_2}}^{a_{2,1}}(t_2)D_{v_{j_1}}^{a_{1,1}}(t_1)U(Z_N(t))\right\vert\leq \frac{C}{N^d},
\end{eqnarray}
where $d$ is the number of different indices in the sequence $j_1,j_2,\dots,j_m$ for which $a_{m,1},\dots, a_{2,1},a_{1,1}$ are strictly positive. Note that the fact that the derivatives are not computed at time $t=0$ but at different times $t_1,t_2,\dots,t_m$, does not change the estimate in an essential way.\\ 
An analogous estimate holds when we replace $U$ by some $\Phi_{j_s}$, namely
\begin{eqnarray}\label{eq: correzMARIO2}
&&\left\vert D_{v_{j_m}}^{a_{m,k}}(t_m)D_{v_{j_{m-1}}}^{a_{m-1,k}}(t_{m-1})\dots D_{v_{j_k}}^{a_{k,k}}(t_k)\Phi_{j_{k-1}}(Z_N(t_{k-1}))\right\vert\leq \frac{C}{N^{d_{k-1}}},
\end{eqnarray}
where $d_{k-1}$ is the number of different indices in the sequence $j_k,\dots,j_m$ which are also different from $j_{k-1}$ and from which $a_{m,k}, \dots,a_{k,k}$ are strictly positive.

As regard to the term in the sum $\sum_{j_1\dots j_m}$ in which all the indices are different (which is the only one of size $O(N^m)$), the constraints (\ref{eq: constraint}) together with estimates (\ref{eq: correzMARIO1}) and (\ref{eq: correzMARIO2}) ensure that the product of derivatives on the right hand side of (\ref{eq: azSTRINGAdiagIII}) is bounded by $1/N^m$. Thus this term is of order one.
Now for each $s=1,\dots, m-1$ consider the $\frac{m!}{s!(m-s)!}$ terms in the sum $\sum_{j_1\dots j_m}$ in which $s$ indices are equal. The sum is bounded by $N^{m-s}$. On the other hand, the constraints (\ref{eq: constraint}) together with (\ref{eq: correzMARIO1}) and (\ref{eq: correzMARIO2}) ensure that the product of derivatives on the right hand side of (\ref{eq: azSTRINGAdiagIII}) is bounded by $1/N^{m-s}$. Thus even these terms are of size one and {\bf i)} is proven.

To prove {\bf ii)} we observe that:
\begin{eqnarray}\label{eq: correzMARIO3}
&& S({\bf \underline{r}}_m,{\bf \underline{t}}_m)U(Z_N(t)) - \hat{S}({\bf \underline{r}}_m,{\bf \underline{t}}_m)U(Z_N(t))
\end{eqnarray}
can be expanded as in (\ref{eq: azSTRINGAdiag}) and (\ref{eq: azSTRINGAdiagIII}). However now we have an extra derivative, arising from the definition of $R_{N}^{(n)}$ (see (\ref{eq: OFFdiagOPt})), which yields an additional $1/N$. We omit the details of the proof which follows the same line of {\bf i)}.
\spaz \hspace{1cm} \hfill $\square$ \newline

In the same way we can also prove the following\\
\\
{\bf Lemma C.2}: \emph{For each $m\geq 0$, $k>0$ and $u\in\mathcal{C}_b^{\infty}(\R^3\times\R^3)$, there exists a constant $C>0$, not depending on $N$, such that:}
\begin{eqnarray}
&&\left\vert \mathcal{D}^{2k}S({\bf \underline{r}}_m,{\bf \underline{t}}_m)
%\hat{T}_N^{(r_m)}(t_m)\dots \hat{T}_{N}^{(r_1)}(t_1)
U(Z_N(t))
\right\vert<C.\ \ \label{eq: Dtag1}
\end{eqnarray}
\emph{where $U(Z_N(t))$ is defined as in (\ref{eq: funzU})}.
%\begin{eqnarray}
%%&&\text{{\bf ii)}}\left\vert \left(u,T_N^{(r_m)}(t_m)\dots T_{N}^{(r_1)}(t_1)\mu_N(t)\right)\right\vert\leq \left\vert \left(u,\hat{T}_N^{(r_m)}(t_m)\dots \hat{T}_{N}^{(r_1)}(t_1)\mu_N(t)\right)\right\vert+O\(\frac{1}{N}\).\nonumber\\
%%%&&\ \ \forall\ u\in\mathcal{C}_b^{\infty}(\R^3\times\R^3),\nonumber\\
%&&\label{eq: streak2O1}
%\end{eqnarray}
\\\\
{\bf Proof:}

First we look at the case $m>0$. Reminding the structure of the operator $\mathcal{D}^{2k}$ (see (\ref{eq: opDstortoGEN})), we are led to consider the term $D_{G,j}^{2s_j}\hat{S}(\underline{\mathbf{r}}_m,\underline{\mathbf{t}}_m)U(Z_N(t))$. We remind that $D_{G,j}^{2s_j}$ is a derivation operator with respect to the variable $z_j$ that acts as specified by (\ref{eq: operatorD_G}). By the expansion (\ref{eq: azSTRINGAdiagIII}) we readily arrive to the bound:
\begin{eqnarray}\label{eq: correzMARIO4}
&&\left\vert D_{G,j}^{2s_j}\hat{S}({\bf \underline{r}}_m,{\bf \underline{t}}_m)
%\hat{T}_N^{(r_m)}(t_m)\dots \hat{T}_{N}^{(r_1)}(t_1)
U(Z_N(t))
\right\vert\leq \frac{C}{N}.
\end{eqnarray}
Indeed by applying $D_{G,j}^{2s_j}$ to (\ref{eq: azSTRINGAdiagIII}) either $j\notin (j_1\dots j_m)$ so that we gain $1/N$ by the extra derivative, or $j\in (j_1\dots j_m)$ so that we reduce the sum $\sum_{j_1\dots j_m}$ by a factor $1/N$. More generally, by the same argument we find:
\begin{eqnarray}\label{eq: correzMARIO5}
&&\left\vert \prod_{j\in I}D_{G,j}^{2s_j}\hat{S}({\bf \underline{r}}_m,{\bf \underline{t}}_m)
%\hat{T}_N^{(r_m)}(t_m)\dots \hat{T}_{N}^{(r_1)}(t_1)
U(Z_N(t))
\right\vert\leq \frac{C}{N^n},
\end{eqnarray}
where $n=\vert I\vert$.\\
Finally by writing the action of the operator $\mathcal{D}^{2k}$ as in (\ref{eq: convINDATstruttura}), we obtain
\begin{eqnarray}\label{eq: correzMARIO6}
&&\left\vert \mathcal{D}^{2k} \hat{S}({\bf \underline{r}}_m,{\bf \underline{t}}_m)
%\hat{T}_N^{(r_m)}(t_m)\dots \hat{T}_{N}^{(r_1)}(t_1)
U(Z_N(t))
\right\vert\leq \sum_{n=1}^{N}\frac{N!}{n!(N-n)!}\sum_{\substack{s_1\dots s_n\\1\leq s_j\leq k\\ \sum_j s_j=k}}\frac{C}{N^n}\leq B^k\sum_{n=1}^{N}\frac{N!}{n!(N-n)!}\frac{C^n}{N^n}\leq \nonumber\\
&&\leq B^k\(1+\frac{C}{N}\)^N\leq C,
\end{eqnarray}
$B, C$ being positive constants not depending on $N$. Again $\mathcal{D}^{2k} \hat{S}({\bf \underline{r}}_m,{\bf \underline{t}}_m)U(Z_N(t))$ is the leading term of $\mathcal{D}^{2k} S({\bf \underline{r}}_m,{\bf \underline{t}}_m)U(Z_N(t))$ for the same reasons we discussed in Lemma C.1.

If $m=0$, the estimates (\ref{eq: correzMARIO4}) and (\ref{eq: correzMARIO5}) follow directly by Proposition 5.2. Thus, even in this case, the proof is concluded by (\ref{eq: correzMARIO6}).
\newline \spaz \hspace{1cm} \hfill $\square$ \newline

The fact that the error term $E_N^1$ (see (\ref{eq: eqPEReta})) and hence $E_N^2$ (see (\ref{eq: eqPERnuDEF})) are $\mathcal{C}^{\infty}_{b}$-weakly vanishing when $N\to\infty$ is an immediate consequence of the following\\\\
{\bf Lemma C.3}: \emph{
%Let $S({\bf \underline{r}}_{I},{\bf \underline{t}}_I)$ and $S({\bf \underline{r}}_{I_N\setminus I},{\bf \underline{t}}_{I_N\setminus I})$ as defined in Section 7. 
Let ${\bf \underline{r}}_{J}$ and ${\bf \underline{t}}_{J}$ be defined as in Section 7, for any $J\subset I_n$ with $I_n=\{1,2,\dots,n\}$.  For any 
%$k> 0$, 
$r
%,n$
\geq 0$ 
%such that $r+\vert {\bf \underline{r}}_{n}\vert=k$
we have:}
\begin{eqnarray}\label{eq: correzMARIO7}
&& \mathcal{D}^{2r} S({\bf \underline{r}}_n,{\bf \underline{t}}_n)\mu_N(z'_1\vert Z_N(t))\mu_N(z'_2\vert Z_N(t))=\nonumber\\
&&=\sum_{0\leq \ell \leq r}\sum_{0\leq m\leq n}\sum_{\substack{I\subset I_n\\ \vert I\vert=m}}\(\mathcal{D}^{2\ell}S({\bf \underline{r}}_{I},{\bf \underline{t}}_{I})
\mu_N(z'_1\vert Z_N(t)) \)\(\mathcal{D}^{2(r-\ell)}S({\bf \underline{r}}_{I_n\setminus I},{\bf \underline{t}}_{I_n\setminus I})\mu_N(z'_2\vert Z_N(t))\)+e_{r,N}\nonumber\\
&&
%\hat{T}_N^{(r_m)}(t_m)\dots \hat{T}_{N}^{(r_1)}(t_1)
\end{eqnarray}
\emph{where}
\begin{eqnarray}\label{eq: correzMARIO8}
&& e_{r,N}\rightarrow 0\ \ as\ N\to\infty\ \ \ \mathcal{C}_b^{\infty}-weakly.
\end{eqnarray}
\\
{\bf Proof:}

It is enough to prove (\ref{eq: correzMARIO7}) and (\ref{eq: correzMARIO8}) replacing each streak $S$ with the corresponding $\hat{S}$, being the difference $S-\hat{S}$ negligible in the limit.

We start by assuming $r=0$.
%$n>0$. 
In that case, testing the left hand side of (\ref{eq: correzMARIO7}) against a product of two test functions $u_1,u_2$, we are led to consider:
\begin{eqnarray}\label{eq: correzMARIO9}
&&\hat{S}({\bf \underline{r}}_n,{\bf \underline{t}}_n)U_1(Z_N(t))U_2(Z_N(t))
\end{eqnarray}
for which we can apply the expansion (\ref{eq: azSTRINGAdiag}).

Proceeding as in the proof of Lemma C.1 (see (\ref{eq: azSTRINGAdiagIII})), we have to consider:
\begin{eqnarray}\label{eq: correzMARIO10}
&& D_{v_{j_m}}^{a_{m,1}}(t_m)\dots D_{v_{j_2}}^{a_{2,1}}(t_2)D_{v_{j_1}}^{a_{1,1}}(t_1)U_1(Z_N(t))U_2(Z_N(t)),
\end{eqnarray}
where $a_{1,1}=r_1+1>0$. Now any contribution of the form
\begin{eqnarray}\label{eq: correzMARIO11}
&& D_{v_{j_1}}^{\alpha}(t_1)U_1(Z_N(t))D_{v_{j_1}}^{\beta}(t_1)U_2(Z_N(t)),
\end{eqnarray}
with $\alpha>0$, $\beta>0$, $\alpha+\beta=a_{1,1}$ is $O\(\frac{1}{N^2}\)$, therefore it is negligible in the limit. The same argument applies to $D_{v_{j_k}}^{a_{k,1}}(t_k)$ whenever $a_{k,1}>0$ . This means that each derivative appearing in $\hat{S}$ either applies to $\mu_N(z'_1\vert Z_N(t))$ or to $\mu_N(z'_2\vert Z_N(t))$ up to an error $e_{0,N}$ vanishing in the limit. This is exactly what (\ref{eq: correzMARIO7}) and (\ref{eq: correzMARIO8}) say for $r=0$.

For $r>0$ 
%and $n>0$
we have to apply $\mathcal{D}^{2r}$ to (\ref{eq: correzMARIO7}) (replacing $S$ by $\hat{S}$) with $r=0$. Clearly $\mathcal{D}^{2r} e_{0,N}$ vanishes in the limit. Moreover:
\begin{eqnarray}\label{eq: correzMARIO12}
&& D_{G,j}^{2s_j}\[\hat{S}({\bf \underline{r}}_I,{\bf \underline{t}}_I)U_1(Z_N(t))\hat{S}({\bf \underline{r}}_{I_n\setminus I},{\bf \underline{t}}_{I_n\setminus I})U_2(Z_N(t))\]=\nonumber\\
&&= \(D_{G,j}^{2s_j}\hat{S}({\bf \underline{r}}_I,{\bf \underline{t}}_I)U_1(Z_N(t))\)\hat{S}({\bf \underline{r}}_{I_n\setminus I},{\bf \underline{t}}_{I_n\setminus I})U_2(Z_N(t))+\nonumber\\
&&+\hat{S}({\bf \underline{r}}_I,{\bf \underline{t}}_I)U_1(Z_N(t))\(D_{G,j}^{2s_j}\hat{S}({\bf \underline{r}}_{I_n\setminus I},{\bf \underline{t}}_{I_n\setminus I})U_2(Z_N(t))\)+O\(\frac{1}{N^2}\)
\end{eqnarray}
By simple algebraic manipulation we finally arrive to (\ref{eq: correzMARIO7}) and (\ref{eq: correzMARIO8}).

%Finally, if $r>0$ and $n=0$ we have simply to deal with
%\begin{eqnarray}\label{eq: correzIO13}
%&&D_{G,j}^{2s_j}\[U_1(Z_N(t))U_2(Z_N(t))\],
%\end{eqnarray}
%then
%\begin{eqnarray}\label{eq: correzIO14}
% D_{G,j}^{2s_j}\[U_1(Z_N(t))U_2(Z_N(t))\]&&= \(D_{G,j}^{2s_j}U_1(Z_N(t))\)U_2(Z_N(t))+\nonumber\\
%&&+U_1(Z_N(t))\(D_{G,j}^{2s_j}U_2(Z_N(t))\)+O\(\frac{1}{N^2}\).
%\end{eqnarray}
%Again, by simple computations we arrive to (\ref{eq: correzMARIO7}) and (\ref{eq: correzMARIO8}).

\end{document}